# Single-Atom Tuning of Structural and Optoelectronic Properties in Halogenated Anthracene-Based Covalent Organic Frameworks

Klaudija Paliušytė[#a], Laura Fuchs[#b], Zehua Xu[a], Kuangjie Liu[a], Kornel Roztocki[c], Shuo Sun[a], Hendrik Zipse[a], Achim Hartschuh[a], Frank Ortmann[b]* and Jenny Schneider[a]*

[a]Department of Chemistry and Center for NanoScience (CeNS), University of Munich (LMU), Munich 81377, Germany
[b]Department of Chemistry, TUM School of Natural Sciences and Atomistic Modeling Center, Data Science Institute, Technische Universität München, 85748 Garching b. München, Germany
[c]Faculty of Chemistry, Adam Mickiewicz University, 61-614 Poznań, Poland

[#]K.P. and L.F. contributed equally.

## Abstract

Strategies for tuning structural and (opto-)electronic properties are fundamental to the rational design of functional materials. Here, a molecular-design approach precisely modulating the optoelectronic properties of covalent organic frameworks (COFs) through single-atom halogen substitution on π-extended anthracene linkers, is presented. Using a Wurster-type tetratopic amine (W–NH$_2$) and a series of anthracene-based dialdehydes bearing H, Cl, Br, or I at the 2-position, a family of imine-linked COFs, W-A-X (X = H, Cl, Br, I) were synthesized, all displaying well-ordered porous structure. The halogen substituent strongly influence framework formation, with brominated COFs forming substantially larger crystalline domains than their chloro- and iodo-functionalized analogues. UV-vis absorption and photoluminescence measurements reveal a systematic redshift across the series (H < Cl < Br < I), demonstrating that a single-atom modification tunes the optical response. Time dependent density functional theory calculations on both isolated fragments and extended COF models attribute these trends to halogen-induced changes in the COF band structure and enable a mechanistic understanding of how a single-atom substitution influences the optoelectronic properties of the extended π-framework. Overall, this study establishes single-atom halogen substitution as a powerful and modular tool for tailoring the structural and optical properties of anthracene-based COFs.

## Introduction

Covalent organic frameworks (COFs) are porous crystalline polymers constructed from a diverse set of molecular organic building blocks, linked through covalent bonds in a periodic arrangement.[1] This structural regularity enables precise control over their properties, resulting in a rich spectrum of functionalities. The ability to design COFs with high structural precision allows for a direct connection between their architecture and (opto-)electronic properties, establishing well-defined structure–property relationships that are critical for the development of novel functional materials. These frameworks can be designed with numerous organic molecules capable of engaging in extensive chemical interactions, further broadening their utility.[2] The choice of molecular building units is therefore pivotal in defining structural, electronic, and chemical characteristics of COFs, directly influencing their applicability in fields such as photocatalysis,[3] photovoltaics,[4] sensing,[5] and gas storage.[6]

Anthracene, an aromatic polycyclic compound composed of three linearly fused benzene rings, is a particularly compelling molecular building block for COF synthesis. In its native form, anthracene absorbs in the UV region and emits in the blue region of the visible spectrum, which limits direct applicability in visible-light-driven technologies. To fully exploit its potential in solar energy conversion systems, a redshift of the optical response is required.[7,8] This can be achieved by integrating anthracene units into extended π-conjugated frameworks such as COFs, or through peripheral chemical functionalization that preserves the integrity of the aromatic core.[9,10]

Owing to the favorable optoelectronic properties of anthracene derivatives such as strong photoluminescence,[11] efficient intramolecular charge transfer[11], and mechanochromic[12] behavior, anthracene offers significant promise as a versatile structural motif for advanced COF design. The anthracene-based COFs reported to date have been explored for catalysis,[13–15] fluorescence quenching,[16,17] sensing,[18,19] supercapacitors,[20] and optoelectronic applications.[21,22] However, the large size and inherent planarity of anthracene, together with the positioning of its functional groups, can hinder the formation of highly crystalline COFs. Consequently, combining anthracene with more flexible co-monomers may enable the synthesis of well-ordered, crystalline frameworks.

In this context, N,N,N′,N′-tetraphenyl-1,4-phenylen (Wurster-type) building blocks offer a particularly attractive option for COF synthesis due to their structural flexibility, which stems from free rotation around single bonds.[23,24] This adaptability allows them to conform to more rigid components such as anthracene, facilitating the formation of well-ordered, crystalline frameworks. Notably, the additional aromatic rings in Wurster-type building blocks enhance π-conjugation, enabling fine-tuning of the (opto-)electronic properties and expanding the design space for modular and functional COF architectures.[23,25] Specifically, the combination of electron-donating Wurster-type units with electron-accepting anthracene moieties can promote the formation of donor-acceptor COFs, a design strategy known to enhance charge carrier separation and improve the efficiency of light-induced processes.[26]

One effective strategy for modulating the properties of COFs is the integration of functional groups or atoms, among which halogen functionalization stands out as particularly impactful. The incorporation of halogens into the COF backbone can enhance charge separation and transfer efficiency owing to their electron-withdrawing characteristics[27,28] or tune the COF's electrostatic potential.[29] Furthermore, halogen-framework interactions have been reported to induce changes in both the electronic structure and the molecular geometry of the framework, making halogenation a powerful strategy for tuning properties of organic porous semiconductors.[30] However, the majority of halogenated COFs described to date are derived from small benzene-based building blocks bearing halogen substituents,[31–37] whereas examples incorporating larger, π-extended halogen units are still limited. To the best of our knowledge, halogenated anthracene units have not previously been incorporated into COFs, although even a single halogen atom is expected to significantly alter orbital energies and the resulting electronic structure of the extended framework.[38]

Here, we introduce a concept for the atom-precise tuning of structural and optoelectronic properties in COFs based on single-atom halogen substitution of π-extended anthracene linkers. By combining a Wurster-type tetratopic amine node (W–NH$_2$) with a series of anthracene-based dialdehydes functionalized at the 2-position with H, Cl, Br, or I, we construct a family of imine-linked anthracene COFs, denoted W-A-X (X = H, Cl, Br, I), that are isostructural yet electronically distinct. This systematic variation of a single substituent enables us to isolate and investigate the influence of minimal atomic-level modifications on framework formation, crystallinity, morphology, porosity, and optoelectronic response. Through a combination of optical spectroscopy, time-resolved photophysical measurements, and density functional theory (DFT) calculations on functionalized molecular building blocks, extended framework fragments, and periodic COF models, we achieve a fundamental understanding of how halogen identity modulates orbital energies, band structure in the extended π-framework. Altogether, this work establishes single-atom halogen substitution as a powerful and modular strategy for engineering the structure–property relationships of anthracene-based COFs, providing a rational design principle for tailoring porous organic semiconductors for optoelectronic and photocatalytic applications.

**Results and discussion**

**Synthesis and structural characterization**

Novel anthracene-based linkers, functionalized with different halogens at the 2-position, were synthesized following a general synthetic route outlined in Scheme S1.[39] The synthesis involves a Diels-Alder reaction between halogenated anthracene precursors (2-X-anthracene (A-X); X = Cl, Br, I) and vinylene carbonate to form a cyclic carbonate intermediate. This intermediate is subsequently converted into the corresponding diol, which is then oxidized to yield the target compound: 2-halogen-9,10-anthracenedialdehyde (A-X-CHO; X = Cl, Br, I). Details on the synthesis are provided in the SI (Figure S1 - Figure S16).

The freshly synthesized halogenated A-X-CHO linkers, along with the commercially available non-halogenated analog, were employed to construct four novel crystalline COFs (W-A-X, where X = H, Cl, Br, I) via a Schiff-base condensation reaction with the electron-rich N,N,N′,N′-tetrakis(4-aminophenyl)-1,4-phenylenediamine (W-NH$_2$) building block (see Figure 1a, Figure S17 - Figure S20). Powder X-ray diffraction (PXRD) analysis reveals well-ordered structures of all synthesized COFs (see Figure 1b-d). All four COFs exhibit prominent diffraction peaks corresponding to the (100), (110), and (210) lattice planes at similar positions. Additionally, (200) and (310) peaks were observed for W-A-H, W-A-Cl, and W-A-Br. Intense and sharp diffraction peaks for W-A-H, W-A-Cl, and W-A-Br COFs establish high crystallinity, while the W-A-I COF features lower crystallinity.

Building on the PXRD characterization, we further investigated the structure of all four W-A-X COFs using DFT simulations. For better comparison of the four models, we assumed hexagonal lattice symmetry with a Kagome lattice structure in all cases. Each unit cell comprises six mono-halogenated anthracenes and three Wurster units per layer, arranged in an eclipsed vertical stacking geometry. Due to the asymmetric structure of the functionalized anthracene building blocks, various orientations and combinations of the halogenated units relative to the smaller and larger pores are possible. Multiple halogen atom arrangements were explored (see Figure S21), yielding the most energetically favorable configuration (shown in Figure 1a), which also exhibited the best agreement with the experimental PXRD data. In this structural model, two halogen atoms per unit cell are oriented toward the smaller trigonal pore (highlighted with pink arrows), while the remaining four halogen atoms are facing the larger hexagonal pore (blue arrows). For comparison, the non-halogenated analog (W-A-H), in which halogen atoms were replaced by hydrogen, was used as a reference structure in the simulations.

Based on the structural DFT models and the PXRD patterns (Figure 1b-e, Table S1 to Table S4), we performed Pawley refinement of the four COFs, adopting P-3 symmetry for W-A-H and P-1 for W-A-X (X = Cl, Br, I), as detailed in the SI. The refined unit cell parameters are: W-A-H ($a = b = 4.236$ nm, $c = 0.411$ nm), W-A-Cl ($a = b = 4.265$ nm, $c = 0.408$ nm), and W-A-Br ($a = b = 4.263$ nm, $c = 0.409$ nm). These COFs exhibited excellent agreement between the simulated and experimental diffraction peaks with $R_p$ values of 4.66%, 5.09%, and 5.64%, respectively. All observed diffraction peaks were successfully indexed to specific lattice planes, confirming the high crystallinity of the synthesized frameworks. The W-A-I ($a = b = 4.263$ nm, $c = 0.408$ nm) COF exhibits slightly lower crystallinity compared to the other three COFs. Nevertheless, its experimental PXRD pattern shows a good agreement with the simulated pattern of the dual-pore hexagonal structural model, with a refinement factor of $R_p = 3.34\%$.

We note that minor additional reflections were observed in the PXRD patterns of the W-A-Cl and W-A-Br COFs. Specifically, diffraction peaks at 5.25° and 5.30°, respectively, could not be assigned to the simulated dual-pore hexagonal model, suggesting the presence of a minor phase impurity in these frameworks, possibly caused by the asymmetry of the halogenated anthracene linker.

To gain deeper insight into the structural properties of the synthesized COFs, transmission electron microscopy (TEM) was performed for W-A-H, -Cl, -Br COFs as representatives of the series. Hereby, the images confirm the Kagome-type hexagonal lattice. The crystalline domain sizes were found to range from 50–100 nm for W-A-H COF, up to 50 nm for W-A-Cl, and as large as 200-400 nm for W-A-Br, see Figure 1f-h. Additional TEM images of all three COFs, recorded at different regions, are shown in Figures S22–S24 and confirm the same trend in crystalline domain size. Furthermore, Figure S25 reveals the presence of a minor additional phase in Br-COF, consistent with the extra signals observed in the PXRD pattern.

The observed differences in crystalline domain sizes can be rationalized by nucleation and growth dynamics. Slower nucleation rates promote the growth of larger domains, as fewer nucleation sites are formed, allowing each crystal domain to grow more extensively. In contrast, faster nucleation rates typically lead to smaller crystallite sizes due to the formation of numerous nucleation sites, which limits individual crystal growth.[40,41] These processes can be influenced both by the synthesis conditions and the intrinsic characteristics of the four anthracene-based linkers. In our synthesis, temperature and reaction time were kept constant, while solvents were adjusted to optimize crystallinity. W-A-H and W-A-Br were obtained in benzyl alcohol/chlorobenzene, whereas W-A-Cl and W-A-I formed in benzyl alcohol/chloroform. Notably, despite identical reaction conditions, W-A-Br develops larger crystalline domains than W-A-H, indicating that linker properties play an important role in directing nucleation and determining crystallite size.

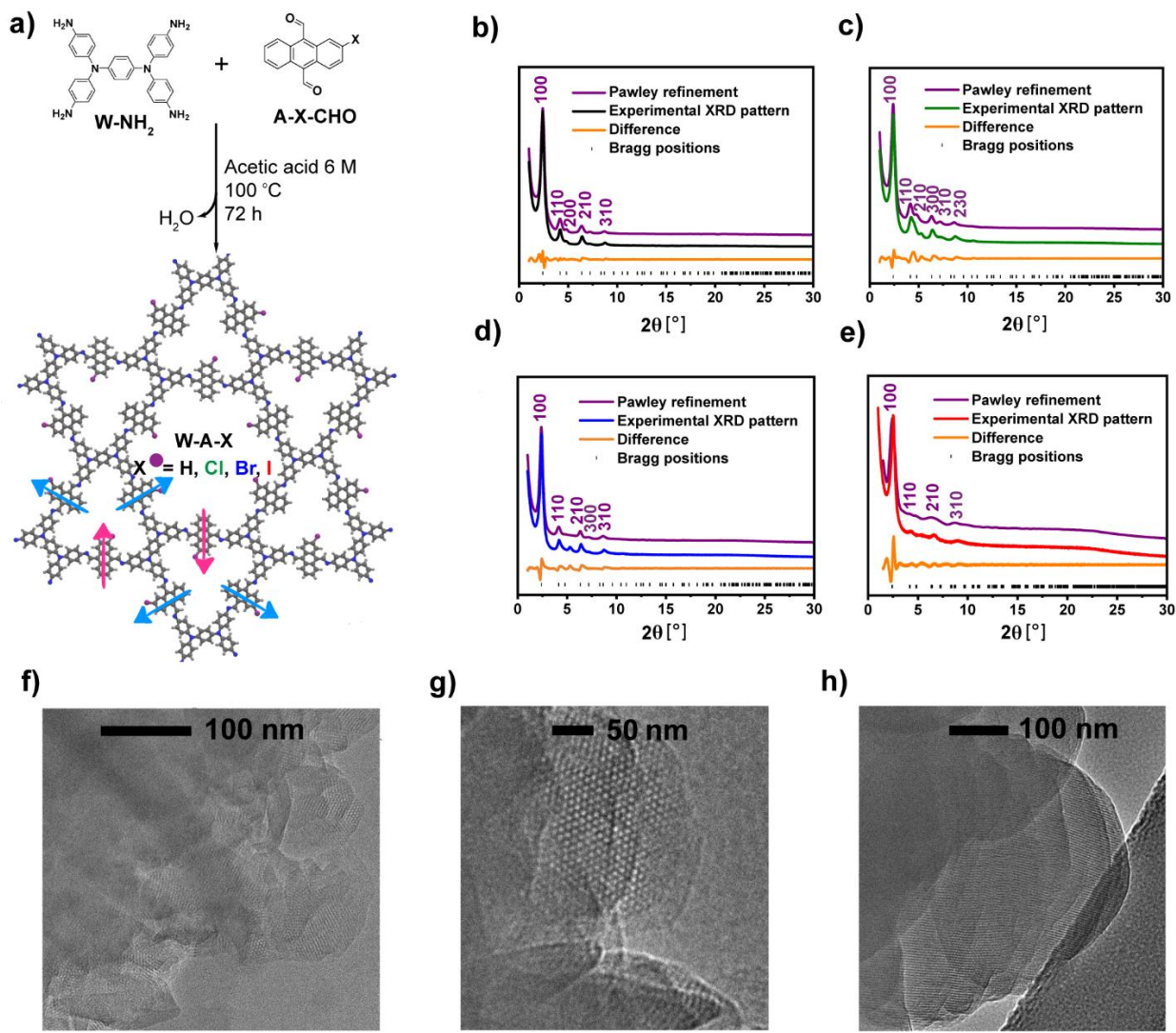

Figure 1. (a) Schematic representation of Schiff-base reaction to obtain W-A-X COFs (X = H, Cl, Br, I). Arrows indicate the orientation of the halogen atoms of the energetically most stable configuration with two halogens facing the smaller trigonal pore (pink) and the remaining four halogens directed toward the larger hexagonal pore shown in blue. Simulated and experimental

PXRD patterns of (b) W-A-H, (c) W-A-Cl, (d) W-A-Br, (e) W-A-I. TEM images of (f) W-A-H, f(g) W-A-Cl, (h) W-A-Br.

Addressing the potential impact of halogenation on COF formation, the introduction of a single halogen atom into the anthracene linker induces polarization and generates regions of distinct electrostatic potential (ESP). This effect can modulate monomer–monomer interactions during nucleation. ESP calculations for the A-X-CHO linkers (X = H, Cl, Br, I) and the W-NH$_2$ node (Figure S26) show that the halogen atoms, particularly Br and I, feature a positive σ-hole opposite the C–X bond (ESP = +0.011, +0.021, and +0.038 a.u. for Cl, Br, and I), enabling directional halogen bonding with electron-rich amino groups.[42–44] Such noncovalent interactions may help stabilize early supramolecular assemblies, slow nucleation, and promote larger crystallite growth.[45,46] As expected from their increasing polarizability (I > Br > Cl), σ-hole strength follows the same trend, thereby enhancing interactions with nucleophiles.[47] This may explain why W-A-Br forms large crystalline domains (200–400 nm), whereas W-A-l does not show a comparable effect. For W-A-I, however, the very strong σ-hole interactions may over-stabilize or redirect early-stage assemblies, slightly perturbing framework ordering and leading to reduced crystallinity.[48–51] In contrast, the moderate σ-hole of bromine appears to act as a more balanced modulator, supporting controlled interactions that favor well-ordered COF growth.[52,53]

The morphologies of the COFs were evaluated via scanning electron microscopy (SEM). SEM images demonstrate that despite the similar crystal structure, morphologies of all four COFs differ (Figure S27). The pristine W-A-H COF consists of small platelets, W-A-Cl COF contains a mixture of platelets and spherical particles, W-A-Br COF is composed mainly of spherical particles and W-A-I COF exhibits rods. The morphology of COFs can be significantly influenced by solvent polarity and electrostatic repulsion, the latter of which may arise from the presence of electronegative halogen atoms.[54,55] These factors affect layer spacing and solvation, possibly leading to different morphologies.

Nitrogen sorption isotherms were recorded to analyze the porosity of all four COFs (Figure S28). A gradual reduction in surface area is observed with increasing atomic radius of the halogen $r$ ($r$(H) < $r$(Cl) < $r$(Br) < $r$(I)): W-A-H 548 m$^2$ g$^{-1}$, W-A-Cl 490 m$^2$ g$^{-1}$, W-A-Br 187 m$^2$ g$^{-1}$ and W-A-I 170 m$^2$ g$^{-1}$. These values are considerably lower than the theoretical network-accessible surface areas per gram calculated from perfect crystal structures using PoreBlazer:[56] W-A-H, 897 m$^2$ g$^{-1}$; W-A-Cl, 806 m$^2$ g$^{-1}$; W-A-Br, 674 m$^2$ g$^{-1}$; and W-A-I, 653 m$^2$ g$^{-1}$ (Table S5). Nevertheless, the experimental pore volumes for W-A-H (0.35 cm$^3$ g$^{-1}$) and W-A-Cl (0.33 cm$^3$ g$^{-1}$) are in good agreement with the theoretical values of 0.364 cm$^3$ g$^{-1}$ and 0.317 cm$^3$ g$^{-1}$, respectively, indicating good crystallinity and stability during activation prior to the N$_2$ adsorption experiment. By contrast, W-A-Br and W-A-I COFs show both substantially lower experimental pore volumes (0.18 cm$^3$ g$^{-1}$ and 0.13 cm$^3$ g$^{-1}$, respectively) and BET surface area than their theoretical counterparts. For the iodine derivative, this is attributed to reduced crystallinity compared with the other three materials (Figure 1e). Furthermore, the activation process (vacuum drying) of COFs may cause partial pore collapse. Due to steric hindrance, this effect may be more pronounced in materials functionalized with larger atoms, such as Br and I, compared to H and Cl,[57,58] leading to a further reduction in the overall BET surface area and accessible pore volume.

Using non-local density functional theory (NLDFT)[59] for slit and cylindrical pores for evaluation of the isotherms, average pore sizes were calculated to be 1.0 nm and 1.7 nm for W-A-H, 1.1 nm and 2.0 nm for W-A-Cl, 1.1 nm and 2.0 nm for W-A-Br, and 0.9 nm and 1.5 nm for W-A-I, respectively. Additionally, porosity parameters were simulated using Zeo++ software[60] (Table S5 and Figure S29), providing theoretical pore sizes and pore volumes. The results indicated pore sizes of 0.7 nm and 2.0 nm for W-A-H, 0.7 nm and 1.9 nm for W-A-Cl, 0.7 nm and 1.9 nm for W-A-Br, and 0.7 nm and 1.8 nm for W-A-I. The deviations between the experimental and theoretical pore sizes can be attributed to the idealized nature of the theoretical models, which assume a perfectly ordered and defect-free COF structure. In reality, structural imperfections, defects, distortions, and variations in the distribution of halogen substituents can occur, leading to differences in the experimentally observed porosity parameters. Additionally, the same Zeo++ software was applied to calculate average unit cell densities, yielding 0.67 g cm$^{-3}$ for W-A-H, 0.74 g cm$^{-3}$ for W-A-Cl, 0.81 g cm$^{-3}$ for W-A-Br, and 0.90 g cm$^{-3}$ for W-A-I. The increase in unit cell density for halogenated COFs compared to the non-halogenated W-A-H COF is attributed to the incorporation of halogen atoms, which increase the overall framework density. A comparison of theoretical and experimental porosity parameters is presented in Table S5.

Fourier Transform Infrared Spectroscopy (FTIR) analysis was conducted to confirm the formation of imine bonds following the condensation reaction of the monomers (Figure S30). All four COFs show an FTIR-band at 1609 cm$^{-1}$ which is assigned to the newly formed imine (C=N) bonds.[61,62] We note that the stretching vibrations of the carbon-halogen bonds are hidden in the fingerprint region (< 600-840 cm$^{-1}$ C-Cl, < 700 cm$^{-1}$ C-Br, < 600 cm$^{-1}$ C-I),[63] therefore these vibration bands are not presented in FTIR spectra.

To further characterize the chemical structure of the four COFs, $^{13}$C cross-polarization magic angle spinning (CPMAS) analysis was performed (Figure S31). All COFs exhibited a distinct peak at 157 ppm, confirming imine bond formation. An additional peak at 93 ppm in the $^{13}$C spectrum of the W-A-I COF indicates the presence of a C-I bond. In contrast, the $^{13}$C NMR signals corresponding to carbons bonded to chlorine (123.14 ppm) and bromine (122.87 ppm) in the W-A-Cl and W-A-Br COFs, respectively, were less distinct due to overlap with other carbon resonances in the 109-140 ppm region of the spectrum.

A good thermal stability of the COFs was confirmed by thermogravimetric analysis (TGA), Figure S32, with decomposition temperatures occurring at 403 °C, 419 °C, 404 °C and 378 °C for W-A-H, W-A-Cl, W-A-Br and W-A-I COFs, respectively.

**Optical properties**

The optical properties of the COFs were examined using UV-vis diffuse reflectance absorption spectroscopy and are presented as the Kubelka-Munk function (F(R)) for solid materials.[64] The optical bandgap energies were determined using Tauc plots (direct transition model), yielding 1.70 eV for W-A-H COF, 1.66 eV for W-A-Cl, 1.62 eV for W-A-Br, and 1.61 eV for W-A-I (Figure 2a, inset). The observed decrease in the optical band gap energy correlates with the increasing size of the halogen atom. [28,31,38,65] In perovskites, this optical change has been attributed to variations in orbital participation, however, a corresponding analysis of halogen effects in COFs has not yet been reported.[66,67]

Additionally, the optical properties of the W-A-H COF were compared to the previously reported[68] Wurster-terephthalaldehyde (W-TA) COF, which is structurally similar to the anthracene moiety but lacks the additional fused benzene rings (Figure S33). UV-vis spectra revealed that the W-A-H COF exhibits a redshift of 97 nm relative to the W-TA COF, along with a significantly reduced optical gap energy (1.70 eV vs. 1.89 eV). The transition towards longer wavelengths and the concomitant reduction in bandgap can be attributed to the augmented conjugation provided by the tri-fused benzene rings present in anthracene.[13,69]

To further investigate the optical properties of the materials, photoluminescence (PL) spectroscopy was performed on the series of W-A-X COFs (Figure 2b). All anthracene-based frameworks exhibited strong emission in the red to near-infrared region, with emission maxima at 755 nm for W-A-H, 757 nm for W-A-Cl, 771 nm for W-A-Br, and 776 nm for W-A-I. Consistent with the trend observed in the UV-vis spectra, the emission profiles display a gradual redshift with increasing halogen atomic size, indicating enhanced π-conjugation and modulated electronic interactions within the framework. Additionally, PL measurements were performed for the W-TA COF to serve as a reference (Figure S34). In contrast to the anthracene-containing systems, W-TA exhibited a markedly blue-shifted emission maximum at 670 nm, consistent with its reduced conjugation.

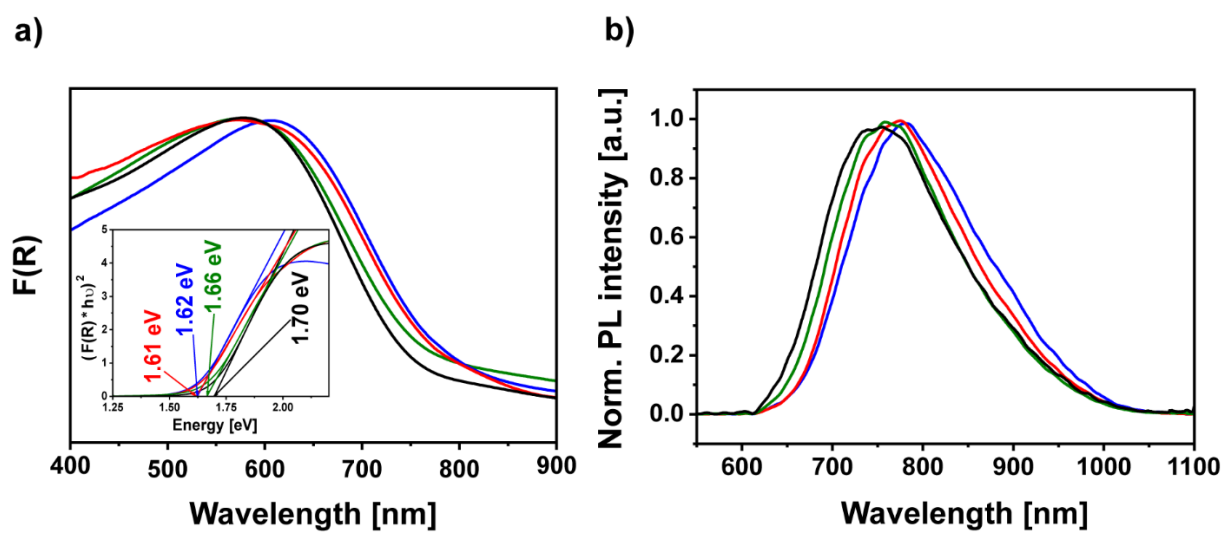

Figure 2. Optical properties of W-A-H (black), W-A-Cl (green), W-A-Br (blue), W-A-I (red): (a) F(R) and Tauc plots (inset), (b) PL spectra.

Transient PL spectroscopy measurements were performed for the W-A-X COFs to investigate the excited-state lifetimes as a function of the halogen atom present in the anthracene linker, with the W-TA COF included as a reference (Figure S35). The W-A-Cl COF exhibits a significantly longer average lifetime of 7 ps, whereas the lifetimes of the other COFs fall below the resolution limit of 3 ps. Chlorine is the most electronegative halogen among those used in these COF systems, and the extended luminescence lifetime can be attributed to its enhanced electron-withdrawing character. The combination of the A-Cl-CHO linker with an electron donor such as W-NH$_2$ enhances donor-acceptor interactions within the COF, thereby stabilizing the excited state and extending its lifetime.[70]

**DFT and TD-DFT calculations**

To understand the observed redshift in UV–vis absorption and the corresponding decrease in optical band gap energy across the W-A-X COF series (X = H, Cl, Br, I), DFT calculations were first performed on the COF models to determine their electronic structures. The band structure of W-A-H (Figure 3b) shows typical Kagome-like band features[71] along the in-plane high-symmetry path (Γ – M – K – Γ) with moderate band widths, while the band dispersions in the out-of-plane direction (Γ – A) are significantly stronger. Consistent with the strategy of tuning anthracene's (opto-)electronic properties through framework integration, the electronic band gap is observed to be relatively low, indicative of semiconducting behavior. The W-A-H COF has a direct electronic band gap of 0.85 eV. The halogenated COF derivatives exhibit the same electronic band features as well as a direct band gap of similar size (W-A-Cl: 0.85 eV, W-A-Br: 0.84 eV, W-A-I: 0.85 eV; see Figure 3e and Figure S36). However, the asymmetric halogenated anthracene moieties break the threefold symmetry that underlies the distinct Kagome-band-like features. This becomes noticeable in the band structure by a gap opening at the Dirac cone at the K point in the Brillouin zone (Figure 3a) in the halogenated COFs, as well as the emergence of small band dispersions for the former flat bands (Figure 3e, highlighted in green).

More detailed analyses of the HOMO and LUMO states show that they are highly localized on the electron-rich donor (Wurster) and electron-deficient acceptor (anthracene) fragments, as exemplarily shown in Figure 3d and 3c for W-A-H, respectively. The same localization tendencies are found in the halogenated COFs (see Figure S36), emphasizing the donor-acceptor character of all four COFs. The energetic comparison of the W-NH$_2$ and (non-)halogenated anthracene (A-X, X = H, Cl, Br, I) building blocks in the gas phase (Figure S37) confirms the profound donor character of W-NH$_2$ with an energetically high HOMO and the acceptor character of the A-X fragments (energetically low lying LUMO).

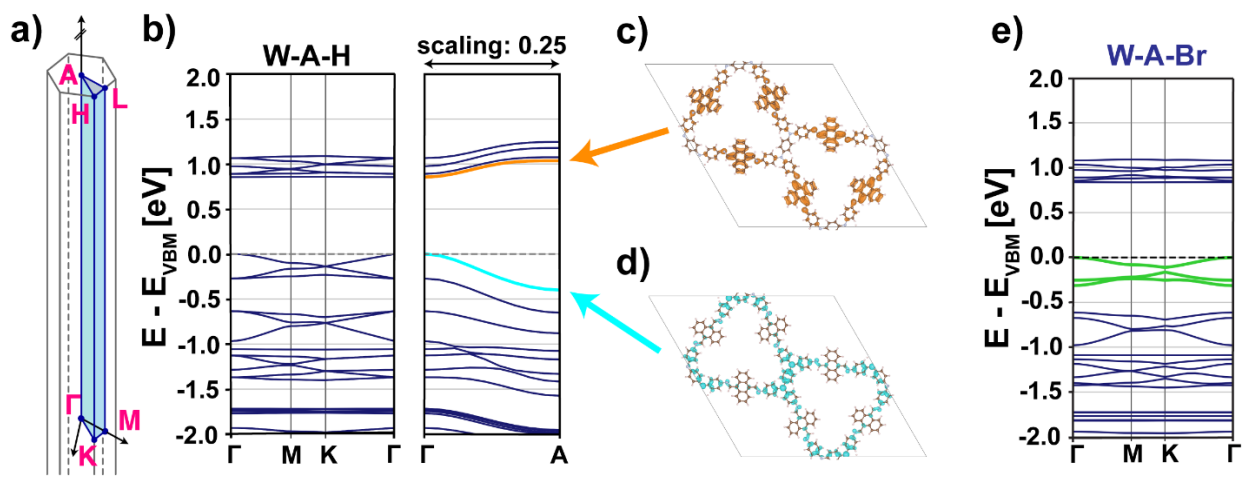

Figure 3. (a) Schematic representation of the high symmetry path in the Brillouin zone of a primitive hexagonal unit cell. (b) Electronic band structure of W-A-H with the partial charge densities of (c) LUMO band in orange and (d) HOMO band in blue. (e) Electronic band structure of W-A-Br with Kagome-like band feature highlighted in green.

After discussing the electronic band gap, we now focus on the optical band gap. Although the reduction in electronic band gaps of the W-A-X COFs compared to the COF building blocks W-$NH_2$ and A-X (Figure S37) may show an analogous trend for optical gaps, this does not allow for a quantitative prediction of the latter. Unfortunately, explicit optical simulations of the W-A-X COFs are computationally not feasible due to size limitations. Therefore, to study the impact of anthracene halogenation on the optical properties of the W-A-X COFs, we simulated the UV-vis absorption spectrum of the halogenated and non-halogenated anthracene molecules A-X (Figure 4a) with TD-DFT (more details in the experimental section of the supporting information). The spectra show a distinct redshift in the absorption maximum from anthracene (A-H) to the halogenated derivatives (A-Cl ≈ A-Br < A-I). These maxima are dominated by transitions between the HOMO and LUMO of the molecules, while other energetically close orbitals have substantially lower weights in these transitions. The energetic shift of the absorption maxima (A-Cl: - 35 meV, A-Br: - 40 meV, A-I: - 59 meV) with respect to anthracene is consistent with the observed trend in the electronic HOMO – LUMO gaps of the halogenated anthracenes with respect to A-H (A-Cl: - 31 meV, A-Br: - 34 meV, A-I: - 49 meV). To extrapolate these findings to the larger COF systems, a systematic extension of anthracene toward a combined anthracene-Wurster fragment of W-A-H, as present in the COF (see Figure 4b), was studied next. Already with phenyl substitution, we observe a redshift of the absorption maxima into the visible region (Figure 4c), accompanied by a significant lowering of the LUMO while the HOMO remains essentially unchanged (more details in the supplementary information Figure S38). A similar, though smaller, redshift trend is observed for the halogenated A-X-2(CN-Ph) fragments, (Figure S39, shifts of optical excitation energy relative to A-H-2(CN-Ph) for Cl: - 13 meV, Br: - 16 meV, I: - 27 meV), indicating that π-extension toward the Wurster fragment modulates the halogenation effect rather than simply adding to it. Based on these trends, we expect the COFs to inherit the fragment absorption properties, namely the strong redshift from π-extension and an additional, smaller halogen-induced redshift that fine-tunes the optical response of the W-A-X COFs.

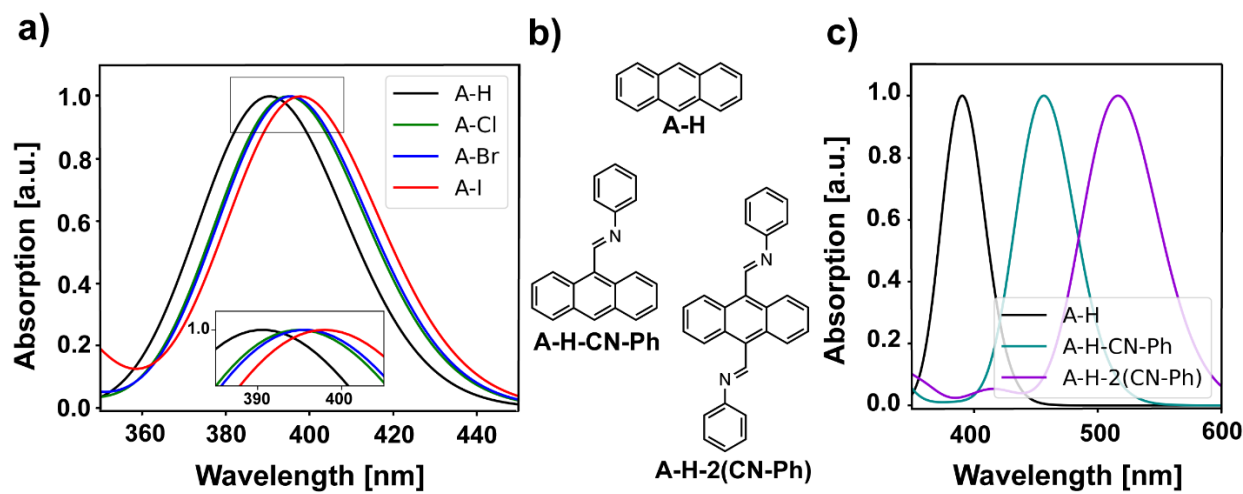

Figure 4. (a) Theoretical absorption spectrum of anthracene (A-H, black) and its halogenated derivatives (A-X, X = Cl (green), Br (blue), I (red)) with an inset showing the absorption maxima. (b) Molecular structure and (c) theoretical absorption spectrum of extended anthracene molecules A-H (black), A-H-CN-Ph (turquoise) and A-H-2(CN-Ph) (purple) simulating the extension toward a combined Wurster-anthracene fragment.

Interestingly, we note that the experimental optical band gaps of the COFs (1.60–1.70 eV) do not coincide with the calculated electronic band gaps (~0.85 eV). Typically, one would expect a significant exciton binding energy to reduce the electronic band gap toward the optical band gap. In the present case, however, the calculated electronic band gap is substantially lower than the observed optical band gap, and a negative exciton binding energy is physically implausible. To gain insight into this apparent discrepancy, we analyzed the spatial localization of the HOMO and LUMO states (see Figure 3d and c). This would indicate a charge-transfer-type transition as the lowest-energy excitation around the electronic band gap; however, such transitions are typically weaker than localized molecular-type excitations.[72] In contrast, our theoretical results from the gradual extension of the anthracene units indicate a predominantly local, anthracene-based excitation rather than a charge-transfer-type transition. Figure S40a presents an orbital-resolved analysis of the electronic bands in W-A-H, distinguishing states originating primarily from the anthracene units from those dominated by contributions of the Wurster nodes. This analysis shows that an anthracene-dominated optical transition requires about 1.91 eV. Accounting for the exciton binding energy, this value is close to and consistent with the experimental optical band gap. Similar conclusions can be drawn for the halogenated COFs (see Figure S40b-c; local halogenated-anthracenes transitions for W-A-Cl: 1.93 eV, W-A-Br: 1.93 eV, W-A-I: 1.93 eV), while the influence of the halogen atoms is more clearly discernible in the optical simulation data.

**Conclusions**

In summary, we present a new strategy for tuning the structural and optical properties of covalent organic frameworks through single-atom halogen substitution of anthracene-based linkers. By systematically synthesizing and characterizing the W-A-X COF series (X = H, Cl, Br, I), we demonstrate that even minimal molecular modifications lead to pronounced changes in crystallinity, morphology, and optical response. Larger, more polarizable halogens such as bromine promote the growth of substantially larger crystalline domains compared to the non-substituted W-A-H COF. UV–vis absorption and photoluminescence measurements reveal a clear redshift across the halogen series, in excellent agreement with DFT and TD-DFT results. Incorporation of anthracene in the framework enhances $\pi$-conjugation, directly impacting the optical response of the W-A-X COFs, while additional mono-halogenation of anthracene provides a further, finely tunable modulation of the absorption maxima. Band structure analyses confirm the donor–acceptor character of the materials and show that local anthracene-based excitations dominate the optical band gap, rather than charge-transfer transitions. In addition, the markedly longer excited-state lifetime observed for W-A-Cl COF underscores the strengthened donor–acceptor interactions imparted by the more electron-withdrawing chlorine substituent. Together, these results establish halogenation as a powerful and modular approach for fine-tuning the optoelectronic properties of COFs through atom-level design. We anticipate that this effective strategy will broaden the design space for functional framework materials and enable the development of COF-based devices with tailored optical characteristics, including photocatalysts and light-management components for next-generation optoelectronic technologies.

## Acknowledgements

We would like to thank the Deutsche Forschungsgemeinschaft for financial support [projects 511287670, 541495916 and the Cluster of Excellence e-conversion (Grant No. EXC 2089/1-390776260)]. Grants for computer time from the Leibniz Supercomputing Centre in Garching are gratefully acknowledged. We further gratefully acknowledge the computing time made available on the high-performance computer Barnard at the NHR Center TUD-ZIH. This center is jointly supported by the Federal Ministry of Education and Research and the state governments participating in the National High-Performance Computing (NHR) joint funding program (http://www.nhr-verein.de/en/our-partners). The authors would like to thank Dr. Markus Döblinger for providing TEM images and valuable discussions, and Dr. Dana D. Medina for her helpful suggestions and insightful discussions.

# Supporting Information

# Single-Atom Tuning of Structural and Optoelectronic Properties in Halogenated Anthracene-Based Covalent Organic Frameworks

Klaudija Paliušytė[#a], Laura Fuchs[#b], Zehua Xu[a], Kuangjie Liu[a], Kornel Roztocki[c], Shuo Sun[a], Hendrik Zipse[a], Achim Hartschuh[a], Frank Ortmann[b]* and Jenny Schneider[a]*

[a]Department of Chemistry and Center for NanoScience (CeNS), University of Munich (LMU), Munich 81377, Germany
[b]Department of Chemistry, TUM School of Natural Sciences, and Atomistic Modeling Center, Munich Data Science Institute, Technische Universität München, 85748 Garching b. München, Germany
[c]Faculty of Chemistry, Adam Mickiewicz University, 61-614 Poznań, Poland

[#]K.P. and L.F. contributed equally.

## 1. Experimental section

**Chemicals.** All materials were purchased from Aldrich, Fluka, Acros, Activate Scientific, or TCI Europe in the common purities purum, puriss, or reagent grade. Materials were used as received without additional purification and handled in air unless noted otherwise. All used solvents were anhydrous and purged with inert gas.

**Powder X-ray diffraction (PXRD) measurements.** PXRD measurements were performed on a Bruker D8 Discover diffractometer using Ni-filtered Cu Kα radiation and a position sensitive LynxEye detector in Bragg-Brentano geometry.

**Nitrogen sorption measurement.** Nitrogen sorption isotherms were recorded on a Quantachrome Autosorb 1 at 77 K within a pressure range from $p/p_0$ = 0.001 to 0.98. Prior to the measurement of the sorption isotherms, the samples were heated for 24 h at 120 °C under turbo-pumped vacuum. For the evaluation of the surface area the BET model was applied between 0.05 and 0.28 $p/p_0$. Pore size distributions were calculated using the NLDFT equilibrium model with a carbon kernel for slit/cylindrical pores.

The **structure models of the COFs** were constructed using the Accelrys Materials Studio software package. For W-A-H, P-3 symmetry and for W-A-Cl, W-A-Br and W-A-I COFs P-1 symmetry was applied. The structure models were optimized using the Forcite module with the Dreiding force-field. Further refinements using the Pawley method were carried out as implemented in the Reflex module of the Materials Studio software. Thompson-Cox-Hastings peak profiles were used, and peak asymmetry was corrected using the Berar-Baldinozzi method.

**Liquid state $^1$H and $^{13}$C nuclear magnetic resonance (NMR) analysis.** Liquid state NMR spectra were recorded on Bruker AV 400 and AV 400 TR spectrometers. Proton chemical shifts are expressed in parts per million ($\delta$ scale) and are calibrated using residual non-deuterated solvent peaks as internal reference (e.g. $CDCl_3$: 7.26 ppm in $^1$H NMR and 77.0 ppm in $^{13}$C NMR).

**Solid state $^{13}$C NMR analysis.** The solid state $^{13}$C cross-polarization magic angle spinning (CP/MAS) spectra were obtained on a Bruker Avance III-500 solid state NMR spectrometer with a 4 mm double resonance MAS probe and at a MAS rate of 10.0 kHz with a contact time of 2-5 ms and a pulse delay of 4 s.

**Fourier-transform infrared (FT-IR) spectra.** FT-IR measurements were performed with a Bruker Vertex 70 FTIR instrument by focusing light of a globar (silicon carbide) as MIR light source through a KBr beam splitter with integrated gold mirrors and an ATR sample stage with a Ge crystal. The spectra were recorded by an $N_2$-cooled MCT detector with a resolution of 2 cm$^{-1}$ and averaged over 1000 scans.

**Thermogravimetric analysis (TGA).** TGA measurements were performed on a Netzsch Jupiter ST 449 C instrument equipped with a Netzsch TASC 414/4 controller. The samples were heated from room temperature to 900 °C under a synthetic air flow (25 mL min$^{-1}$) at a heating rate of 10 K min$^{-1}$.

**Scanning electron microscopy (SEM) images.** SEM images were recorded with an FEI Helios NanoLab G3 UC scanning electron microscope equipped with a field emission gun operated at 3 kV. Prior to the measurements, the samples were sputtered with carbon.

**Transmission electron microscopy (TEM) images.** TEM images were recorded with an FEI Titan Themis 60 - 300 equipped with a field emission gun operated at 300 kV.

**Ultraviolet-Vis-infrared (UV-vis) absorption spectra.** The UV-vis spectra were recorded on a Perkin-Elmer Lambda 1050 spectrometer equipped with a 150 mm integrating sphere with InGaAs detector. Diffuse reflectance spectra were collected with a Praying Mantis (Harrick) accessory and were referenced to barium sulfate powder as white standard. The specular reflection of the sample surface was removed from the signal using apertures that allow only light scattered at angles > 20° to pass.

**Steady-state photoluminescence (PL) and time-correlated single-photon counting (TCSPC).** A home-built confocal laser scanning microscope (CLSM) setup was used for characterizing the photoluminescence of powder of W-A-X COFs (X = H, Cl, Br, I) and their linkers A-X-CHO (X = H, Cl, Br, I). The samples were measured in the epi-direction using an air objective (0.85 NA, Fluor 40, NIKON). A beamsplitter (MELLES GRIOT 03BTL005) and a 490 nm long-pass filter were utilized to separate the laser from the photoluminescence (PL) light. Excitation was provided by a sub-picosecond laser (iChrome TOPTICA) operating at 476 nm with repetition rate of 40 MHz. The detection system was divided into two components. The first part featured an avalanche photodiode (APD, type: MPD PDM, with a detector size of 50 × 50 μm), which was used in combination with time-correlated single-photon counting (TCSPC) electronics (BECKER UND HICKEL) to measure time-resolved PL transients. The second part comprised a spectrometer (ANDOR SHAMROCK SRi303) connected to a CCD camera (ANDOR NEWTON DU920) for capturing spectra. The data were recorded using a customized LABVIEW (National Instruments) program that integrated the manufacturers' software with our specific measurement requirements. Further data processing and analysis, including extracting PL spectra and TCSPC transients, were performed using MATLAB (MATHWORKS).

**Density functional Theory (DFT)**. Theoretical calculations of structural and electronic properties of the four COFs and their fragments were performed with Vienna Ab initio Simulation Package (VASP),[1–4] utilizing the projector-augmented wave (PAW) method[5,6] in combination with the Perdew-Burke-Ernzerhof (PBE) exchange correlation functional[7] and periodic boundary conditions. For the gas-phase calculations for all building blocks and their extended fragments, a vacuum of at least 5 Å in each direction was applied. For the COF structures, we employed the Becke-Johnson damping variant of DFT-D3[8] to correct the van der Waals (vdW) dispersion. Geometry optimization of the atomic positions and the lattice parameters of the molecular structures were optimized in an alternating fashion with multiple steps, where an energy convergence value of $10^{-6}$ eV and a kinetic energy cutoff of 400 eV (for atom relaxation) and 520 eV (for lattice relaxation, only for 2D COFs) were used.

The electronic band structure was computed along the high symmetry paths in the Brillouin zone of the primitive hexagonal unit cell. Each segment along the high symmetry paths $\Gamma - M - K - \Gamma - A$ was sampled by 10 points. To account for the opening of the band gap and to estimate the band gap at the Hybrid-DFT level (Heyd–Scuseria–Ernzerhof (HSE06) functional),[9] we employed a scissors shift[10] along the high symmetry path.

**Time-Dependent Density Functional Theory (TD-DFT).** For calculations of optical absorption data on the building blocks and the extended fragments, we used TD-DFT[11] with the HSE06 exchange correlation functional and a triple zeta basis set[12], as implemented in the Gaussian16 software package[13]. To account for spectral broadening, a Gaussian broadening with a finite width ($\sigma = 0.2$ eV) was applied. Since our focus lies on the relative, chemically induced shifts rather than absolute excitation energies, we did not apply any additional empirical offset, as is sometimes done.[14]

## 2. Linker synthesis

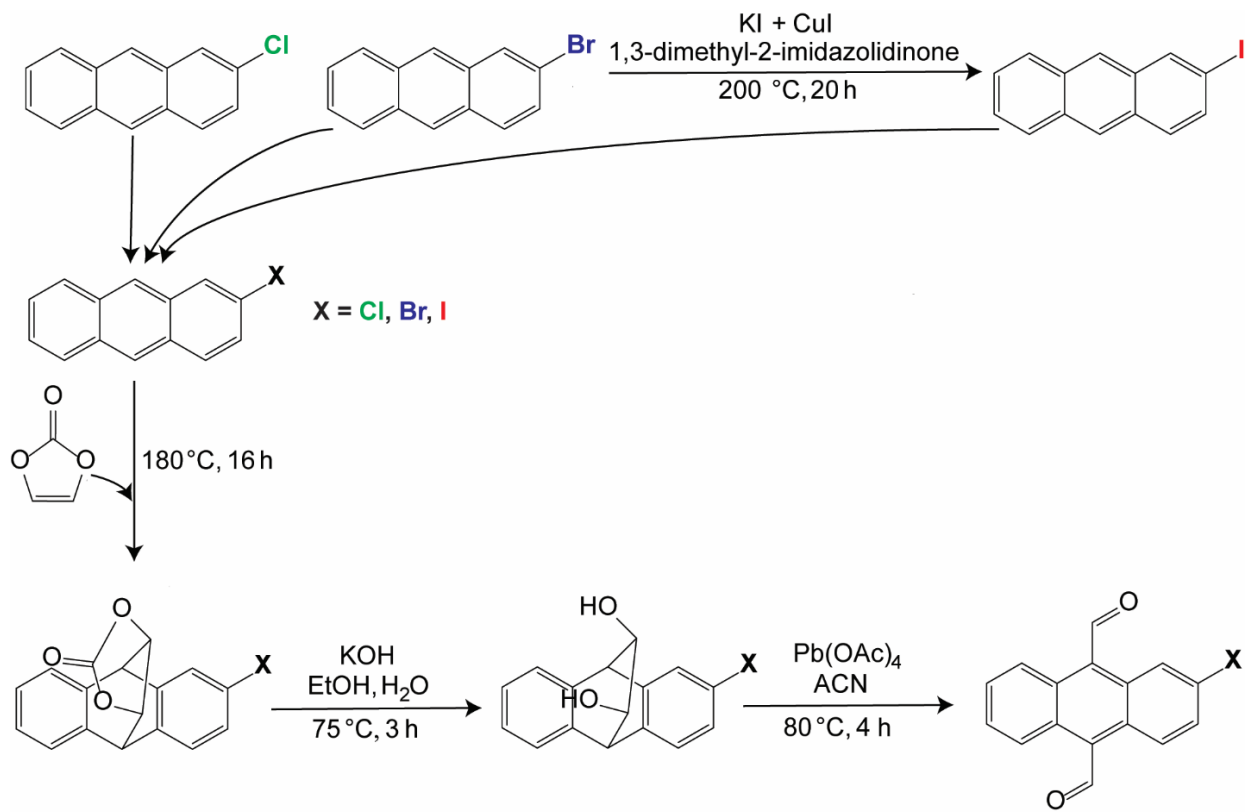

Scheme S1. Synthetic route to 2-halogen-9,10-anthracenedialdehyde (A-X-CHO, X = Cl, Br, I) linkers.

### 2.1. 2-Chloro-9,10-dihydro-9,10-[4,5]epidioxoloanthracen-13-one (A-Cl-epO)

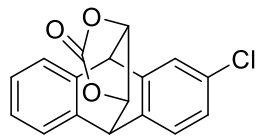

The following procedure was adapted from the previously published literature.[15] 2-Chloroanthracene (1.175 g, 5.541 mmol, 1 equiv.) and vinylene carbonate (3.484 g, 40.452 mmol, 7.3 equiv.) were heated under reflux with stirring for 18 hours, slowly forming a dark brown solution. The consumption of 2-chloroanthracene was monitored by thin-layer chromatography (CH$_2$Cl$_2$/hexane 1:49 , R=0.30). The mixture underwent rotary evaporation under high vacuum to remove the excess vinylene carbonate, providing 2-chloro-9,10-dihydro-9,10-

[4,5]epidioxoloanthracen-13-one (A-Cl-epO) as a light-brown solid (1.62 g, 5.43 mmol, 98.1 %). The product was used for the further reaction without additional purification.

$^1$H NMR (400 MHz, CDCl$_3$) δ 7.35 – 7.11 (m, 7H), 4.86 – 4.75 (m, 2H), 4.67 – 4.58 (m, 2H).

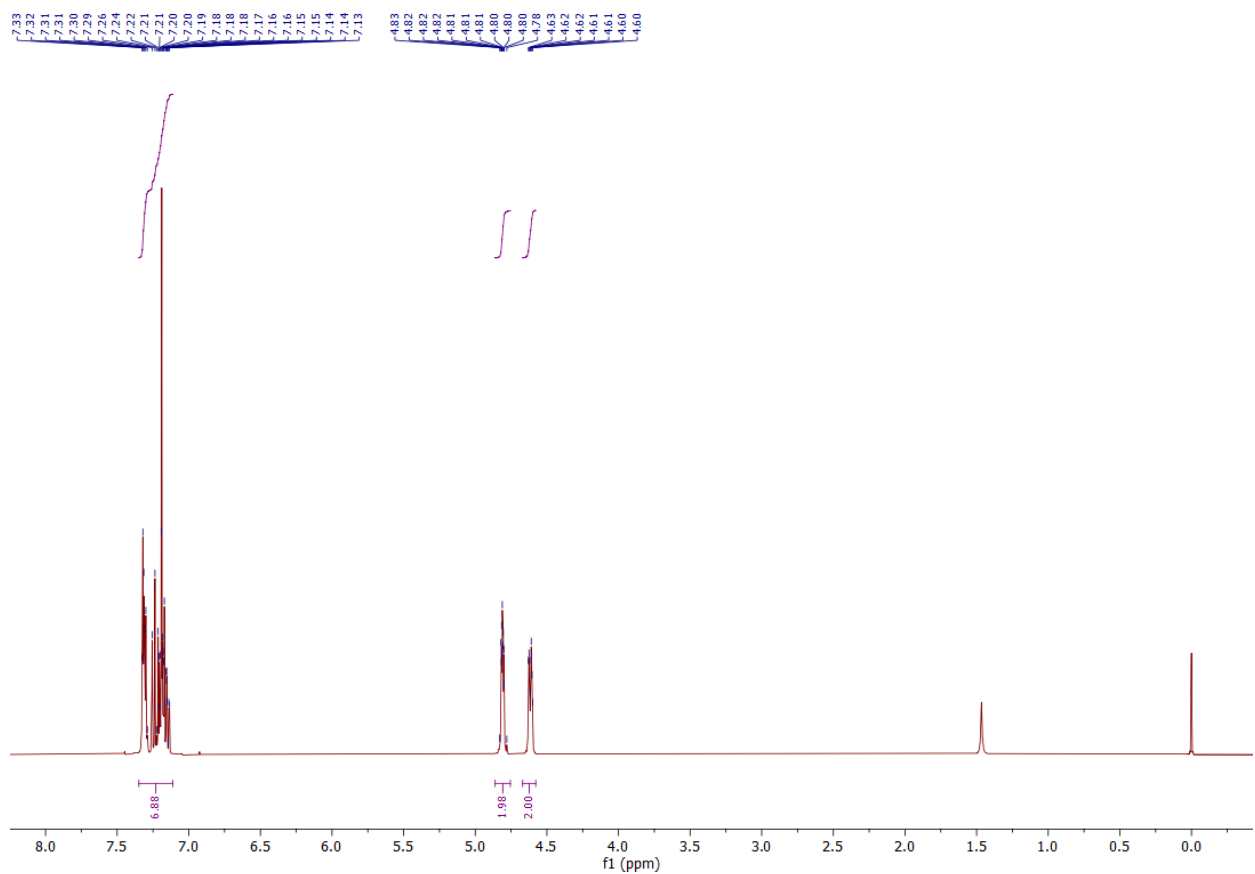

Figure S1. $^1$H NMR spectra (400 MHz in CDCl$_3$) of A-Cl-epO.

## 2.2. 2-Chloro-9,10-dihydro-9,10-ethanoanthracene-11,12-diol (A-Cl-(OH)$_2$)

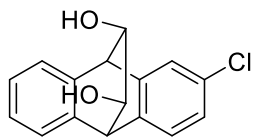

The following procedure was adapted from the previously published literature.[15] Solid potassium hydroxide (1.175 g, 5.225 mmol, 4 equiv.), deionized water (19.60 mL), and absolute ethanol (2.1 mL) were added to the A-Cl-epO (1.568 g, 5.225 mmol, 1 equiv.). The solution was stirred at 75° C for 3 hours. The consumption of the A-Cl-epO was monitored through thin-layer chromatography (100% $CH_2Cl_2$, R=0.30). Afterwards, the solution underwent rotary evaporation under reduced pressure to remove the ethanol and roughly half of the water volume. Additional water (39.2 mL) was added to the solution and the solution was stirred at room temperature for one hour, resulting in the formation of a light-tan solid. The contents were vacuum-filtered and then washed with deionized water. The vacuum-filtration receiving flask was changed and the solid was washed with ethyl acetate through the filter paper. The ethyl acetate was removed through rotary evaporation, leaving a yellow solid residue. The product was purified through column chromatography (hexane/ethyl acetate 1:1, R=0.55 & R=0.50), providing two isomers of 2-chloro-9,10-dihydro-9,10-ethanoanthracene-11,12-diol (A-Cl-$(OH)_2$) as a white solid (1.006 g, 3.70 mmol, 70.8 %).

$^1$H NMR (400 MHz,$CHCl_3$) δ (ppm): 7.30 (d, $J$ = 2.0 Hz, 1H), 7.27 – 7.19 (m, 3H), 7.15 – 7.07 (m, 3H), 4.36 – 4.29 (m, 2H), 4.04 – 3.97 (m, 2H)

$^1$H NMR (400 MHz,$CHCl_3$) δ (ppm): 7.30 (dd, $J$ = 5.4, 3.3 Hz, 2H), 7.26 – 7.04 (m, 5H), 4.33 (dd, $J$ = 7.3, 2.5 Hz, 2H), 4.03 – 3.95 (m, 2H)

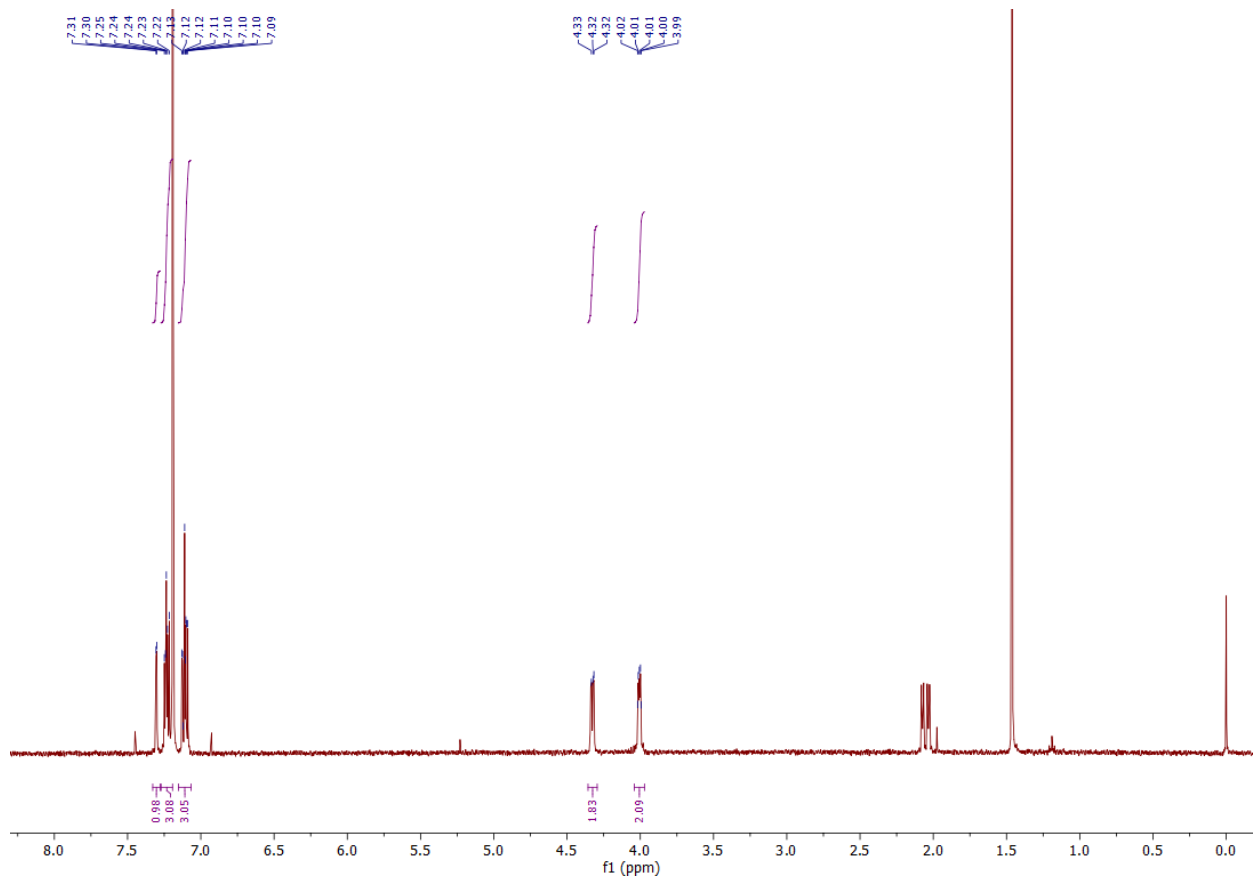

Figure S2. $^1$H NMR spectra (400 MHz in CDCl$_3$) of A-Cl-(OH)$_2$ (isomer 1).

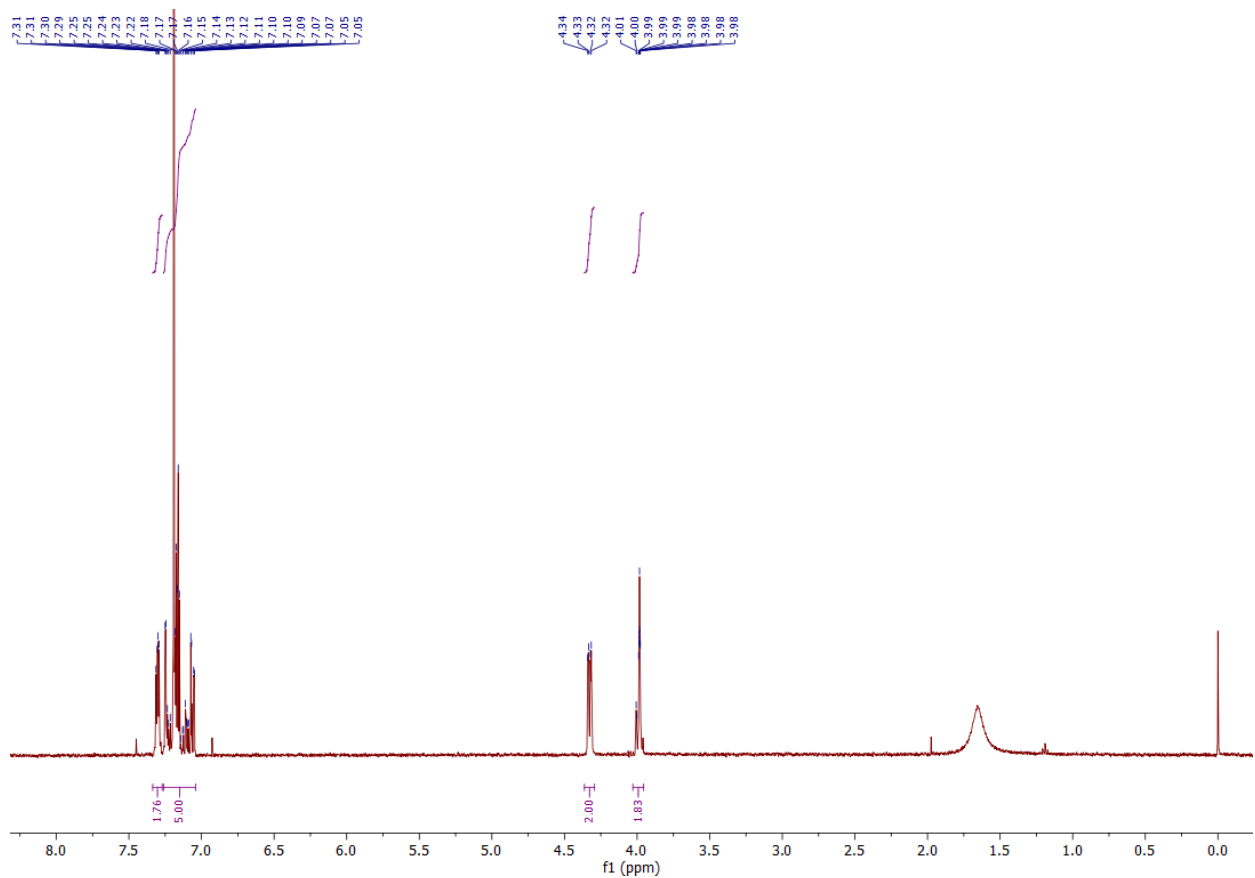

Figure S3. $^1$H NMR spectra (400 MHz in CDCl$_3$) of A-Cl-(OH)$_2$ (isomer 2).

### 2.3. 2-Chloroanthracene-9,10-dicarbaldehyde (A-Cl-CHO)

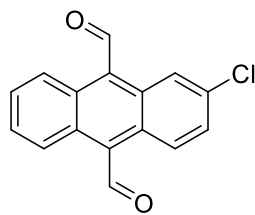

The following procedure was adapted from the previously published literature.[15] The A-Cl-(OH)$_2$ (0.995 g, 3,66 mmol, 1 equiv.) was dissolved in anhydrous acetonitrile (63 mL) and heated to reflux with stirring, forming a transparent, light-tan solution. Lead tetraacetate (3.243 g, 7.316 mmol, 2 equiv.) was added portion-wise to the solution over a period of 5 minutes, quickly turning the solution opaque and brown in color. The solution was then stirred at reflux for 4 hours, resulting in the formation of a dark-orange precipitate within the brown solution. The consumption of the A-Cl-(OH)$_2$ was monitored through thin-layer chromatography (CH$_2$Cl$_2$/ethyl acetate 8:2, R=0.50 & R=0.33). After the reaction was complete, the reaction mixture was concentrated via rotary evaporation and gave a burgundy residue. 10% Aqueous sodium carbonate (70 mL) was added to the residue, allowing the residue to be suspended in the aqueous solution. The organic contents were extracted from the aqueous layer using dichloromethane. The opaque, yellow organic layer was dried over anhydrous sodium sulfate, filtered, and concentrated to give a yellow-orange solid residue. The residue was purified by column chromatography (hexane/$CH_2Cl_2$/ethyl acetate/hexane 10:1:1, R=0.3), providing 2-chloro-9,10-anthracenedialdehyde (A-Cl-CHO) as an orange solid (0.630 g, 2.350 mmol). Yield 64.26 %.

$^1$H NMR (400 MHz, $CDCl_3$) δ 11.39 (s, 1H), 11.37 (s, 1H), 8.80 (d, J = 2.1 Hz, 1H), 8.71 – 8.60 (m, 3H), 7.72 – 7.62 (m, 2H), 7.57 (dd, J = 9.5, 2.1 Hz, 1H).

$^{13}$C NMR (101 MHz, $CDCl_3$) δ 193.88, 193.40, 135.30, 133.05, 132.79, 131.33, 130.41, 129.62, 128.98, 128.55, 128.06, 126.13, 124.19, 123.94, 123.14.

HRMS-EI: Measured (*m/z*): 268.03. Theoretical (*m/z*): 268.70.

Elemental analysis: Measured C (69.06 %), H (3.23 %), Cl (13.20 %). Theoretical: C (71.52 %), H (3.80 %), Cl (13.19 %).

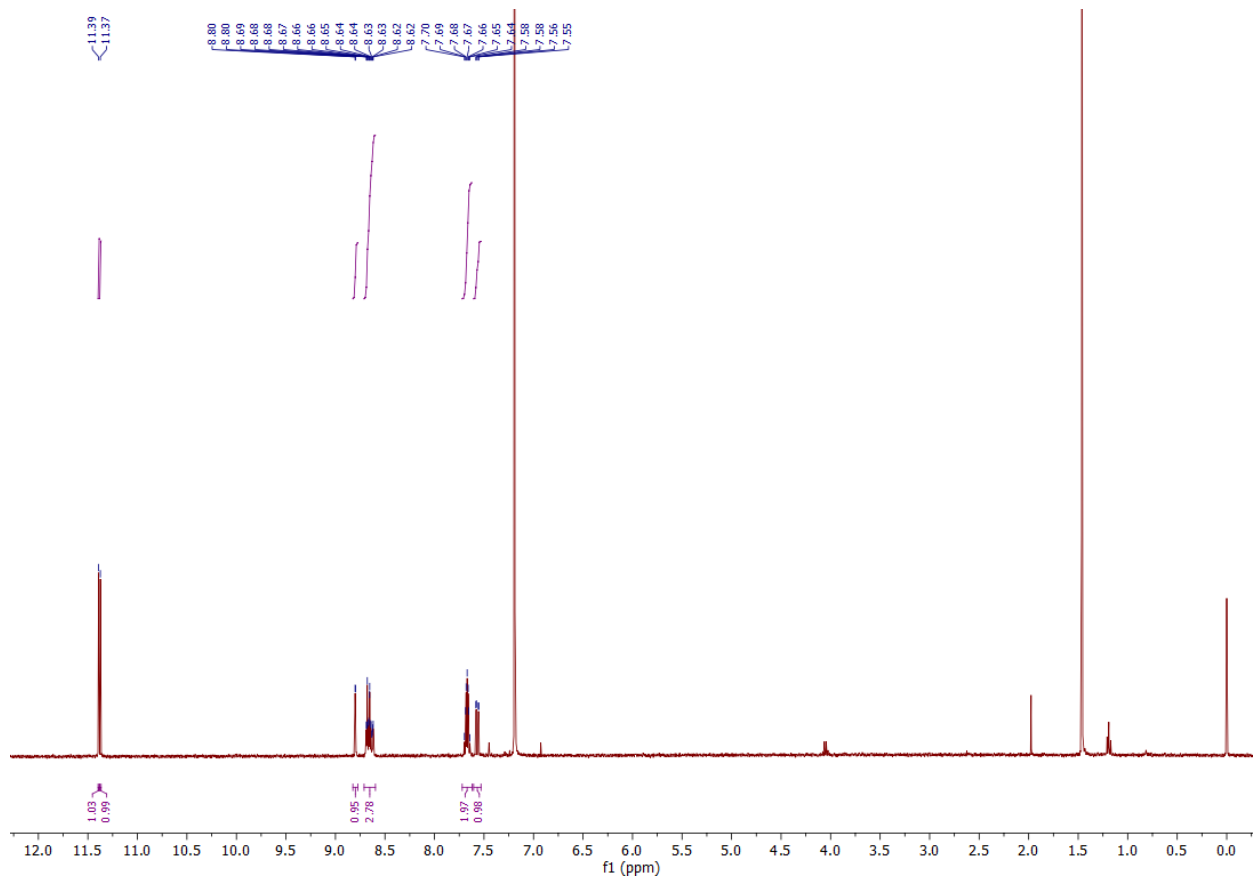

Figure S4. $^1$H NMR spectra (400 MHz in CDCl$_3$) of A-Cl-CHO.

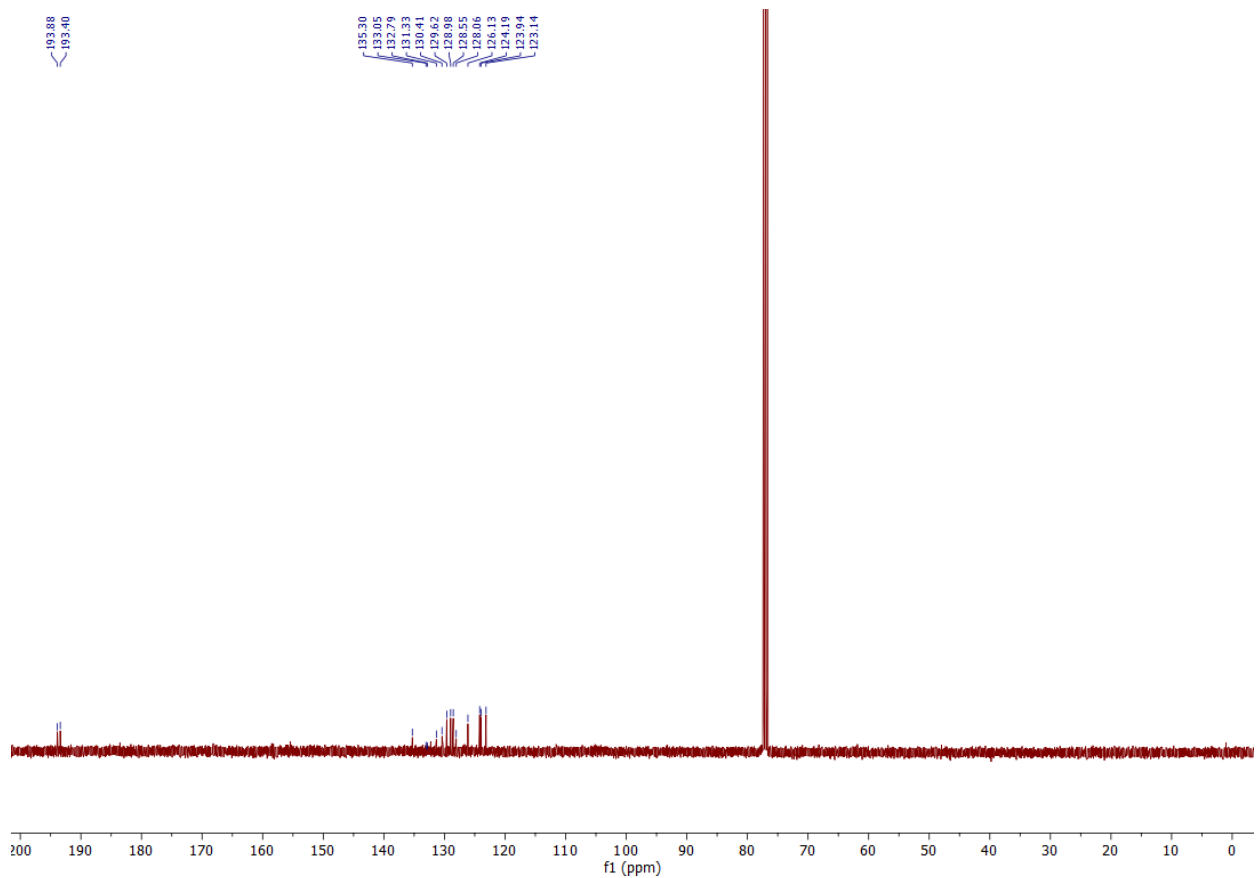

Figure S5. $^{13}$C NMR spectra (101 MHz in CDCl$_3$) of A-Cl-CHO.

### 2.4. 2-Bromo-9,10-dihydro-9,10-[4,5]epidioxoloanthracen-13-one (A-Br-epO)

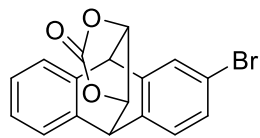

The following procedure was adapted from the previously published literature.[15] 2-Bromoanthracene (2.4 g, 9.39 mmol, 1 equiv.) and vinylene carbonate (5.9 g, 58.56 mmol, 7.3 equiv.) were heated under reflux with stirring for 18 hours, slowly forming a dark brown solution. The consumption of 2-bromoanthracene was monitored by thin-layer chromatography (CH$_2$Cl$_2$/hexane 1:49 , R=0.30). The mixture underwent rotary evaporation under high vacuum to remove the excess vinylene carbonate, providing the as 2-bromo-9,10-dihydro-9,10-[4,5]epidioxoloanthracen-13-one (A-Br-epO) light-brown solid (3.18 g, 9.30 mmol, 66.6%). The product was used for the further reaction without additional purification.

¹H NMR (400 MHz, CDCl₃) δ 7.46 (dd, J = 3.6, 1.9 Hz, 1H), 7.35 – 7.26 (m, 3H), 7.23 – 7.14 (m, 3H), 4.81 (dt, J = 3.7, 1.8 Hz, 2H), 4.61 (dq, J = 7.1, 1.6 Hz, 2H).

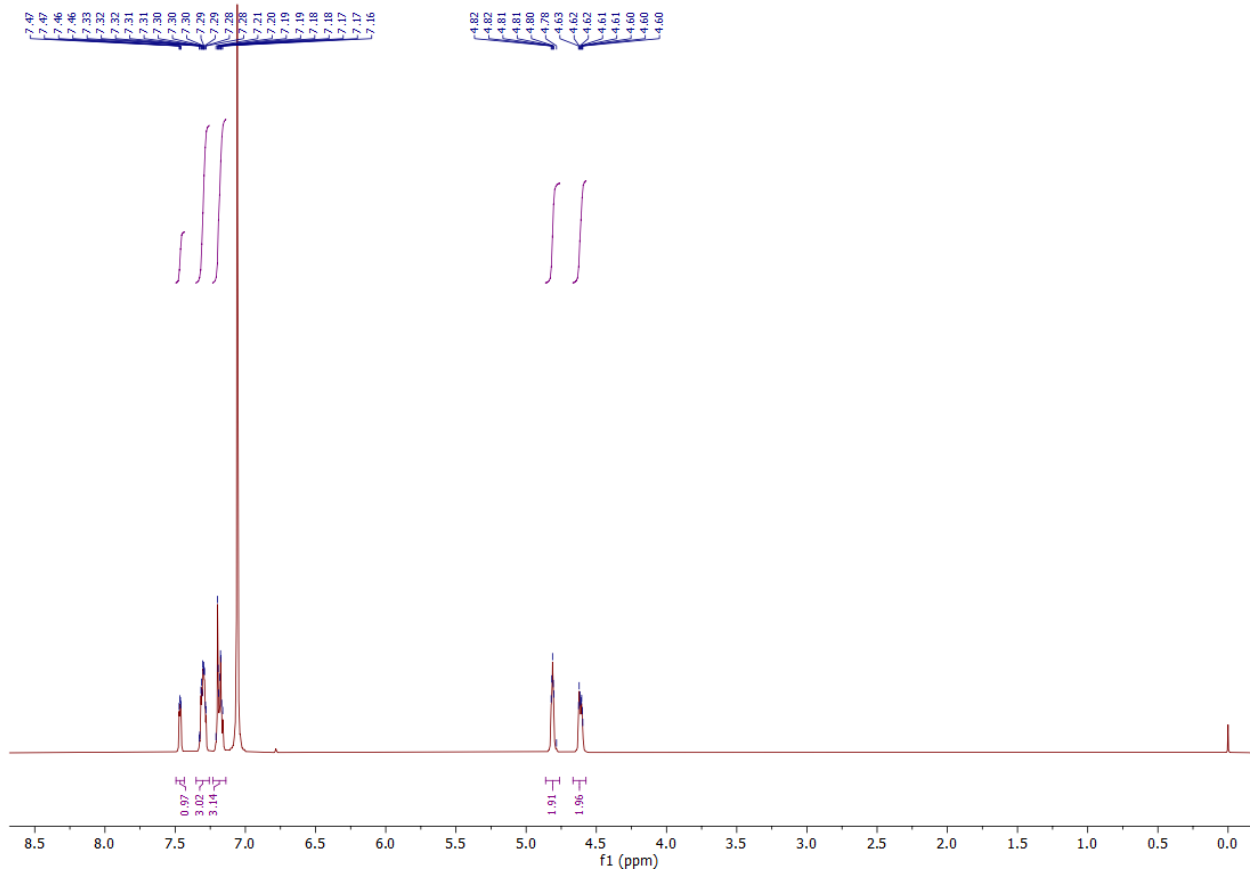

Figure S6. ¹H NMR spectra (400 MHz in CDCl₃) of A-Br-epO.

### 2.5. 2-Bromo-9,10-dihydro-9,10-ethanoanthracene-11,12-diol (A-Br-(OH)₂)

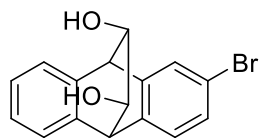

The following procedure was adapted from the previously published literature.[15] Solid potassium hydroxide (2.56 g, 45.59 mmol, 5.1 equiv.), deionized water (42.6 mL), and absolute ethanol (4.5 mL) were added to the A-Br-epO (3.06 g, 8.94 mmol, 1 equiv.). The solution was stirred at 75° C for 3 hours. The consumption of the A-Br-epO was monitored through thin-layer chromatography (100% CH₂Cl₂, R=0.30). Afterwards, the solution underwent rotary evaporation under reduced pressure to remove the ethanol and roughly half of the water volume. Additional water (85 mL) was added to the solution and the solution was stirred at room temperature for one hour, resulting in the formation of a light-tan solid. The contents were vacuum-filtered and then washed with deionized water. The vacuum-filtration receiving flask was changed and the solid was washed with ethyl acetate through the filter paper. The ethyl acetate was removed through rotary evaporation, leaving a yellow solid residue. The product was purified through column chromatography (CH$_2$Cl$_2$/ethyl acetate 9:1, R=0.30 & R=0.15), providing two isomers of 2-bromo-9,10-dihydro-9,10-ethanoanthracene-11,12-diol (A-Br-(OH)$_2$) as a white solid (2.01 g, 79%).

$^1$H NMR (400 MHz, CDCl$_3$) δ 7.40 (d, $J$ = 1.9 Hz, 1H), 7.34 – 7.06 (m, 6H), 4.32 (dd, $J$ = 6.1, 2.4 Hz, 2H), 4.01 – 3.94 (m, 2H).

Elemental analysis calculated: C (60.59 %), H (4.07 %). Theoretical: C (60.59 %). H (4.13 %).

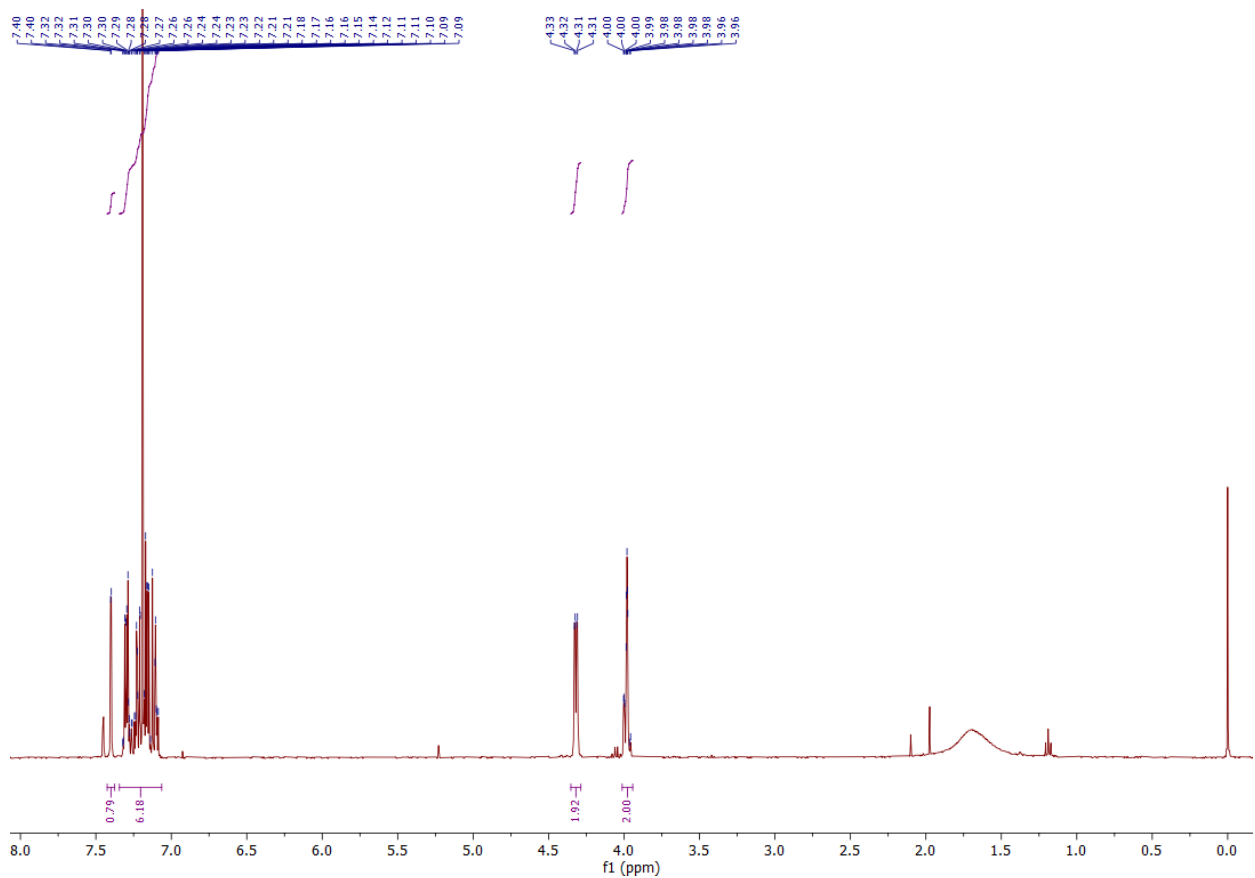

Figure S7. $^1$H NMR spectra (400 MHz in CDCl$_3$) of A-Br-(OH)$_2$ (mixture of two isomers).

## 2.6. 2-Bromoanthracene-9,10-dicarbaldehyde (A-Br-CHO)

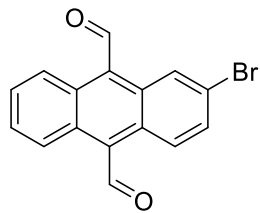

The following procedure was adapted from the previously published literature.[15] A-Br-(OH)$_2$ (2.00 g, mmol, 1 equiv.) was dissolved in anhydrous acetonitrile (127 mL) and heated to reflux with stirring, forming a transparent, light-tan solution. Lead tetraacetate (5.595 g, mmol, 2 equiv.) was added portion-wise to the solution over a period of 5 minutes, quickly turning the solution opaque and brown in color. The solution was then stirred at reflux for 4 hours, resulting in the formation of a dark-orange precipitate within the brown solution. The consumption of the A-Br-(OH)$_2$ was monitored through thin-layer chromatography (CH$_2$Cl$_2$/ethyl acetate 8:2, R=0.50 & R=0.33). After the reaction was complete, the reaction mixture was concentrated via rotary evaporation and gave a burgundy residue. 10% Aqueous sodium carbonate (130 mL) was added to the residue, allowing the residue to be suspended in the aqueous solution. The organic contents were extracted from the aqueous layer using dichloromethane. The opaque, yellow organic layer was dried over anhydrous sodium sulfate, filtered, and concentrated to give a yellow-orange solid residue. The residue was purified by column chromatography (ethyl acetate/hexane 2:8, R=0.3), providing 2-bromo-9,10-anthracenedialdehyde (A-Br-CHO) as an orange solid (1.71 g, 5.465 mmol). Yield 86.35 %.

$^1$H NMR (400 MHz, CDCl$_3$) δ 11.39 (s, 1H), 11.37 (s, 1H), 8.97 (dd, $J$ = 1.9, 0.5 Hz, 1H), 8.71 – 8.55 (m, 3H), 7.73 – 7.63 (m, 3H).

$^{13}$C NMR (101 MHz, CDCl$_3$) δ 192.75, 192.40, 131.21, 130.92, 130.14, 129.63, 129.38, 129.23, 127.95, 127.57, 127.11, 125.50, 124.96, 123.15, 122.97, 122.87.

HRMS-EI: Measured ($m/z$): 311.75. Theoretical ($m/z$): 313.99.

Elemental analysis: Measured C (60.44 %), H (2.98 %). Theoretical: C (61.37 %), H (2.90 %).

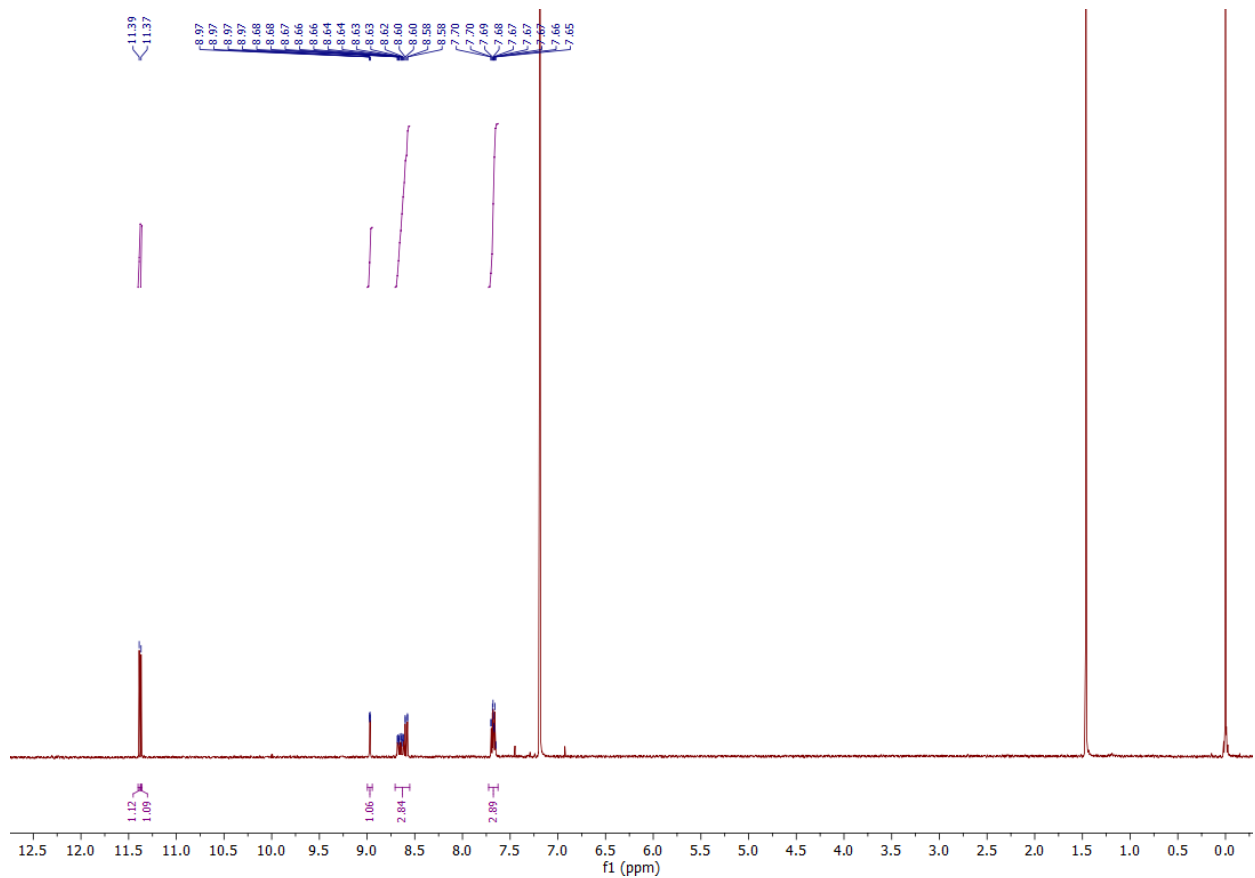

Figure S8. $^1$H NMR spectra (400 MHz in CDCl$_3$) of A-Br-CHO.

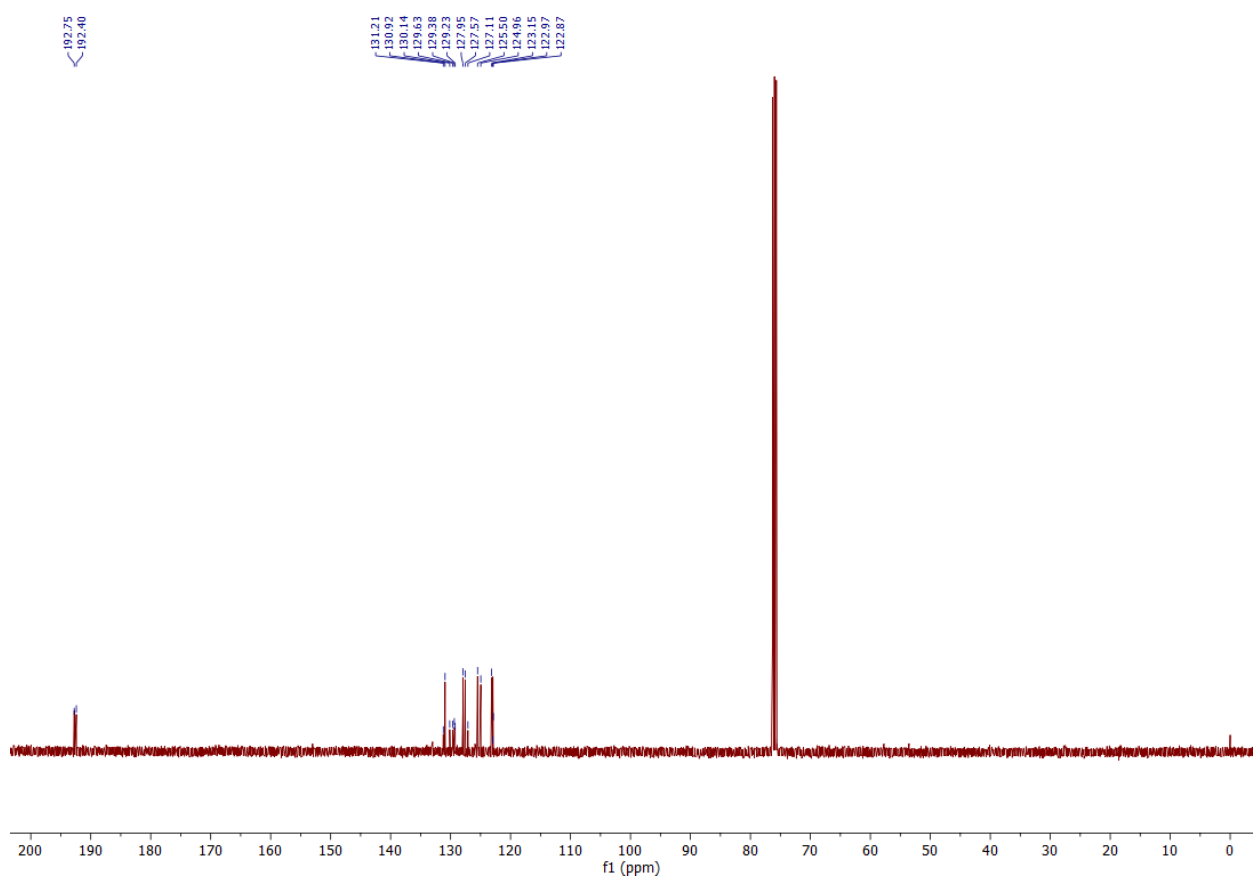

Figure S9. $^{13}$C NMR spectra (101 MHz in CDCl$_3$) of A-Br-CHO.

### 2.7. 2-Iodoantracene (A-I)

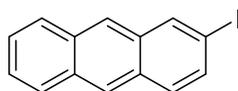

The following procedure was adapted from the previously published literature[16]. A mixture of 2-bromoanthracene (2.582 g, 10.09 mmol), KI (15.12 g, 91.09 mmol, 9.03 equiv.) and CuI (5.934 g, 31.16 mmol, 3.09 equiv.) in 38.7 ml 1,3-dimethyl-2-imidazolidinone was placed in 100 ml flask. The mixture was purged with N$_2$ and heated with vigorous stirring at 200 °C for 20 hours. After cooling to room temperature, brine and ice were added. The reaction vessel was placed in an ice bath for several hours, then precipitated inorganic salts were removed by filtration. The mixture underwent rotary evaporation under high vacuum to remove the excess of 1,3-dimethyl-2-imidazolidinone. The solid residue was extracted with DCM and the product was purified through column chromatography (hexane/CHCl$_3$/ethyl acetate 5:3:1, R=0.9), providing 2-iodoanthracene (A-I) as a yellow solid (1.466 g, 4.82 mmol). Yield 47.8 %.

$^1$H NMR (400 MHz, CDCl$_3$) δ 8.36 (dt, $J$ = 1.5, 0.8 Hz, 1H), 8.31 (s, 1H), 8.23 (s, 1H), 7.96 – 7.89 (m, 2H), 7.70 – 7.64 (m, 1H), 7.58 (dd, $J$ = 9.0, 1.6 Hz, 1H), 7.47 – 7.38 (m, 2H).

$^{13}$C NMR (101 MHz, CDCl$_3$) δ 136.71, 133.49, 132.77, 131.79, 131.52, 129.76, 129.48, 128.07, 126.38, 126.04, 125.70, 125.17, 125.03, 91.03.

HRMS-EI: Measured (*m/z*): 303.97. Theoretical (*m/z*): 303.97.

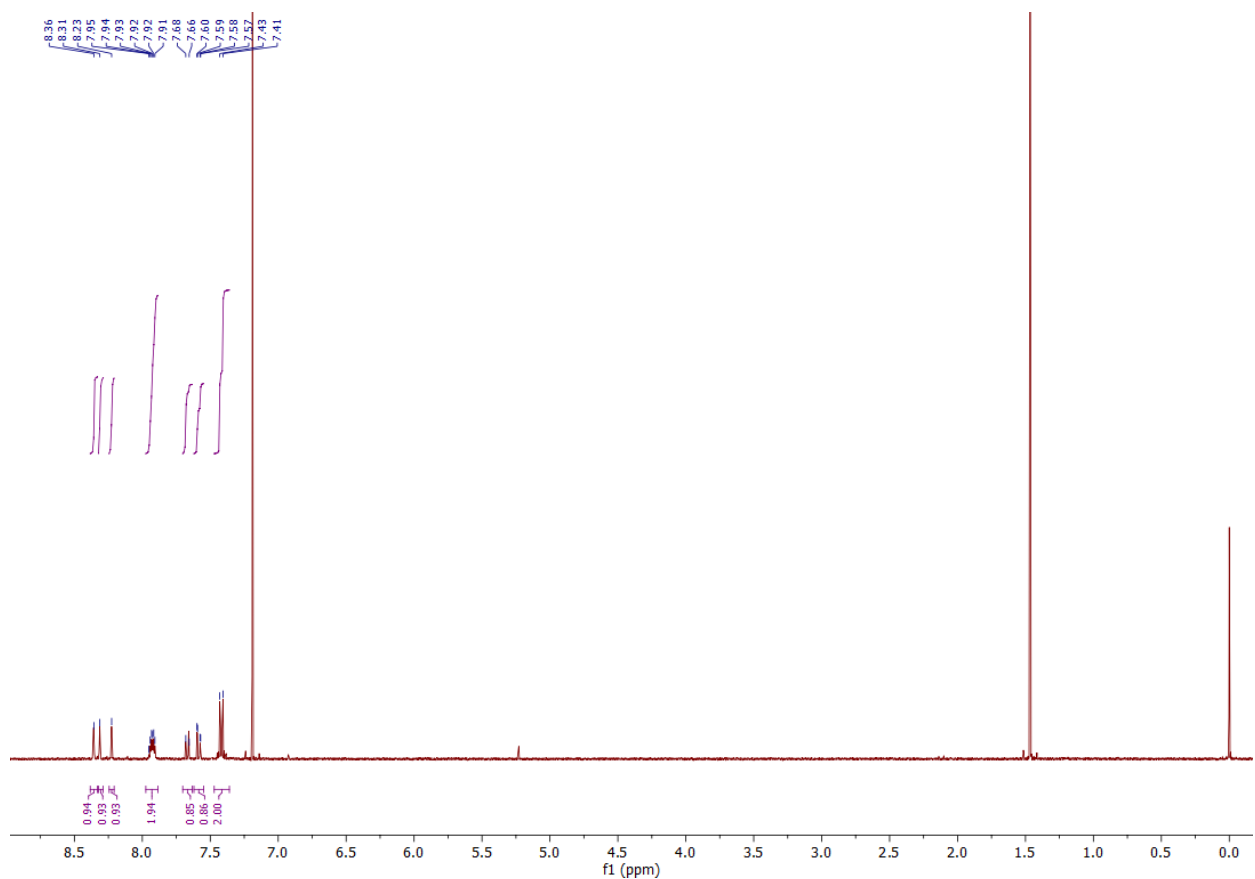

Figure S10. $^1$H NMR spectra (400 MHz in CDCl$_3$) of A-I.

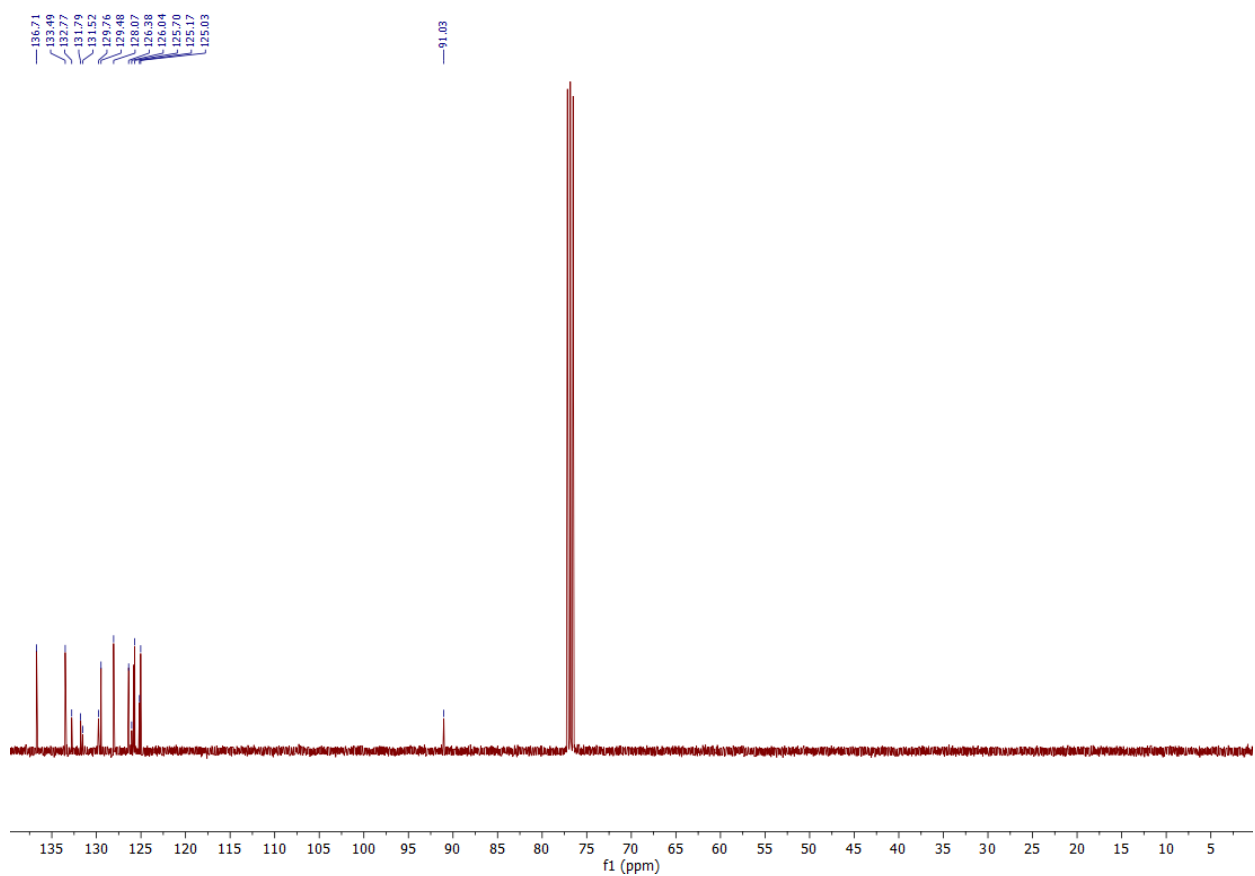

Figure S11. $^{13}$C NMR spectra (101 MHz in CDCl$_3$) of A-I.

## 2.8. 2-Iodo-9,10-dihydro-9,10-[4,5]epidioxoloanthracen-13-one (A-I-epO)

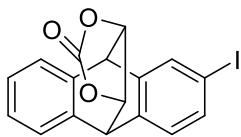

The following procedure was adapted from the previously published literature.[15] A-I (1.446 g, 4.76 mmol, 1 equiv.) and vinylene carbonate (2.99 g, 34.73 mmol, 7.3 equiv.) were heated under reflux with stirring for 18 hours, slowly forming a dark brown solution. The consumption of A-I was monitored by thin-layer chromatography (CH$_2$Cl$_2$/hexane 1:49 , R=0.30). The mixture underwent rotary evaporation under high vacuum to remove the excess vinylene carbonate, providing the 2-iodo-9,10-dihydro-9,10-[4,5]epidioxoloanthracen-13-one (A-I-epO) as a dark-brown solid (1.73 g, 92.8 %). The product was used for the further reaction without additional purification.

¹H NMR (400 MHz, CDCl₃) δ 7.64 (s, 1H), 7.54 – 7.41 (m, 1H), 7.34 – 7.23 (m, 2H), 7.16 (dd, J = 9.2, 4.0 Hz, 2H), 7.03 (dd, J = 7.7, 2.9 Hz, 1H), 4.83 – 4.70 (m, 2H), 4.56 (d, J = 9.4 Hz, 2H).

HRMS-EI: Measured (*m/z*): 389.96. Theoretical (*m/z*): 391.99.

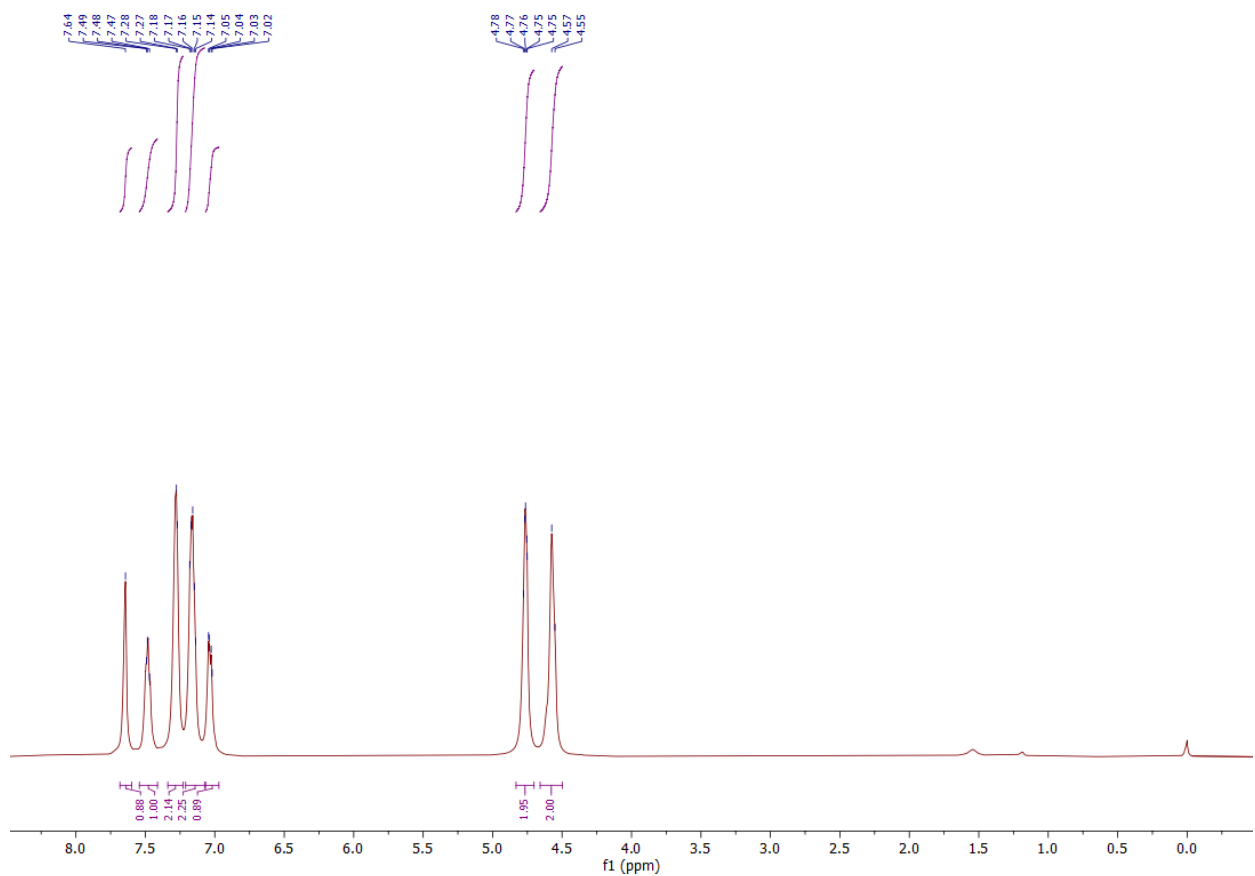

Figure S12. ¹H NMR spectra (400 MHz in CDCl₃) of A-I-epO.

### 2.9. 2-Iodo-9,10-dihydro-9,10-ethanoanthracene-11,12-diol (A-I-(OH)₂)

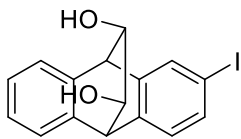

The following procedure was adapted from the previously published literature.[15] Solid potassium hydroxide (0.985 g, 17.6 mmol, 5.1 equiv.), deionized water (16.4 mL), and absolute ethanol (1.75 mL) were added to the A-I-epO (1.72 g, 4.39 mmol, 1 equiv.). The solution was stirred at 75° C for 3 hours. The consumption of the A-I-epO intermediate was monitored through thin-layer chromatography (100% CH$_2$Cl$_2$, R=0.30). Afterwards, the solution underwent rotary evaporation under reduced pressure to remove the ethanol and roughly half of the water volume. Additional water (33 mL) was added to the solution and the solution was stirred at room temperature for one hour, resulting in the formation of light-tan solid. The contents were vacuum-filtered and then washed with deionized water. The vacuum-filtration receiving flask was changed and the solid was washed with ethyl acetate through the filter paper. The ethyl acetate was removed through rotary evaporation, leaving a yellow solid residue. The product was purified through column chromatography (CH$_2$Cl$_2$/ethyl acetate 8:1, R=0.30 & R=0.15), providing a mixture of two isomers of 2-iodo-9,10-dihydro-9,10-ethanoanthracene-11,12-diol (A-I-(OH)$_2$) as a white-yellowish solid (1.189 g, 3.25 mmol), Yield 74.0 %.

$^1$H NMR (400 MHz, CDCl$_3$) δ 7.65 (d, $J$ = 1.7 Hz, 1H), 7.48 (dd, $J$ = 7.8, 1.7 Hz, 1H), 7.23 (dd, $J$ = 5.3, 3.3 Hz, 2H), 7.12 – 7.03 (m, 3H), 4.30 (dd, $J$ = 7.3, 2.6 Hz, 2H), 4.07 – 3.96 (m, 2H).

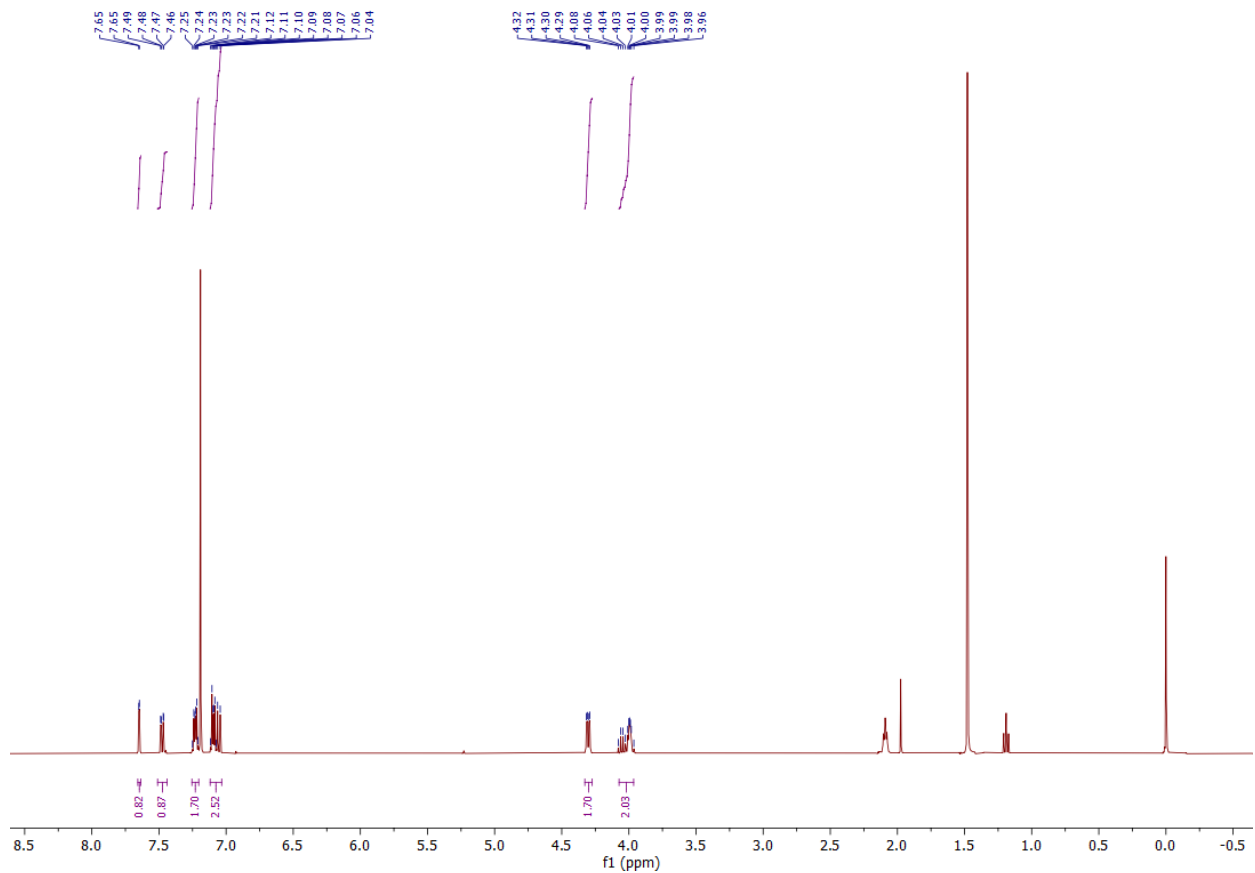

Figure S13. $^1$H NMR spectra (400 MHz in CDCl$_3$) of A-I-(OH)$_2$.

## 2.10. 2-Iodoanthracene-9,10-dicarbaldehyde (A-I-CHO)

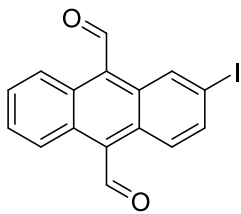

The following procedure was adapted from the previously published literature.[15] A-I-(OH)$_2$ (0.159 g, 0.43 mmol, 1 equiv.) was dissolved in anhydrous acetonitrile (10 mL) and heated to reflux with stirring, forming a transparent, light-tan solution. Lead tetraacetate (0.380 g, 0.86 mmol, 2 equiv.) was added portion-wise to the solution over a period of 5 minutes, quickly turning the solution opaque and brown in color. The solution was then stirred at reflux for 4 hours, resulting in the formation of a dark-orange precipitate within the brown solution. The consumption of the diol intermediates were monitored through thin-layer chromatography (CH$_2$Cl$_2$/ethyl acetate 8:2, R=0.50 & R=0.33). After the reaction was complete, the reaction mixture was concentrated via rotary evaporation and gave a burgundy residue. 10% Aqueous sodium carbonate (30 mL) was added to the residue, allowing the residue to be suspended in the aqueous solution. The organic contents were extracted from the aqueous layer using dichloromethane. The opaque, yellow organic layer was dried over anhydrous sodium sulfate, filtered, and concentrated to give a yellow-orange solid residue. The residue was purified by column chromatography (ethyl acetate/hexane 2:8, R=0.3), providing 2-iodo-9,10-anthracenedialdehyde (A-I-CHO) as an orange solid (0.414 g, 1.15 mmol). Yield 36.0 %.

$^1$H NMR (400 MHz, CDCl$_3$) δ 11.35 (s, 1H), 11.33 (s, 1H), 9.13 (d, $J$ = 1.7 Hz, 1H), 8.62 (ddt, $J$ = 10.8, 7.1, 3.3 Hz, 2H), 8.39 (d, $J$ = 9.4 Hz, 1H), 7.82 (dd, $J$ = 9.4, 1.7 Hz, 1H), 7.68 – 7.62 (m, 2H).

$^{13}$C NMR (101 MHz, CDCl$_3$) δ 192.68, 192.52, 135.82, 132.31, 131.10, 129.88, 129.80, 129.47, 129.19, 127.90, 127.67, 127.35, 124.55, 123.13, 123.09, 95.10.

HRMS-EI: Measured (*m/z*): 360.02. Theoretical (*m/z*): 359.96.

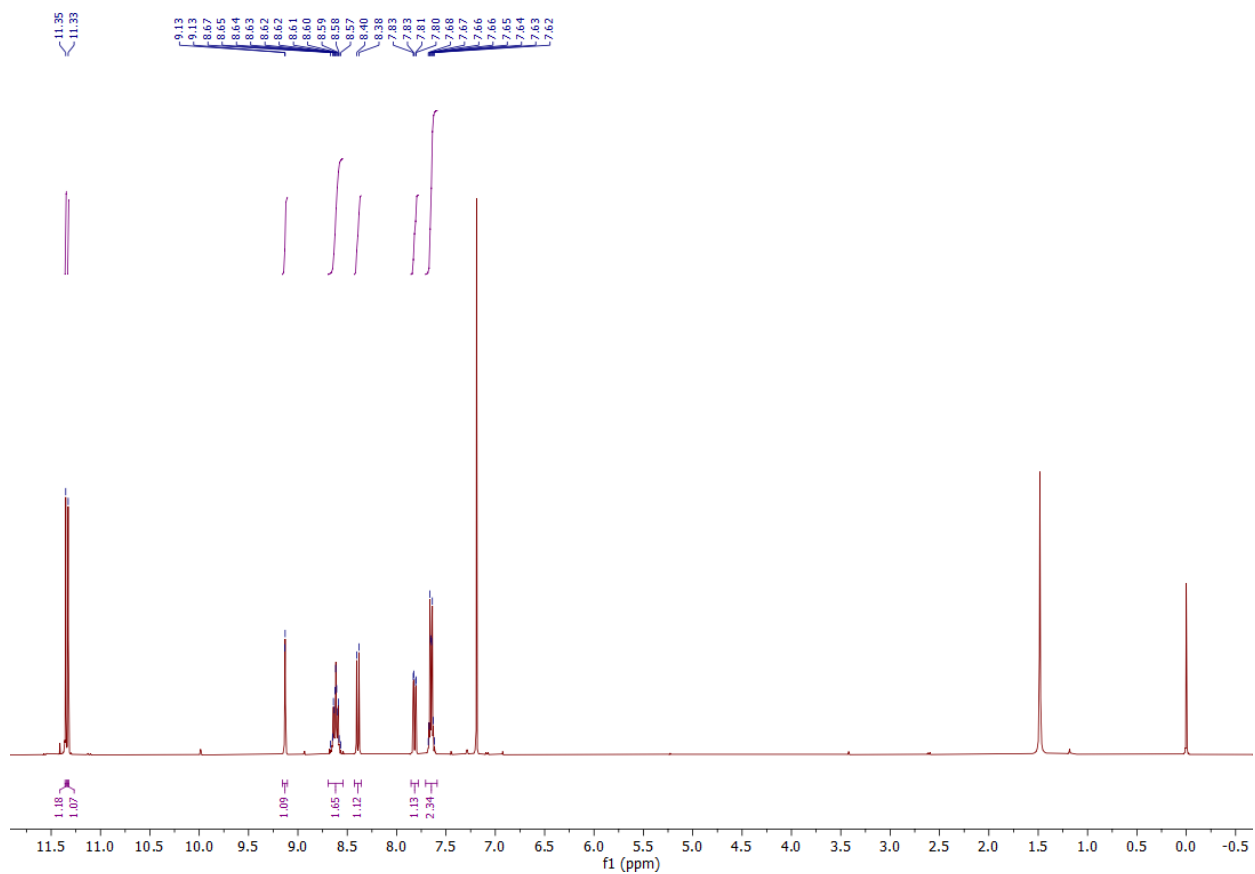

Figure S14. $^1$H NMR spectra (400 MHz in CDCl$_3$) of A-I-CHO.

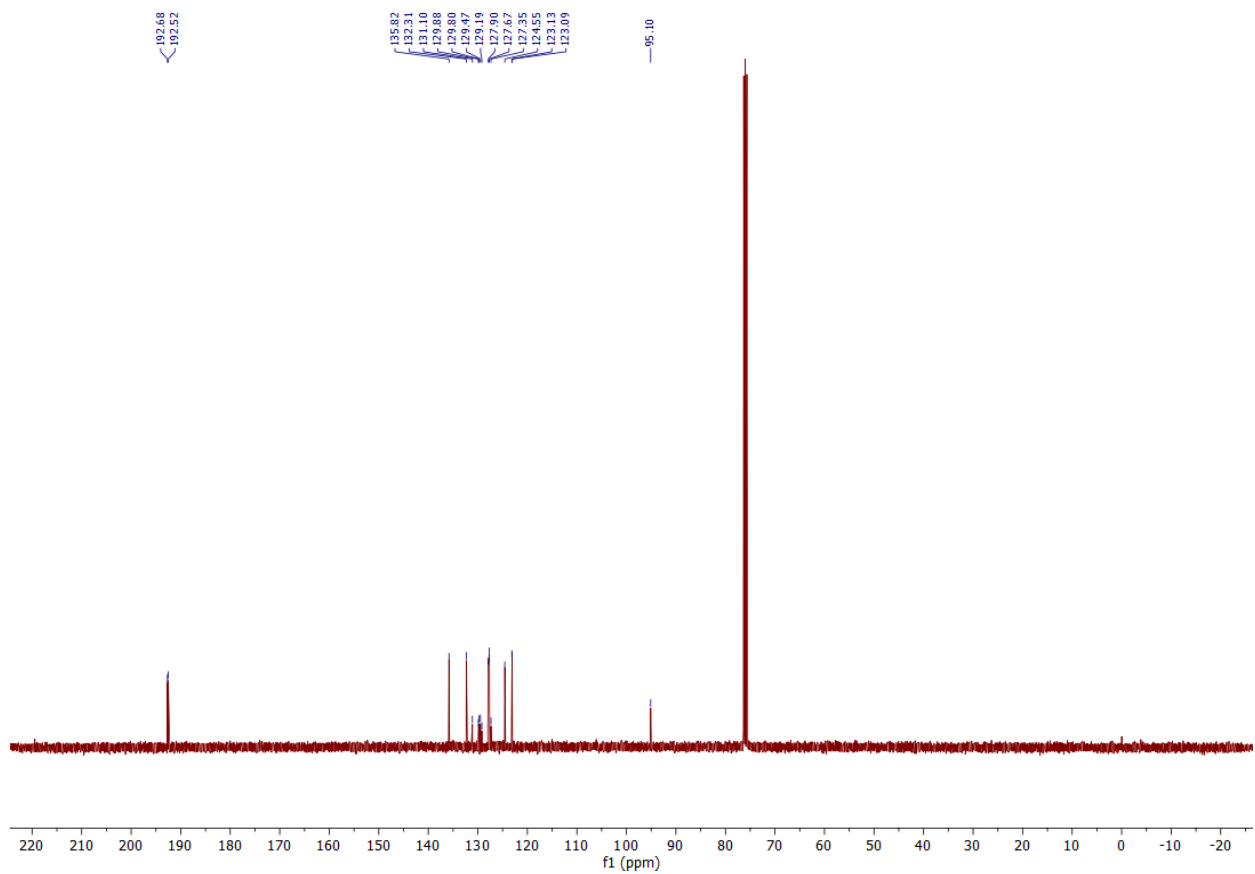

Figure S15. $^{13}$C NMR spectra (101 MHz in CDCl$_3$) of A-I-CHO.

## 2.11. FT-IR analysis of linkers

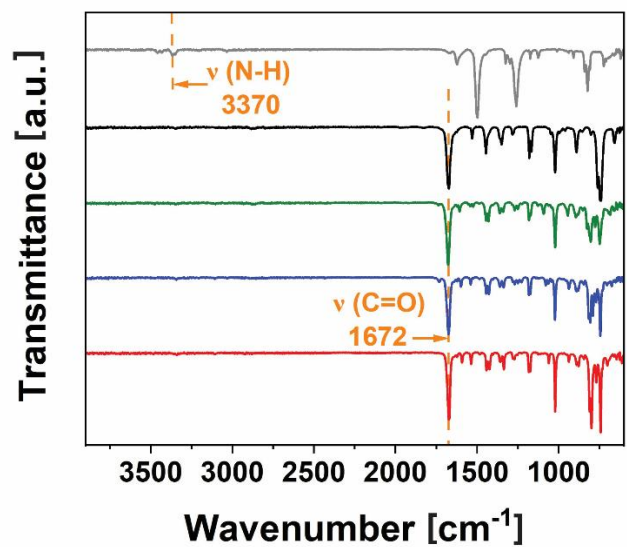

Figure S16. FT-IR spectra of N,N,N',N'-tetrakis(4-aminophenyl)-1,4-phenylenediamine (W-NH$_2$) (grey), anthracene-9,10-dicarbaldehyde (A-H-CHO) (black), A-Cl-CHO (green), A-Br-CHO (blue) and A-I-CHO (red). The vibration of the N–H group in the W–NH$_2$ moiety is visible at 3370 cm$^{-1}$.[17] The carbonyl stretch of the A–X–CHO linker appears at 1672 cm$^{-1}$.[18,19]

## 3. COF synthesis

### 3.1. W-A-H COF synthesis

W-NH$_2$ (9.22 mg, 19.52 µmol, 1.0 eq.) and A-H-CHO (9.14 mg, 39.05 µmol, 2.0 eq.) were filled into a 6 mL pyrex tube followed by the addition of chlorobenzene (400 µL), benzyl alcohol (BnOH) (400 µL), and 6 M acetic acid (100 µL). The tube was sealed and the reaction mixture was heated at 100 °C for 3 d. After cooling to room temperature, the resulting dark red precipitate was suction filtered, Soxhlet-extracted with dry THF, and dried under reduced pressure.

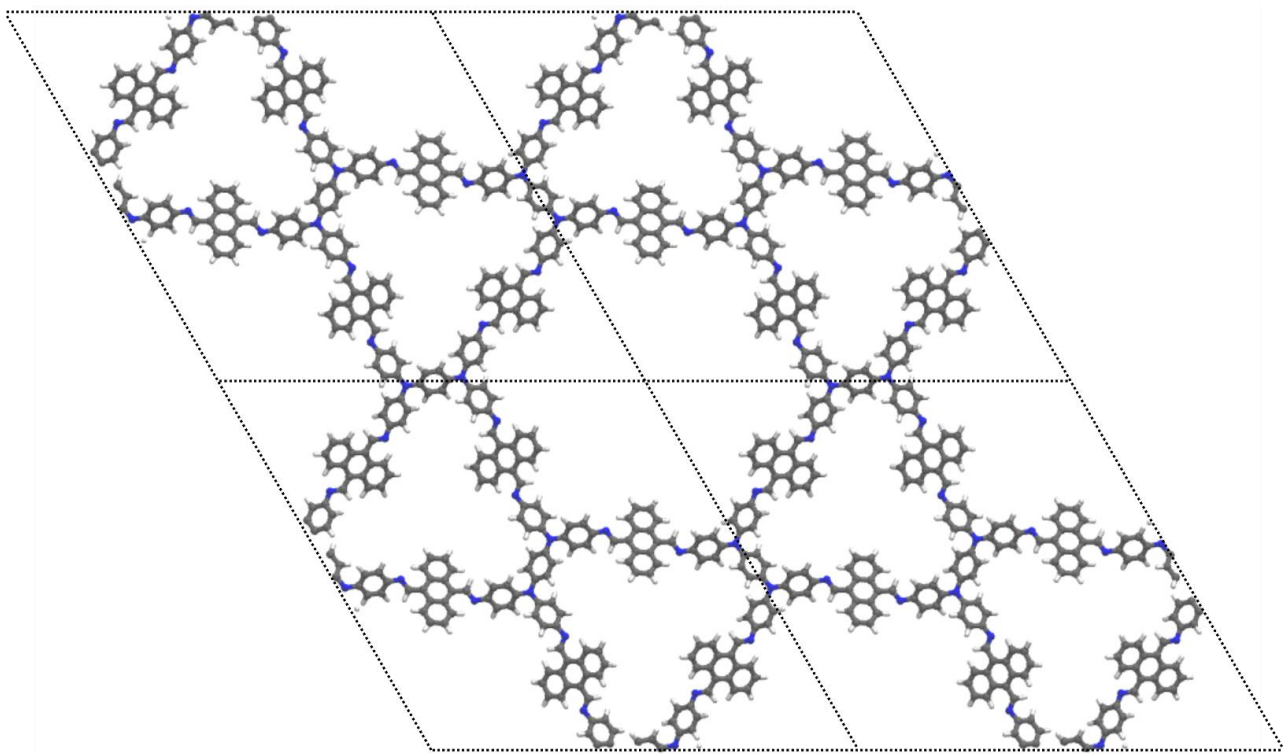

Figure S17. Simulated structure of W-A-H COF.

### 3.2. W-A-Cl COF synthesis

W-NH$_2$ (9.22 mg, 19.52 μmol, 1.0 eq.) and A-Cl-CHO (10.47 mg, 39.05 μmol, 2.0 eq.) were filled into a 6 ml pyrex tube, followed by the addition of CHCl$_3$ (400 μL), BnOH (400 μL), and 6 M acetic acid (100 μL). The tube was sealed and the reaction mixture was heated at 100 °C for 3 d. After cooling to room temperature, the resulting dark red precipitate was suction filtered, Soxhlet-extracted with dry THF, and dried under reduced pressure.

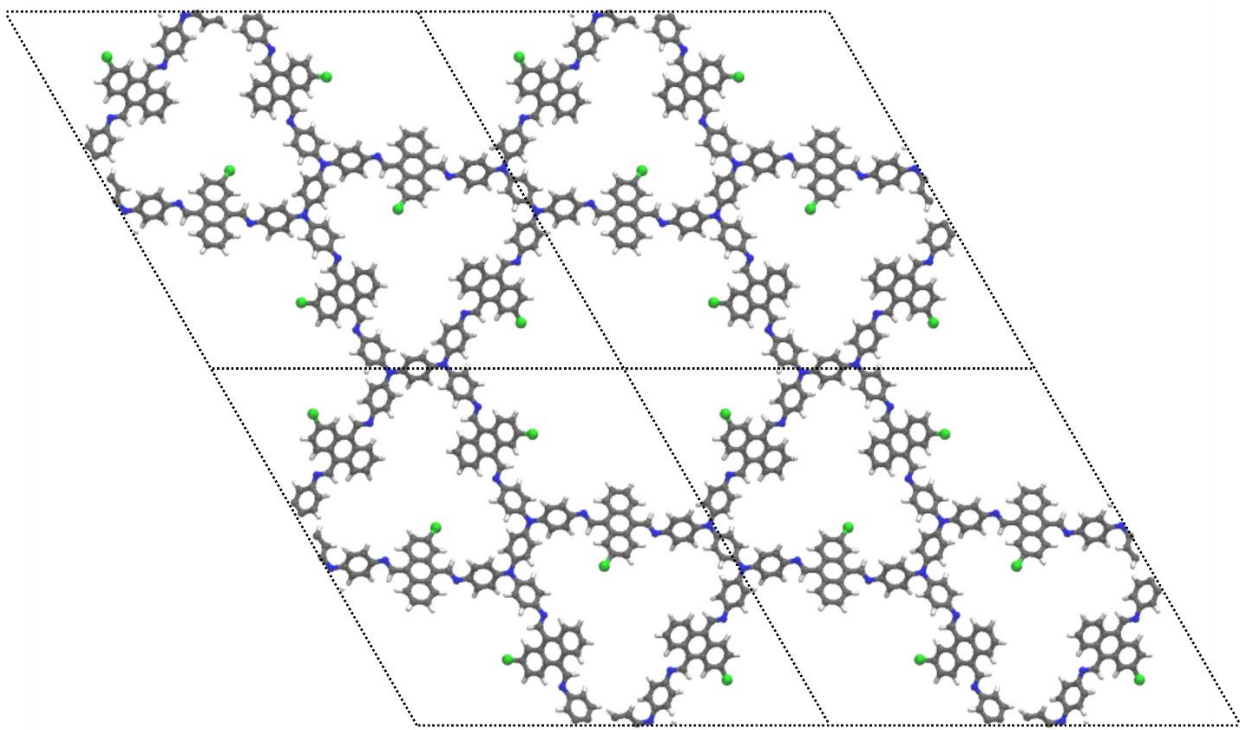

Figure S18. Simulated structure of W-A-Cl.

### 3.3. W-A-Br COF synthesis

W-NH$_2$ (9.22 mg, 19.52 μmol, 1.0 eq.) and A-Br-CHO (12.18 mg, 39.05 μmol, 2.0 eq.) were filled into a 6 ml pyrex tube, followed by the addition of chlorobenzene (400 μL), BnOH (400 μL), and 6 M acetic acid (100 μL). The tube was sealed and the reaction mixture was heated at 100 °C for 3 d. After cooling to room temperature, the resulting dark red precipitate was suction filtered, Soxhlet-extracted with dry THF, and dried under reduced pressure.

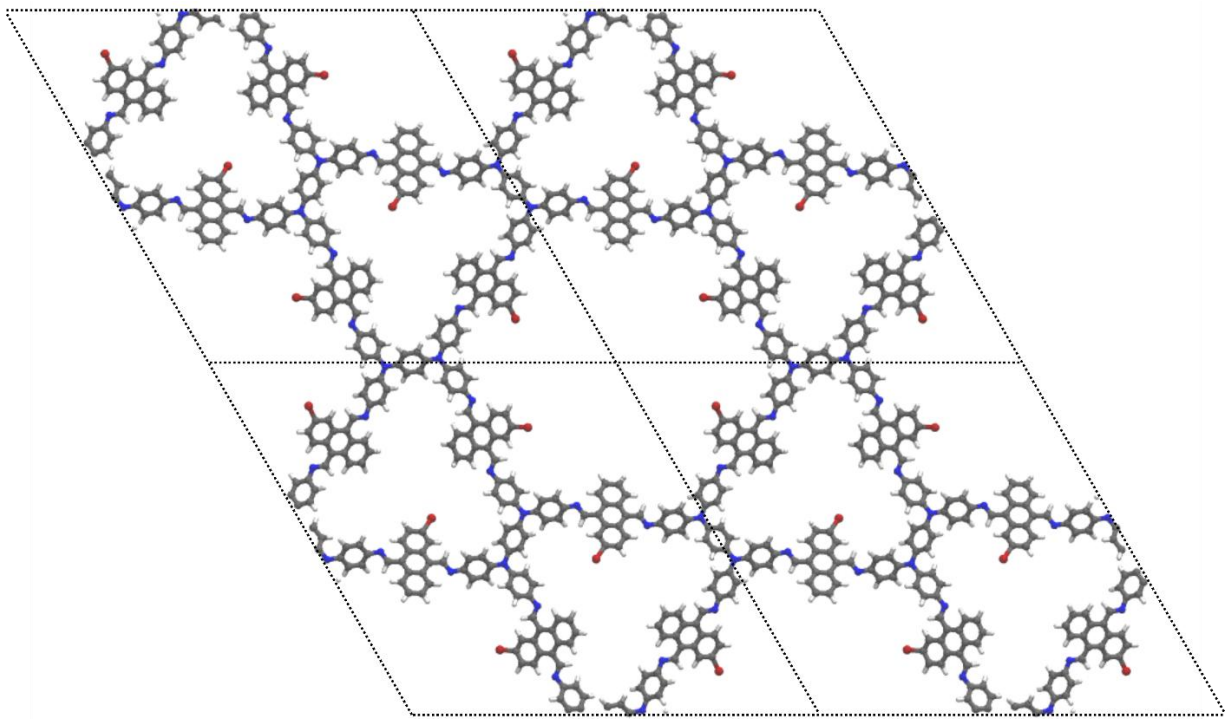

Figure S19. Simulated structure of W-A-Br COF.

### 3.4. W-A-I COF synthesis

W-NH$_2$ (9.22 mg, 19.52 μmol, 1.0 eq.) and A-I-CHO (14.40 mg, 39.05 μmol, 2.0 eq.) were filled into a 6 ml pyrex tube, followed by the addition of CHCl$_3$ (400 μL), BnOH (400 μL), and 6 M acetic acid (100 μL). The tube was sealed and the reaction mixture was heated at 100 °C for 3 d. After cooling to room temperature, the resulting dark red precipitate was suction filtered, Soxhlet-extracted with dry THF, and dried under reduced pressure.

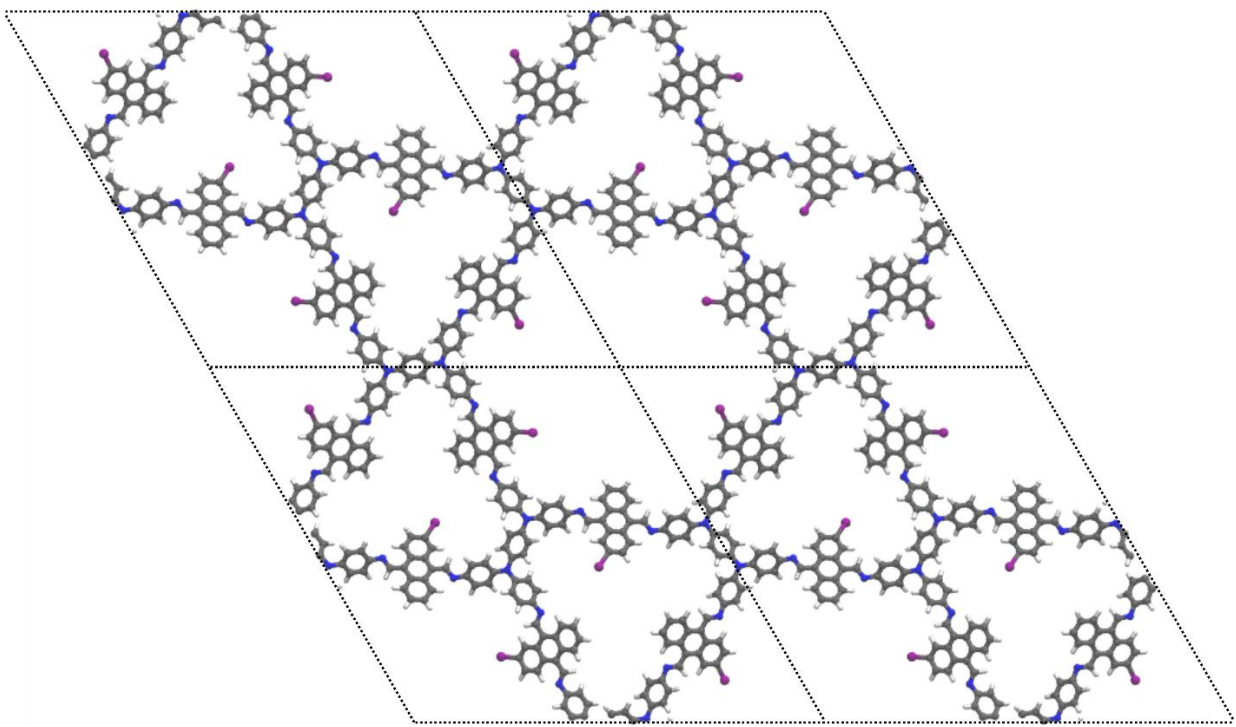

Figure S20. Simulated structure of W-A-I COF.

### 3.5. W-TA COF synthesis

The synthesis procedure for W-TA COF was adapted from the literature.[20] W-NH$_2$ (18.00 mg, 40.00 μmol, 1.0 eq.) and terephthalaldehyde (TA) (10.20 mg, 80.00 μmol, 2.0 eq.) were filled into a 6 ml pyrex tube, followed by the addition of mesitylene (1500 μL), BnOH (1500 μL), and 6 M acetic acid (150 μL). The tube was sealed and the reaction mixture was heated at 100 °C for 3 d. After cooling to room temperature, the resulting dark red precipitate was suction filtered, Soxhlet-extracted with dry THF, and dried under reduced pressure.

## 4. Structural analysis

### 4.1. W-A-H COF unit cell

$a = b = 4.236$ nm, $c = 0.411$ nm

$\alpha = \beta = 90.00°$, $\gamma = 120°$

Table S1. Fractional coordinates of W-A-H with symmetry P-3.

| | | | |
|---|---|---|---|
| C1 | 0.70734 | 0.56751 | 0.60940 |
| C2 | 0.52990 | 0.53593 | 0.50241 |
| C3 | 0.53125 | 0.50786 | 0.67043 |
| C4 | 0.49792 | 0.52722 | 0.33204 |
| C5 | 0.59545 | 0.57758 | 0.51776 |
| C6 | 0.60421 | 0.55308 | 0.37057 |
| C7 | 0.63887 | 0.55808 | 0.39287 |
| C8 | 0.66658 | 0.58820 | 0.56048 |
| C9 | 0.65819 | 0.61366 | 0.69219 |
| C10 | 0.62337 | 0.60823 | 0.67529 |
| C11 | 0.95223 | 0.55381 | 0.50720 |
| C12 | 0.94484 | 0.57836 | 0.67222 |
| C13 | 0.91017 | 0.57340 | 0.68044 |
| C14 | 0.88109 | 0.54336 | 0.52481 |
| C15 | 0.88826 | 0.51799 | 0.37159 |
| C16 | 0.92312 | 0.52334 | 0.35874 |
| C17 | 0.84118 | 0.56510 | 0.52090 |
| C18 | 0.80702 | 0.56468 | 0.54597 |
| C19 | 0.77596 | 0.53626 | 0.70500 |
| C20 | 0.74362 | 0.53883 | 0.73748 |
| C21 | 0.74221 | 0.56919 | 0.60540 |
| C22 | 0.77359 | 0.59808 | 0.45408 |
| C23 | 0.80592 | 0.59547 | 0.42049 |
| C24 | 0.77629 | 0.50581 | 0.84592 |
| C25 | 0.74663 | 0.47906 | 0.00730 |
| C26 | 0.71502 | 0.48163 | 0.04268 |
| C27 | 0.71387 | 0.51088 | 0.91449 |
| C28 | 0.77356 | 0.62899 | 0.32065 |
| C29 | 0.80341 | 0.65610 | 0.16437 |
| C30 | 0.83491 | 0.65340 | 0.12595 |
| C31 | 0.83579 | 0.62376 | 0.24725 |
| H1 | 0.68372 | 0.53991 | 0.62531 |
| H2 | 0.55546 | 0.51377 | 0.80818 |
| H3 | 0.49613 | 0.54824 | 0.19620 |
| H4 | 0.58347 | 0.52995 | 0.23539 |
| H5 | 0.64517 | 0.53934 | 0.26614 |
| H6 | 0.67954 | 0.63722 | 0.81915 |
| H7 | 0.61749 | 0.62806 | 0.78314 |
| H8 | 0.96684 | 0.60149 | 0.79646 |
| H9 | 0.90505 | 0.59211 | 0.82215 |
| H10 | 0.86592 | 0.49463 | 0.25129 |
| H11 | 0.92802 | 0.50371 | 0.23224 |
| H12 | 0.86523 | 0.59228 | 0.49733 |

| H13 | 0.80086 | 0.50421 | 0.82365 |
|---|---|---|---|
| H14 | 0.74796 | 0.45641 | 0.11737 |
| H15 | 0.69198 | 0.46111 | 0.18171 |
| H16 | 0.68991 | 0.51313 | 0.96071 |
| H17 | 0.74905 | 0.63066 | 0.34629 |
| H18 | 0.80235 | 0.67921 | 0.06184 |
| H19 | 0.85816 | 0.67424 | 0.99131 |
| H20 | 0.85983 | 0.62169 | 0.19996 |
| N1 | 0.55972 | 0.57173 | 0.50399 |
| N2 | 0.70180 | 0.59478 | 0.59326 |
| N3 | 0.84581 | 0.53720 | 0.52106 |

## 4.2. W-A-Cl COF unit cell

$a = b = 4.265$ nm, $c = 0.408$ nm

$\alpha = \beta = 90.00°$, $\gamma = 120°$

Table S2. Fractional coordinates of W-A-Cl with symmetry P-1.

| C1 | 0.43306 | 0.14027 | 0.59845 |
|---|---|---|---|
| C2 | 0.85976 | 0.29280 | 0.59807 |
| C3 | 0.70704 | 0.56673 | 0.59648 |
| C4 | 0.47008 | 0.46413 | 0.49627 |
| C6 | 0.53587 | 0.00592 | 0.49643 |
| C8 | 0.99405 | 0.52993 | 0.49670 |
| C10 | 0.53118 | 0.50771 | 0.67036 |
| C11 | 0.49231 | 0.02349 | 0.67014 |
| C12 | 0.97655 | 0.46881 | 0.67023 |
| C13 | 0.50199 | 0.47269 | 0.66659 |
| C14 | 0.52732 | 0.02928 | 0.66658 |
| C15 | 0.97073 | 0.49801 | 0.66692 |
| C16 | 0.59543 | 0.57729 | 0.51861 |
| C17 | 0.42270 | 0.01820 | 0.51835 |
| C18 | 0.98187 | 0.40453 | 0.51781 |
| C19 | 0.60395 | 0.55273 | 0.36941 |
| C20 | 0.44718 | 0.05126 | 0.36863 |
| C21 | 0.94879 | 0.39595 | 0.36854 |
| C22 | 0.63851 | 0.55757 | 0.38892 |
| C23 | 0.44227 | 0.08095 | 0.38802 |
| C24 | 0.91909 | 0.36135 | 0.38801 |
| C25 | 0.66639 | 0.58758 | 0.55566 |
| C26 | 0.41227 | 0.07879 | 0.55496 |
| C27 | 0.92125 | 0.33350 | 0.55468 |
| C28 | 0.65824 | 0.61307 | 0.68999 |
| C29 | 0.38686 | 0.04516 | 0.68992 |
| C30 | 0.95490 | 0.34171 | 0.68915 |
| C31 | 0.62348 | 0.60780 | 0.67596 |
| C32 | 0.39221 | 0.01573 | 0.67621 |
| C33 | 0.98435 | 0.37650 | 0.67522 |
| C34 | 0.44599 | 0.39847 | 0.50353 |
| C35 | 0.60153 | 0.04745 | 0.50377 |
| C36 | 0.95248 | 0.55401 | 0.50476 |
| C37 | 0.42136 | 0.36650 | 0.66712 |
| C38 | 0.63353 | 0.05478 | 0.66696 |
| C39 | 0.94512 | 0.57860 | 0.66890 |
| C40 | 0.42625 | 0.33678 | 0.67442 |
| C41 | 0.66327 | 0.08938 | 0.67409 |
| C42 | 0.91048 | 0.57367 | 0.67677 |
| C43 | 0.45632 | 0.33777 | 0.51964 |
| C44 | 0.66229 | 0.11848 | 0.51949 |
| C45 | 0.88140 | 0.54362 | 0.52196 |
| C46 | 0.48175 | 0.37033 | 0.36781 |
| C47 | 0.62972 | 0.11136 | 0.36781 |
| C48 | 0.88855 | 0.51824 | 0.36946 |
| C49 | 0.47646 | 0.39984 | 0.35563 |
| C50 | 0.60019 | 0.07656 | 0.35585 |
| C51 | 0.92337 | 0.52356 | 0.35673 |
| C52 | 0.43469 | 0.27616 | 0.51592 |
| C53 | 0.72392 | 0.15849 | 0.51605 |
| C54 | 0.84140 | 0.56525 | 0.51712 |
| C55 | 0.43533 | 0.24256 | 0.53961 |
| C56 | 0.75750 | 0.19274 | 0.53954 |
| C57 | 0.80715 | 0.56462 | 0.53963 |
| C58 | 0.46392 | 0.24001 | 0.69571 |
| C59 | 0.76004 | 0.22387 | 0.69574 |
| C60 | 0.77605 | 0.53604 | 0.69510 |
| C61 | 0.46161 | 0.20522 | 0.72628 |
| C62 | 0.79481 | 0.25637 | 0.72597 |
| C63 | 0.74344 | 0.53813 | 0.72245 |
| C64 | 0.43136 | 0.17343 | 0.59383 |
| C65 | 0.82661 | 0.25794 | 0.59335 |
| C66 | 0.74191 | 0.56847 | 0.59195 |
| C67 | 0.40241 | 0.17582 | 0.44448 |
| C68 | 0.82424 | 0.22662 | 0.44382 |
| C69 | 0.77333 | 0.59751 | 0.44324 |
| C70 | 0.40459 | 0.21056 | 0.41515 |
| C71 | 0.78951 | 0.19403 | 0.41472 |
| C72 | 0.80585 | 0.59524 | 0.41267 |
| C73 | 0.49431 | 0.27082 | 0.83616 |
| C74 | 0.72924 | 0.22341 | 0.83683 |
| C75 | 0.77642 | 0.50568 | 0.83693 |
| C76 | 0.52120 | 0.26805 | 0.99596 |
| C77 | 0.73199 | 0.25305 | 0.99664 |
| C78 | 0.74663 | 0.47827 | 0.98988 |
| C79 | 0.51884 | 0.23399 | 0.02985 |
| C80 | 0.76603 | 0.28481 | 0.02989 |
| C81 | 0.71476 | 0.48048 | 0.01316 |
| C82 | 0.48968 | 0.20356 | 0.90162 |
| C83 | 0.79645 | 0.28610 | 0.90123 |
| C84 | 0.71333 | 0.50971 | 0.89017 |
| C85 | 0.37161 | 0.14490 | 0.30980 |
| C86 | 0.85514 | 0.22677 | 0.30859 |
| C87 | 0.77313 | 0.62826 | 0.30885 |
| C88 | 0.34396 | 0.14715 | 0.16048 |
| C89 | 0.85291 | 0.19689 | 0.15906 |
| C90 | 0.80297 | 0.65551 | 0.15429 |
| C91 | 0.34639 | 0.18129 | 0.13275 |
| C92 | 0.81879 | 0.16515 | 0.13168 |
| C93 | 0.83465 | 0.65314 | 0.11899 |
| C94 | 0.37593 | 0.21206 | 0.24990 |
| C95 | 0.78802 | 0.16389 | 0.24930 |
| C96 | 0.83572 | 0.62366 | 0.24143 |
| Cl1 | 0.67712 | 0.44583 | 0.20520 |
| Cl2 | 0.31157 | 0.18418 | 0.94356 |
| Cl3 | 0.81592 | 0.12746 | 0.94238 |
| H1 | 0.46064 | 0.14425 | 0.61416 |
| H2 | 0.85579 | 0.31640 | 0.61371 |
| H3 | 0.68343 | 0.53914 | 0.60877 |
| H4 | 0.55533 | 0.51348 | 0.80786 |
| H5 | 0.48655 | 0.04189 | 0.80754 |
| H6 | 0.95817 | 0.44465 | 0.80773 |
| H7 | 0.50372 | 0.45159 | 0.80120 |
| H8 | 0.54842 | 0.05210 | 0.80132 |
| H9 | 0.94791 | 0.49628 | 0.80191 |
| H10 | 0.58307 | 0.52971 | 0.23475 |
| H11 | 0.47019 | 0.05342 | 0.23358 |
| H12 | 0.94660 | 0.41681 | 0.23383 |
| H13 | 0.64457 | 0.53876 | 0.26083 |
| H14 | 0.46098 | 0.10582 | 0.25924 |
| H15 | 0.89421 | 0.35521 | 0.25954 |
| H16 | 0.67970 | 0.63654 | 0.81646 |
| H17 | 0.36338 | 0.04312 | 0.81656 |
| H18 | 0.95694 | 0.32027 | 0.81580 |
| H19 | 0.61779 | 0.62764 | 0.78603 |
| H20 | 0.37243 | 0.99020 | 0.78683 |
| H21 | 0.00989 | 0.38224 | 0.78546 |
| H22 | 0.39821 | 0.36532 | 0.79098 |
| H23 | 0.63472 | 0.03281 | 0.79055 |
| H24 | 0.96709 | 0.60173 | 0.79281 |
| H25 | 0.40740 | 0.31290 | 0.81440 |
| H26 | 0.68716 | 0.09439 | 0.81377 |
| H27 | 0.90541 | 0.59245 | 0.81753 |
| H28 | 0.50516 | 0.37140 | 0.24861 |
| H29 | 0.62869 | 0.13370 | 0.24821 |
| H30 | 0.86622 | 0.49487 | 0.24975 |
| H31 | 0.49617 | 0.42441 | 0.23052 |
| H32 | 0.57562 | 0.07171 | 0.23071 |
| H33 | 0.92824 | 0.50389 | 0.23090 |
| H34 | 0.40749 | 0.27294 | 0.49376 |
| H35 | 0.72716 | 0.13451 | 0.49435 |
| H36 | 0.86537 | 0.59246 | 0.49398 |
| H37 | 0.49576 | 0.29691 | 0.81482 |
| H38 | 0.70317 | 0.19873 | 0.81605 |
| H39 | 0.80120 | 0.50440 | 0.82049 |
| H40 | 0.54381 | 0.29206 | 0.10534 |
| H41 | 0.70799 | 0.25162 | 0.10677 |
| H42 | 0.74754 | 0.45542 | 0.09888 |
| H43 | 0.53949 | 0.23158 | 0.16737 |
| H44 | 0.76843 | 0.30786 | 0.16726 |
| H45 | 0.48764 | 0.17743 | 0.94622 |
| H46 | 0.82257 | 0.31022 | 0.94540 |
| H47 | 0.68869 | 0.51082 | 0.92831 |
| H48 | 0.37020 | 0.11881 | 0.33065 |
| H49 | 0.88122 | 0.25147 | 0.32911 |
| H50 | 0.74850 | 0.62971 | 0.33229 |
| H51 | 0.32068 | 0.12320 | 0.05880 |
| H52 | 0.87685 | 0.19759 | 0.05682 |
| H53 | 0.80179 | 0.67850 | 0.05109 |
| H54 | 0.85792 | 0.67412 | 0.98623 |
| H55 | 0.37694 | 0.23774 | 0.20976 |
| H56 | 0.76234 | 0.13920 | 0.20938 |
| H57 | 0.85994 | 0.62187 | 0.19684 |
| N1 | 0.44021 | 0.42839 | 0.49344 |
| N3 | 0.57160 | 0.01177 | 0.49375 |
| N5 | 0.98818 | 0.55980 | 0.49420 |
| N7 | 0.70156 | 0.59403 | 0.58582 |
| N8 | 0.40578 | 0.10751 | 0.58453 |
| N9 | 0.89254 | 0.29829 | 0.58436 |
| N10 | 0.46251 | 0.30870 | 0.51550 |
| N11 | 0.69137 | 0.15374 | 0.51519 |

| N12 | 0.84612 | 0.53743 | 0.51799 |

## 4.3. W-A-Br COF unit cell

$a = b = 4.263$ nm, $c = 0.409$ nm

$\alpha = \beta = 90.00°, \gamma = 120°$

Table S3. Fractional coordinates of W-A-Br with symmetry P-1.

| Br1 | 0.67315 | 0.44214 | 0.21732 |
| Br2 | 0.30794 | 0.18448 | 0.93193 |
| Br3 | 0.81566 | 0.12360 | 0.93048 |
| C1 | 0.43303 | 0.14022 | 0.60062 |
| C2 | 0.85982 | 0.29285 | 0.60076 |
| C3 | 0.70711 | 0.56692 | 0.59855 |
| C4 | 0.47009 | 0.46412 | 0.49630 |
| C6 | 0.53589 | 0.00595 | 0.49637 |
| C8 | 0.99402 | 0.52992 | 0.49652 |
| C10 | 0.53119 | 0.50773 | 0.67069 |
| C11 | 0.49228 | 0.02347 | 0.67052 |
| C12 | 0.97656 | 0.46881 | 0.67063 |
| C13 | 0.50199 | 0.47269 | 0.66696 |
| C14 | 0.52731 | 0.02929 | 0.66691 |
| C15 | 0.97071 | 0.49800 | 0.66714 |
| C16 | 0.59543 | 0.57735 | 0.51898 |
| C17 | 0.42265 | 0.01815 | 0.51872 |
| C18 | 0.98192 | 0.40454 | 0.51838 |
| C19 | 0.60400 | 0.55279 | 0.37018 |
| C20 | 0.44715 | 0.05124 | 0.36927 |
| C21 | 0.94881 | 0.39595 | 0.36937 |
| C22 | 0.63859 | 0.55767 | 0.39009 |
| C23 | 0.44223 | 0.08093 | 0.38904 |
| C24 | 0.91910 | 0.36133 | 0.38932 |
| C25 | 0.66644 | 0.58774 | 0.55683 |
| C26 | 0.41219 | 0.07873 | 0.55607 |
| C27 | 0.92131 | 0.33350 | 0.55621 |
| C28 | 0.65824 | 0.61323 | 0.69074 |
| C29 | 0.38676 | 0.04507 | 0.69065 |
| C30 | 0.95499 | 0.34172 | 0.69030 |
| C31 | 0.62346 | 0.60791 | 0.67631 |
| C32 | 0.39213 | 0.01563 | 0.67658 |
| C33 | 0.98444 | 0.37652 | 0.67589 |
| C34 | 0.44604 | 0.39846 | 0.50396 |
| C35 | 0.60155 | 0.04751 | 0.50403 |
| C36 | 0.95242 | 0.55398 | 0.50483 |
| C37 | 0.42139 | 0.36649 | 0.66771 |
| C38 | 0.63355 | 0.05483 | 0.66728 |
| C39 | 0.94509 | 0.57859 | 0.66910 |
| C40 | 0.42631 | 0.33677 | 0.67555 |
| C41 | 0.66329 | 0.08945 | 0.67487 |
| C42 | 0.91045 | 0.57367 | 0.67732 |
| C43 | 0.45642 | 0.33779 | 0.52107 |
| C44 | 0.66228 | 0.11856 | 0.52059 |
| C45 | 0.88134 | 0.54360 | 0.52270 |
| C46 | 0.48187 | 0.37035 | 0.36908 |
| C47 | 0.62971 | 0.11145 | 0.36884 |
| C48 | 0.88846 | 0.51818 | 0.37010 |
| C49 | 0.47655 | 0.39986 | 0.35640 |
| C50 | 0.60018 | 0.07664 | 0.35645 |
| C51 | 0.92328 | 0.52350 | 0.35706 |
| C52 | 0.43476 | 0.27617 | 0.51763 |
| C53 | 0.72391 | 0.15856 | 0.51752 |
| C54 | 0.84144 | 0.56534 | 0.51798 |
| C55 | 0.43537 | 0.24256 | 0.54142 |
| C56 | 0.75751 | 0.19281 | 0.54120 |
| C57 | 0.80723 | 0.56479 | 0.54157 |
| C58 | 0.46396 | 0.24000 | 0.69789 |
| C59 | 0.76004 | 0.22393 | 0.69793 |
| C60 | 0.77614 | 0.53622 | 0.69774 |
| C61 | 0.46162 | 0.20519 | 0.72860 |
| C62 | 0.79484 | 0.25643 | 0.72852 |
| C63 | 0.74353 | 0.53833 | 0.72501 |
| C64 | 0.43135 | 0.17340 | 0.59597 |
| C65 | 0.82664 | 0.25799 | 0.59581 |
| C66 | 0.74200 | 0.56868 | 0.59448 |
| C67 | 0.40239 | 0.17578 | 0.44632 |
| C68 | 0.82427 | 0.22667 | 0.44578 |
| C69 | 0.77344 | 0.59774 | 0.44566 |
| C70 | 0.40462 | 0.21055 | 0.41687 |
| C71 | 0.78952 | 0.19411 | 0.41625 |
| C72 | 0.80594 | 0.59543 | 0.41463 |
| C73 | 0.49436 | 0.27082 | 0.83851 |
| C74 | 0.72923 | 0.22346 | 0.83912 |
| C75 | 0.77652 | 0.50585 | 0.83964 |
| C76 | 0.52124 | 0.26803 | 0.99856 |
| C77 | 0.73198 | 0.25311 | 0.99926 |
| C78 | 0.74666 | 0.47832 | 0.99155 |
| C79 | 0.51886 | 0.23394 | 0.03260 |
| C80 | 0.76605 | 0.28486 | 0.03281 |
| C81 | 0.71482 | 0.48057 | 0.01315 |
| C82 | 0.48968 | 0.20351 | 0.90425 |
| C83 | 0.79649 | 0.28616 | 0.90416 |
| C84 | 0.71334 | 0.50982 | 0.89185 |
| C85 | 0.37156 | 0.14486 | 0.31184 |
| C86 | 0.85520 | 0.22681 | 0.31082 |
| C87 | 0.77324 | 0.62851 | 0.31138 |
| C88 | 0.34385 | 0.14709 | 0.16269 |
| C89 | 0.85299 | 0.19689 | 0.16131 |
| C90 | 0.80307 | 0.65575 | 0.15632 |
| C91 | 0.34639 | 0.18127 | 0.13565 |
| C92 | 0.81883 | 0.16523 | 0.13432 |
| C93 | 0.83473 | 0.65334 | 0.12041 |
| C94 | 0.37592 | 0.21209 | 0.25178 |
| C95 | 0.78800 | 0.16391 | 0.25073 |
| C96 | 0.83580 | 0.62384 | 0.24285 |
| H1 | 0.46060 | 0.14416 | 0.61681 |
| H2 | 0.85588 | 0.31647 | 0.61698 |
| H3 | 0.68349 | 0.53931 | 0.61051 |
| H4 | 0.55535 | 0.51353 | 0.80847 |
| H5 | 0.48650 | 0.04185 | 0.80823 |
| H6 | 0.95821 | 0.44464 | 0.80842 |
| H7 | 0.50372 | 0.45159 | 0.80184 |
| H8 | 0.54842 | 0.05210 | 0.80191 |
| H9 | 0.94790 | 0.49627 | 0.80230 |
| H10 | 0.58313 | 0.52972 | 0.23559 |
| H11 | 0.47018 | 0.05342 | 0.23418 |
| H12 | 0.94661 | 0.41680 | 0.23452 |
| H13 | 0.64468 | 0.53885 | 0.26247 |
| H14 | 0.46095 | 0.10584 | 0.26053 |
| H15 | 0.89419 | 0.35516 | 0.26113 |
| H16 | 0.67970 | 0.63673 | 0.81724 |
| H17 | 0.36326 | 0.04300 | 0.81731 |
| H18 | 0.95705 | 0.32029 | 0.81706 |
| H19 | 0.61772 | 0.62775 | 0.78597 |
| H20 | 0.37233 | 0.99007 | 0.78684 |
| H21 | 0.01002 | 0.38229 | 0.78577 |
| H22 | 0.39820 | 0.36530 | 0.79114 |
| H23 | 0.63475 | 0.03283 | 0.79046 |
| H24 | 0.96710 | 0.60174 | 0.79274 |
| H25 | 0.40744 | 0.31288 | 0.81553 |
| H26 | 0.68719 | 0.09446 | 0.81448 |
| H27 | 0.90539 | 0.59246 | 0.81817 |
| H28 | 0.50530 | 0.37142 | 0.25012 |
| H29 | 0.62867 | 0.13382 | 0.24948 |
| H30 | 0.86610 | 0.49479 | 0.25055 |

| H31 | 0.49627 | 0.42444 | 0.23119 |
| H32 | 0.57560 | 0.07179 | 0.23127 |
| H33 | 0.92814 | 0.50381 | 0.23118 |
| H34 | 0.40756 | 0.27298 | 0.49559 |
| H35 | 0.72714 | 0.13456 | 0.49599 |
| H36 | 0.86545 | 0.59252 | 0.49368 |
| H37 | 0.49583 | 0.29693 | 0.81703 |
| H38 | 0.70314 | 0.19878 | 0.81810 |
| H39 | 0.80132 | 0.50459 | 0.82372 |
| H40 | 0.54387 | 0.29205 | 0.10804 |
| H41 | 0.70798 | 0.25168 | 0.10939 |
| H42 | 0.74763 | 0.45545 | 0.09950 |
| H43 | 0.53950 | 0.23152 | 0.17037 |
| H44 | 0.76845 | 0.30792 | 0.17045 |
| H45 | 0.48762 | 0.17736 | 0.94899 |
| H46 | 0.82262 | 0.31027 | 0.94864 |
| H47 | 0.68868 | 0.51097 | 0.92854 |
| H48 | 0.37011 | 0.11874 | 0.33288 |
| H49 | 0.88130 | 0.25150 | 0.33173 |
| H50 | 0.74861 | 0.62998 | 0.33529 |
| H51 | 0.32054 | 0.12306 | 0.06204 |
| H52 | 0.87702 | 0.19766 | 0.06016 |
| H53 | 0.80189 | 0.67875 | 0.05320 |
| H54 | 0.85800 | 0.67430 | 0.98718 |
| H55 | 0.37701 | 0.23783 | 0.21283 |
| H56 | 0.76227 | 0.13926 | 0.21168 |
| H57 | 0.86001 | 0.62202 | 0.19782 |
| N1 | 0.44023 | 0.42837 | 0.49346 |
| N3 | 0.57163 | 0.01182 | 0.49365 |
| N5 | 0.98813 | 0.55978 | 0.49391 |
| N7 | 0.70163 | 0.59422 | 0.58760 |
| N8 | 0.40571 | 0.10745 | 0.58606 |
| N9 | 0.89259 | 0.29830 | 0.58637 |
| N10 | 0.46261 | 0.30871 | 0.51708 |
| N11 | 0.69136 | 0.15384 | 0.51643 |
| N12 | 0.84608 | 0.53746 | 0.51917 |

## 4.4. W-A-I COF unit cell

$a = b = 4.263$ nm, $c = 0.408$ nm

$α = β = 90.00°, γ = 120°$

Table S4. Fractional coordinates of W-A-I with symmetry P-1.

| C1 | 0.43346 | 0.14046 | 0.59726 |
| C2 | 0.85957 | 0.29298 | 0.59642 |
| C3 | 0.70711 | 0.56680 | 0.59298 |
| C4 | 0.47008 | 0.46414 | 0.49659 |
| C6 | 0.53586 | 0.00592 | 0.49644 |
| C8 | 0.99406 | 0.52993 | 0.49648 |
| C10 | 0.53119 | 0.50771 | 0.67007 |
| C11 | 0.49231 | 0.02348 | 0.67021 |
| C12 | 0.97654 | 0.46883 | 0.67019 |
| C13 | 0.50200 | 0.47270 | 0.66663 |
| C14 | 0.52731 | 0.02927 | 0.66666 |
| C15 | 0.97072 | 0.49803 | 0.66667 |
| C16 | 0.59542 | 0.57726 | 0.51772 |
| C17 | 0.42278 | 0.01825 | 0.51808 |
| C18 | 0.98178 | 0.40454 | 0.51772 |
| C19 | 0.60393 | 0.55270 | 0.36842 |
| C20 | 0.44730 | 0.05128 | 0.36821 |
| C21 | 0.94874 | 0.39604 | 0.36804 |
| C22 | 0.63849 | 0.55755 | 0.38731 |
| C23 | 0.44248 | 0.08102 | 0.38751 |
| C24 | 0.91900 | 0.36147 | 0.38712 |
| C25 | 0.66639 | 0.58758 | 0.55354 |
| C26 | 0.41253 | 0.07895 | 0.55447 |
| C27 | 0.92108 | 0.33358 | 0.55375 |
| C28 | 0.65824 | 0.61305 | 0.68830 |
| C29 | 0.38707 | 0.04535 | 0.68960 |
| C30 | 0.95468 | 0.34171 | 0.68874 |
| C31 | 0.62349 | 0.60777 | 0.67490 |
| C32 | 0.39233 | 0.01586 | 0.67600 |
| C33 | 0.98417 | 0.37647 | 0.67524 |
| C34 | 0.44599 | 0.39849 | 0.50382 |
| C35 | 0.60152 | 0.04741 | 0.50373 |
| C36 | 0.95256 | 0.55410 | 0.50446 |
| C37 | 0.42136 | 0.36651 | 0.66717 |
| C38 | 0.63352 | 0.05475 | 0.66670 |
| C39 | 0.94526 | 0.57873 | 0.66839 |
| C40 | 0.42628 | 0.33680 | 0.67432 |
| C41 | 0.66324 | 0.08936 | 0.67401 |
| C42 | 0.91065 | 0.57387 | 0.67624 |
| C43 | 0.45637 | 0.33784 | 0.51966 |
| C44 | 0.66222 | 0.11845 | 0.51979 |
| C45 | 0.88153 | 0.54384 | 0.52179 |
| C46 | 0.48178 | 0.37041 | 0.36792 |
| C47 | 0.62966 | 0.11130 | 0.36807 |
| C48 | 0.88863 | 0.51842 | 0.36934 |
| C49 | 0.47646 | 0.39990 | 0.35585 |
| C50 | 0.60015 | 0.07651 | 0.35594 |
| C51 | 0.92342 | 0.52368 | 0.35653 |
| C52 | 0.43487 | 0.27625 | 0.51650 |
| C53 | 0.72382 | 0.15855 | 0.51701 |
| C54 | 0.84158 | 0.56551 | 0.51683 |
| C55 | 0.43564 | 0.24271 | 0.53870 |
| C56 | 0.75735 | 0.19287 | 0.53894 |
| C57 | 0.80733 | 0.56488 | 0.53928 |
| C58 | 0.46426 | 0.24020 | 0.69443 |
| C59 | 0.75986 | 0.22400 | 0.69469 |
| C60 | 0.77623 | 0.53631 | 0.69488 |
| C61 | 0.46200 | 0.20543 | 0.72460 |
| C62 | 0.79462 | 0.25654 | 0.72430 |
| C63 | 0.74361 | 0.53837 | 0.72109 |
| C64 | 0.43179 | 0.17363 | 0.59193 |
| C65 | 0.82641 | 0.25814 | 0.59123 |
| C66 | 0.74202 | 0.56863 | 0.58974 |
| C67 | 0.40285 | 0.17599 | 0.44214 |
| C68 | 0.82405 | 0.22684 | 0.44137 |
| C69 | 0.77344 | 0.59767 | 0.44116 |
| C70 | 0.40498 | 0.21070 | 0.41363 |
| C71 | 0.78935 | 0.19424 | 0.41334 |
| C72 | 0.80598 | 0.59544 | 0.41150 |
| C73 | 0.49460 | 0.27102 | 0.83521 |
| C74 | 0.72906 | 0.22350 | 0.83616 |
| C75 | 0.77657 | 0.50597 | 0.83731 |
| C76 | 0.52150 | 0.26828 | 0.99501 |
| C77 | 0.73180 | 0.25313 | 0.99587 |
| C78 | 0.74667 | 0.47842 | 0.98848 |
| C79 | 0.51919 | 0.23424 | 0.02860 |
| C80 | 0.76582 | 0.28491 | 0.02867 |
| C81 | 0.71471 | 0.48051 | 0.00951 |
| C82 | 0.49008 | 0.20381 | 0.90002 |
| C83 | 0.79624 | 0.28624 | 0.89959 |
| C84 | 0.71340 | 0.50985 | 0.88738 |
| C85 | 0.37210 | 0.14511 | 0.30680 |
| C86 | 0.85491 | 0.22700 | 0.30537 |
| C87 | 0.77318 | 0.62835 | 0.30579 |
| C88 | 0.34431 | 0.14728 | 0.15887 |
| C89 | 0.85273 | 0.19704 | 0.15734 |
| C90 | 0.80300 | 0.65558 | 0.15106 |
| C91 | 0.34659 | 0.18138 | 0.13323 |
| C92 | 0.81865 | 0.16519 | 0.13228 |
| C93 | 0.83471 | 0.65326 | 0.11664 |
| C94 | 0.37620 | 0.21217 | 0.24968 |
| C95 | 0.78787 | 0.16400 | 0.24931 |
| C96 | 0.83583 | 0.62386 | 0.24008 |
| H1 | 0.46100 | 0.14438 | 0.61338 |
| H2 | 0.85564 | 0.31660 | 0.61253 |
| H3 | 0.68353 | 0.53918 | 0.60332 |
| H4 | 0.55536 | 0.51348 | 0.80728 |
| H5 | 0.48655 | 0.04188 | 0.80761 |
| H6 | 0.95815 | 0.44467 | 0.80759 |
| H7 | 0.50375 | 0.45161 | 0.80122 |
| H8 | 0.54841 | 0.05208 | 0.80145 |
| H9 | 0.94791 | 0.49630 | 0.80148 |
| H10 | 0.58303 | 0.52965 | 0.23424 |
| H11 | 0.47028 | 0.05337 | 0.23317 |
| H12 | 0.94664 | 0.41694 | 0.23336 |
| H13 | 0.64450 | 0.53869 | 0.25940 |
| H14 | 0.46123 | 0.10586 | 0.25863 |
| H15 | 0.89416 | 0.35540 | 0.25838 |
| H16 | 0.67972 | 0.63651 | 0.81462 |
| H17 | 0.36363 | 0.04338 | 0.81628 |
| H18 | 0.95664 | 0.32024 | 0.81544 |
| H19 | 0.61781 | 0.62760 | 0.78537 |
| H20 | 0.37252 | 0.99037 | 0.78683 |
| H21 | 0.00967 | 0.38215 | 0.78595 |
| H22 | 0.39820 | 0.36529 | 0.79097 |
| H23 | 0.63474 | 0.03278 | 0.79003 |
| H24 | 0.96725 | 0.60185 | 0.79218 |
| H25 | 0.40741 | 0.31290 | 0.81387 |
| H26 | 0.68714 | 0.09436 | 0.81331 |
| H27 | 0.90563 | 0.59269 | 0.81678 |
| H28 | 0.50519 | 0.37150 | 0.24870 |
| H29 | 0.62861 | 0.13364 | 0.24861 |
| H30 | 0.86628 | 0.49506 | 0.24975 |
| H31 | 0.49616 | 0.42447 | 0.23085 |
| H32 | 0.57558 | 0.07165 | 0.23084 |
| H33 | 0.92824 | 0.50398 | 0.23073 |
| H34 | 0.40762 | 0.27294 | 0.49654 |
| H35 | 0.72714 | 0.13461 | 0.49729 |
| H36 | 0.86554 | 0.59271 | 0.49299 |
| H37 | 0.49599 | 0.29708 | 0.81406 |
| H38 | 0.70302 | 0.19880 | 0.81572 |
| H39 | 0.80136 | 0.50472 | 0.82220 |
| H40 | 0.54408 | 0.29229 | 0.10464 |
| H41 | 0.70781 | 0.25168 | 0.10624 |
| H42 | 0.74781 | 0.45562 | 0.09579 |
| H43 | 0.53985 | 0.23185 | 0.16612 |
| H44 | 0.76820 | 0.30796 | 0.16598 |
| H45 | 0.48808 | 0.17770 | 0.94436 |
| H46 | 0.82234 | 0.31038 | 0.94338 |
| H47 | 0.68881 | 0.51116 | 0.92188 |
| H48 | 0.37073 | 0.11903 | 0.32647 |
| H49 | 0.88098 | 0.25173 | 0.32453 |
| H50 | 0.74852 | 0.62974 | 0.32857 |
| H51 | 0.32112 | 0.12315 | 0.05858 |
| H52 | 0.87685 | 0.19800 | 0.05643 |
| H53 | 0.80177 | 0.67851 | 0.04701 |
| H54 | 0.85796 | 0.67423 | 0.98371 |
| H55 | 0.37739 | 0.23797 | 0.21297 |
| H56 | 0.76208 | 0.13937 | 0.21301 |
| H57 | 0.86008 | 0.62211 | 0.19612 |
| I1 | 0.66838 | 0.43786 | 0.22862 |
| I2 | 0.30377 | 0.18485 | 0.91586 |
| I3 | 0.81519 | 0.11890 | 0.91510 |
| N1 | 0.44021 | 0.42841 | 0.49403 |
| N3 | 0.57158 | 0.01174 | 0.49382 |
| N5 | 0.98823 | 0.55982 | 0.49392 |
| N7 | 0.70158 | 0.59407 | 0.58353 |
| N8 | 0.40613 | 0.10772 | 0.58366 |
| N9 | 0.89231 | 0.29840 | 0.58286 |
| N10 | 0.46263 | 0.30881 | 0.51502 |
| N11 | 0.69125 | 0.15373 | 0.51526 |
| N12 | 0.84628 | 0.53769 | 0.51821 |

## 5. Simulations of different halogen positions

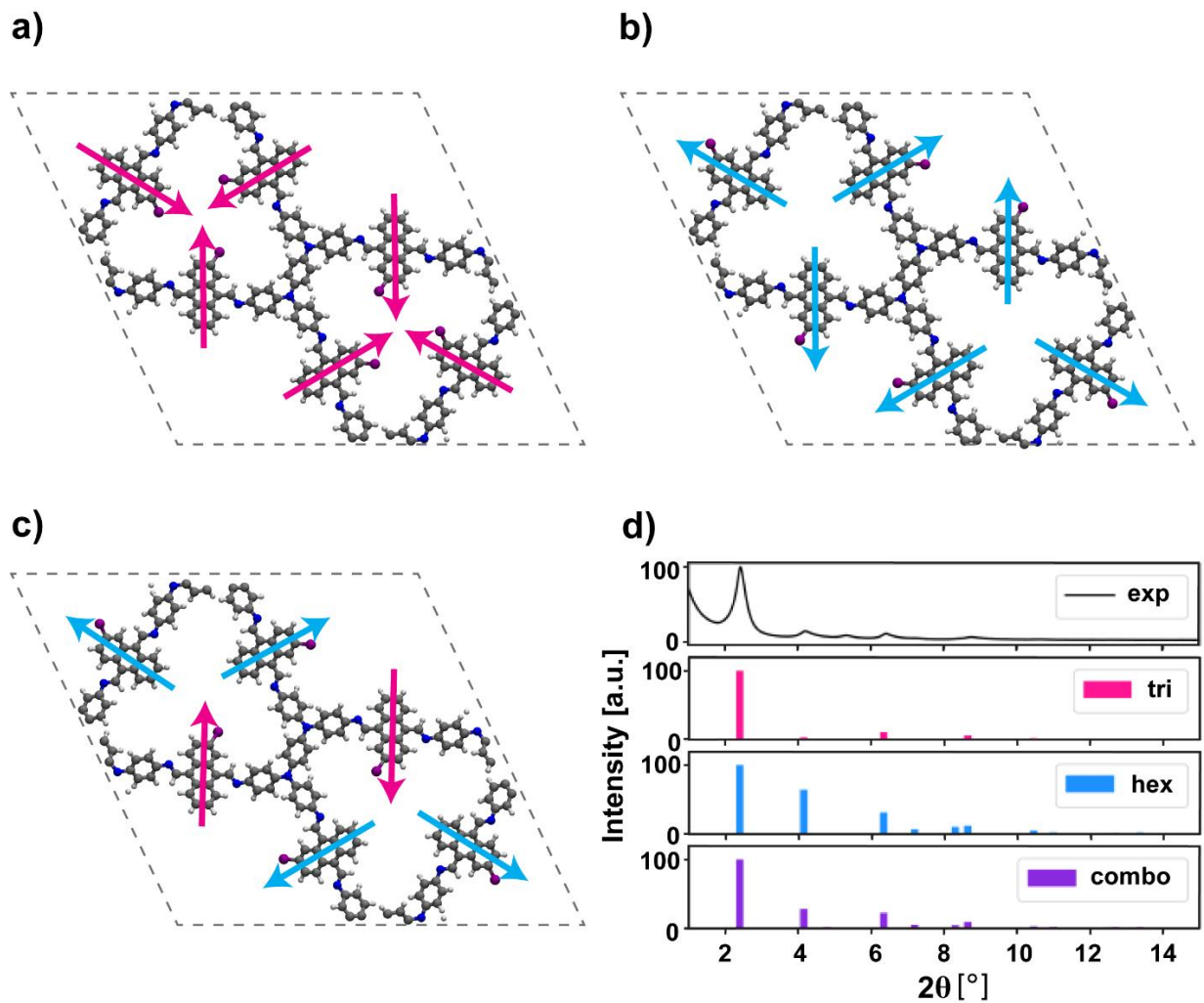

Figure S21. Simulated structures of W-A-Br COF when halogens atoms are facing (a) exclusively into the trigonal pore (pink arrows), (b) facing exclusively into the hexagonal pore (blue arrows) and (c) with a combination of two halogens facing trigonal pores and four halogens facing the hexagonal pores per unit cell. (d) Experimental PXRD pattern for W-A-Br COF (black) and simulated PXRD patterns for simulated structures of all halogens facing into the trigonal pores (pink) and into the hexagonal pores (blue), as well as the simulated PXRD pattern for the combinational structure with halogens facing both types of pores (purple).

## 6. TEM images

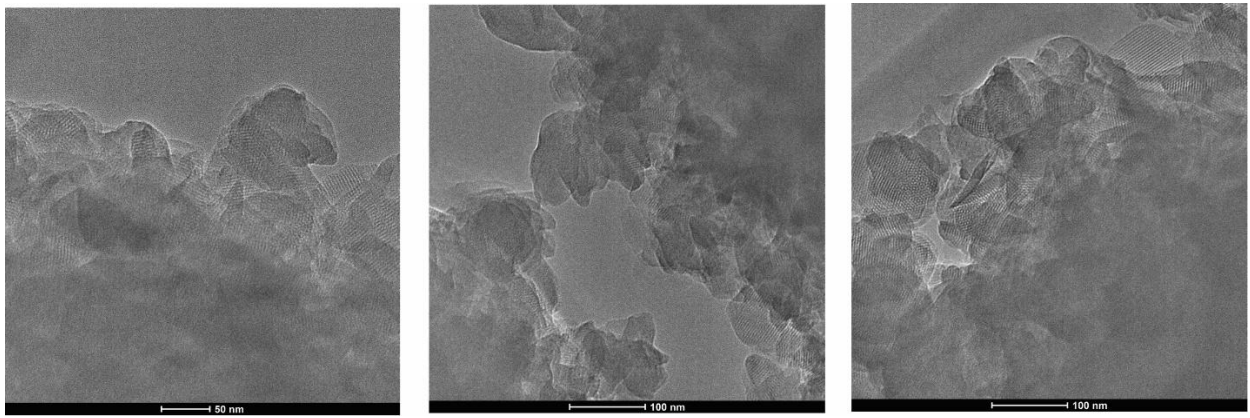

Figure S22. TEM images of W-A-H COF.

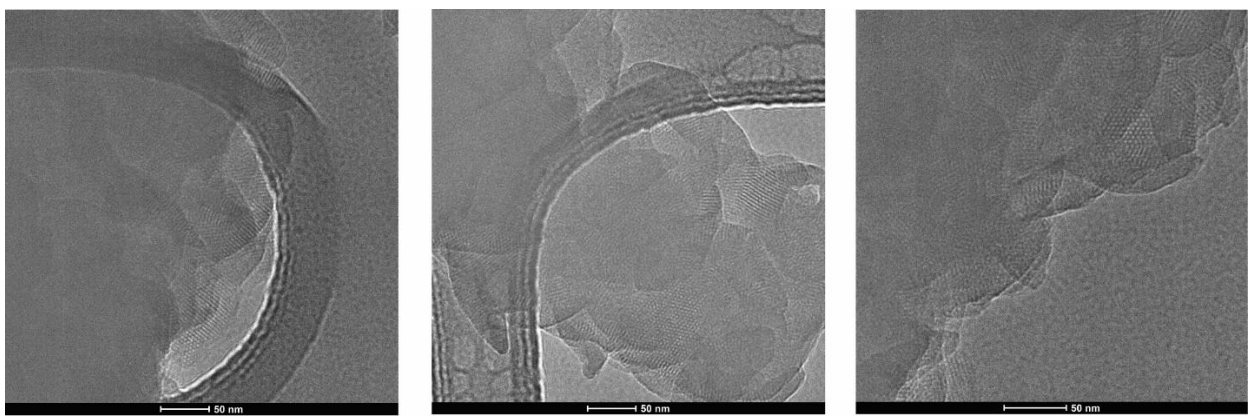

Figure S23. TEM images of W-A-Cl COF.

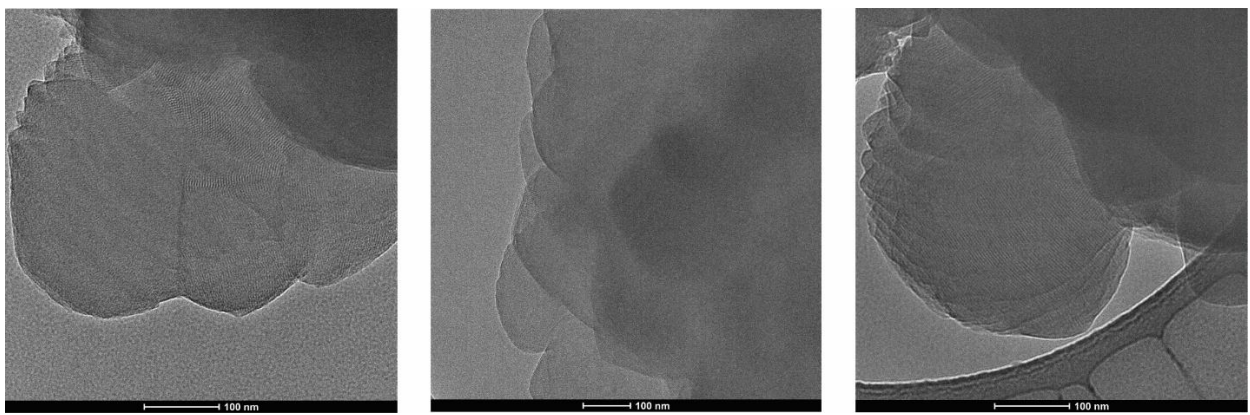

Figure S24. TEM images of W-A-Br COF.

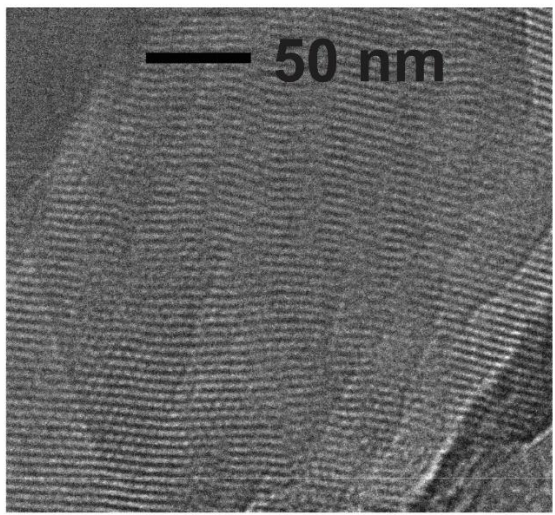
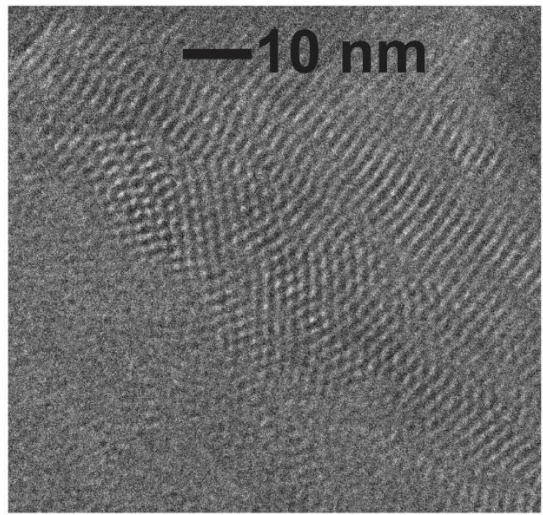

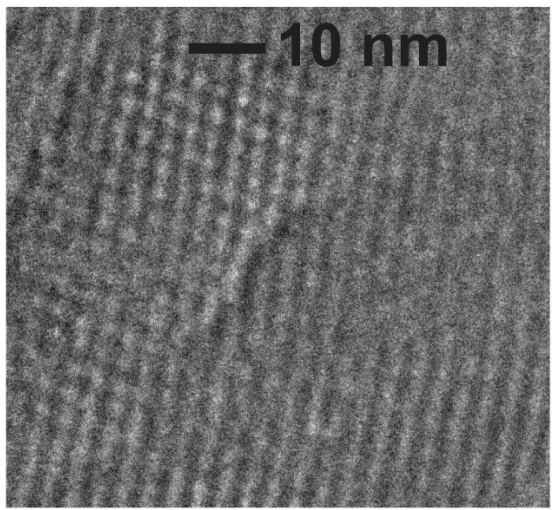
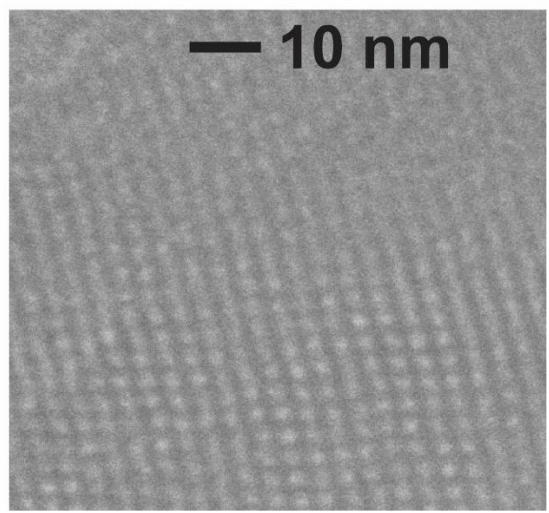

Figure S25. TEM images of W-A-Br COF showing (a) dominant Kagome structure with 120° angle and (b) distorted crystalline structure and revealing a lattice angles of approximately 100°.

## 7. ESP calculations

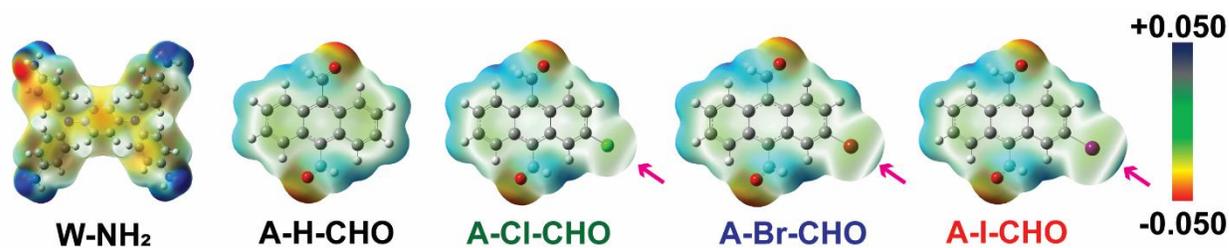

Figure S26. ESP maps superimposed on the M06-2X/def2-SVPP optimized geometry of W-NH$_2$ and anthracene-based linkers A-X-CHO (X = H, Cl, Br, I). The ESP values in atomic units (a.u.) are mapped onto the total electron density surface with an isovalue of 0.001 highlighting electron-rich (red) and electron-deficient (blue) regions. Introduction of Cl, Br, and I induces progressively larger σ-holes (positive ESP regions) opposite the C-X bonds (highlighted with pink arrows).

## 8. SEM images

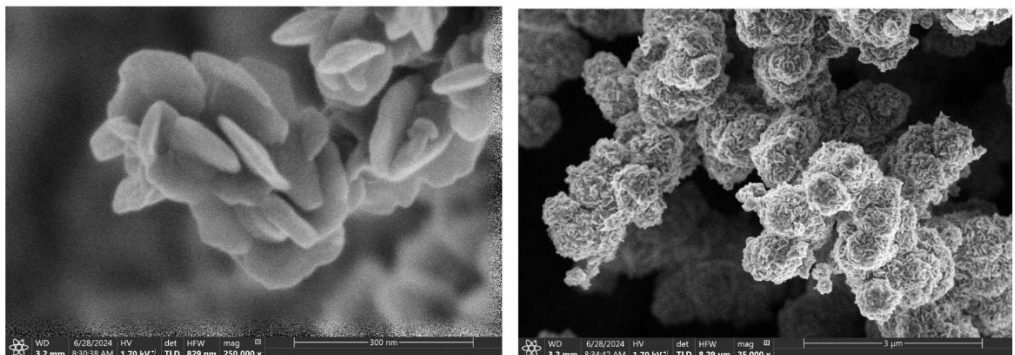
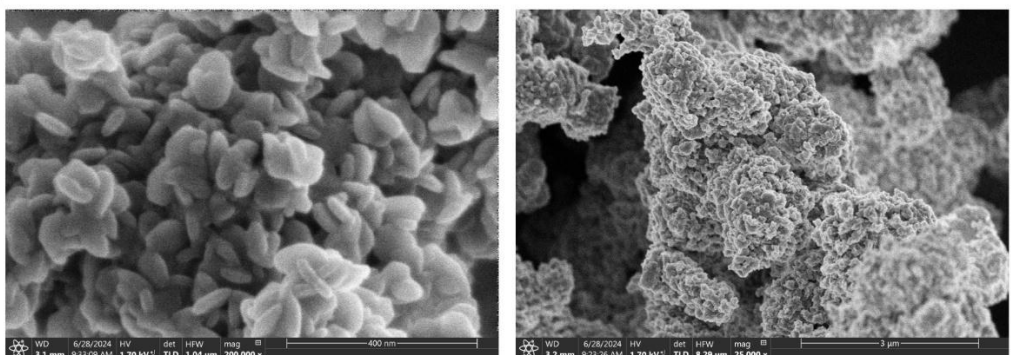
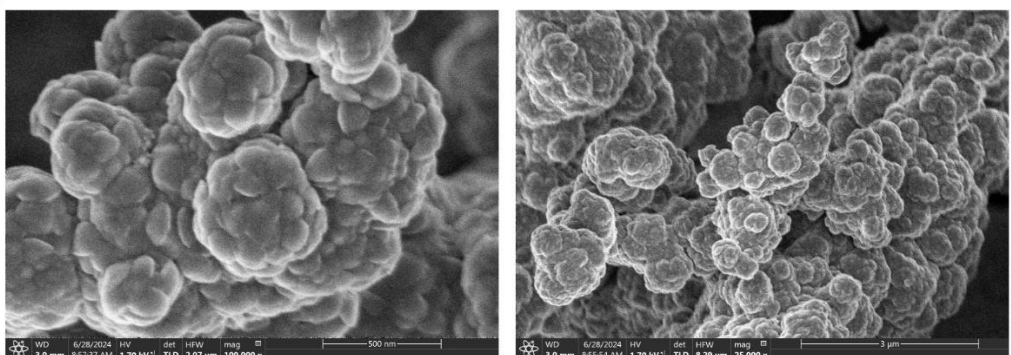
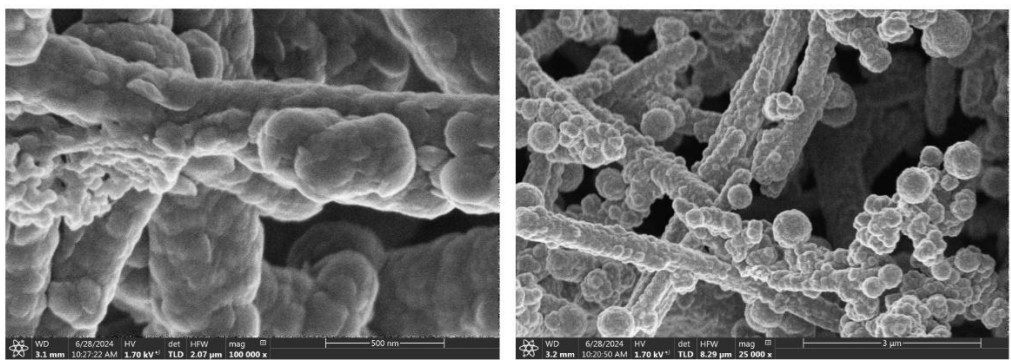

Figure S27. SEM images of (a) W-A-H, (b) W-A-Cl, (c) W-A-Br, (d) W-A-I.

## 9. Sorption and porosity parameters

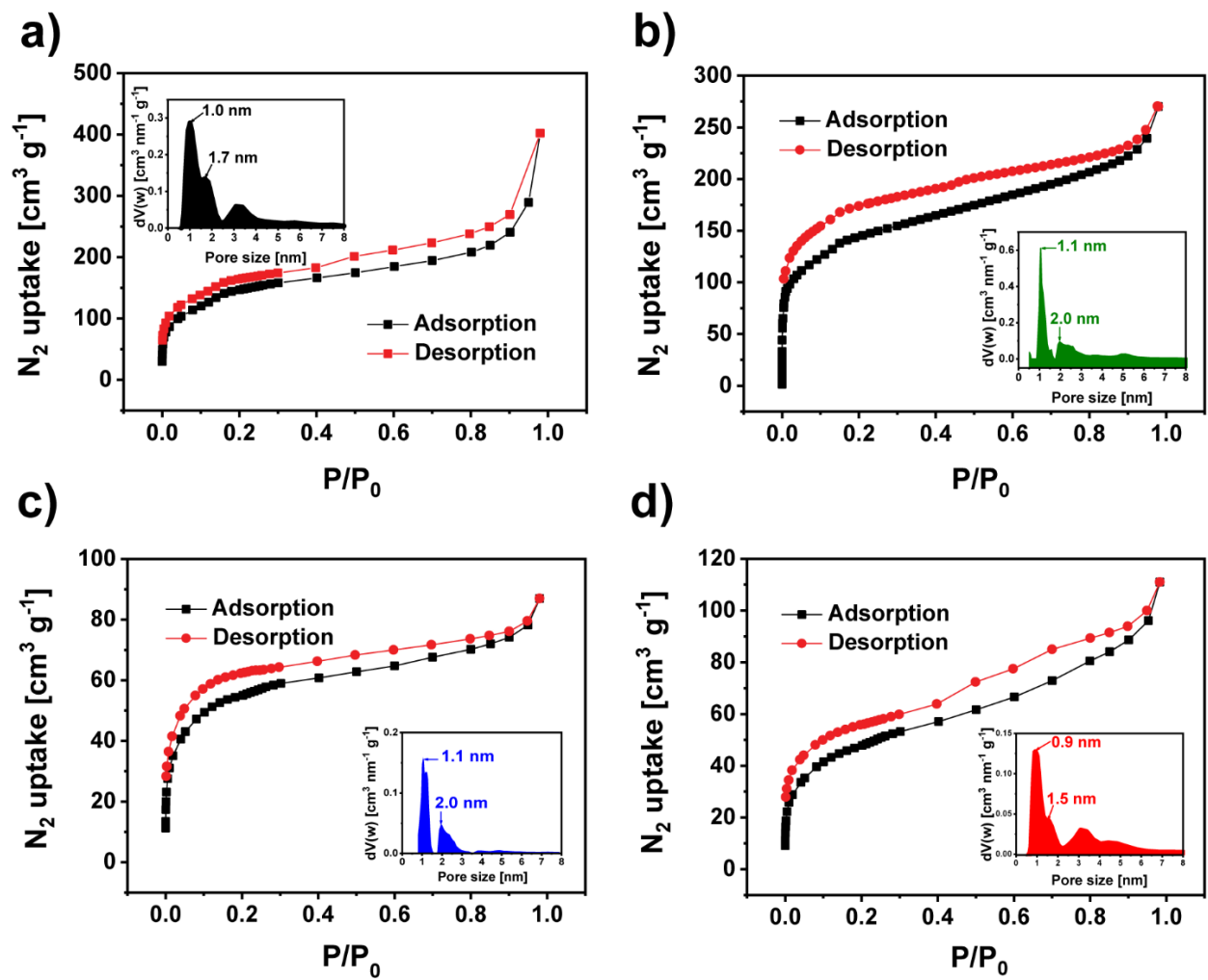

Figure S28. Nitrogen sorption isotherms and pore size distributions (insets) of the COFs (a) W-A-H, (b) W-A-Cl, (c) W-A-Br, (d) W-A-I.

Table S5. Theoretical (*Zeo++[21] and **PoreBlazer v4.0[22]) and experimental geometric porosity parameters of the COFs (probe radius corresponds to $N_2$; 0.18 nm). *** Experimental pore volume was calculated based on the average uptake from the desorption branch in the relative pressure range $P/P_0$ = 0.60-0.80.

| COF | Pore | *Maximum pore diameter ($m_{pd}$) (nm) | *Pore window size ($p_{ws}$) (nm) | **Theoretical pore volume ($V_{pt}$) (cm³ g⁻¹) | Experimental pore volume*** ($V_{et}$) (cm³ g⁻¹) | *Unit cell density (g cm⁻³) | Experimental BET surface area (m² g⁻¹) | Theoretical network-accessible surface area (m² g⁻¹) |
|---|---|---|---|---|---|---|---|---|
65

| | | | | | | | | |
|---|---|---|---|---|---|---|---|---|
| W-A-H | Trigonal | 0.73 | 0.70 | 0.364 | 0.35 | 0.67 | 548 | 897 |
| | Hexagonal | 2.08 | 2.07 | | | | | |
| W-A-Cl | Trigonal | 0.68 | 0.62 | 0.317 | 0.33 | 0.74 | 490 | 806 |
| | Hexagonal | 1.88 | 1.86 | | | | | |
| W-A-Br | Trigonal | 0.69 | 0.63 | 0.293 | 0.11 | 0.81 | 187 | 674 |
| | Hexagonal | 1.87 | 1.84 | | | | | |
| W-A-I | Trigonal | 0.66 | 0.61 | 0.260 | 0.13 | 0.90 | 170 | 653 |
| | Hexagonal | 1.84 | 1.80 | | | | | |

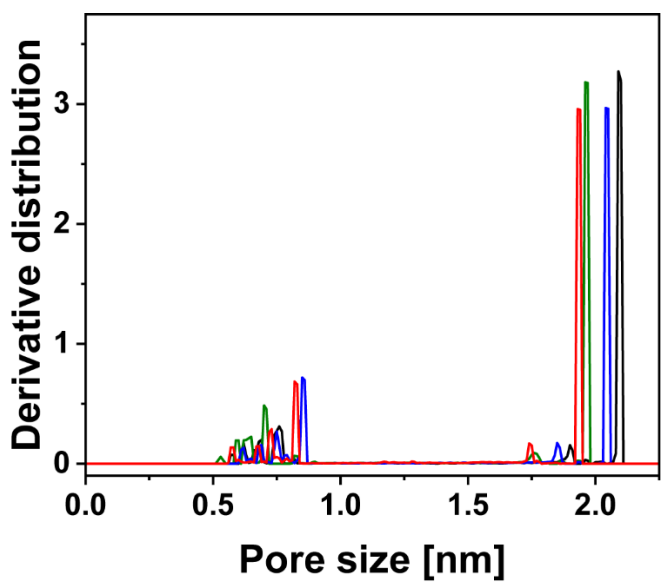

Figure S29. Simulated pore size distribution of the COFs W-A-H (black), W-A-Cl (green), W-A-Br (blue) and W-A-I (red) using Zeo++.[21]

## 10. FT-IR analysis of COFs

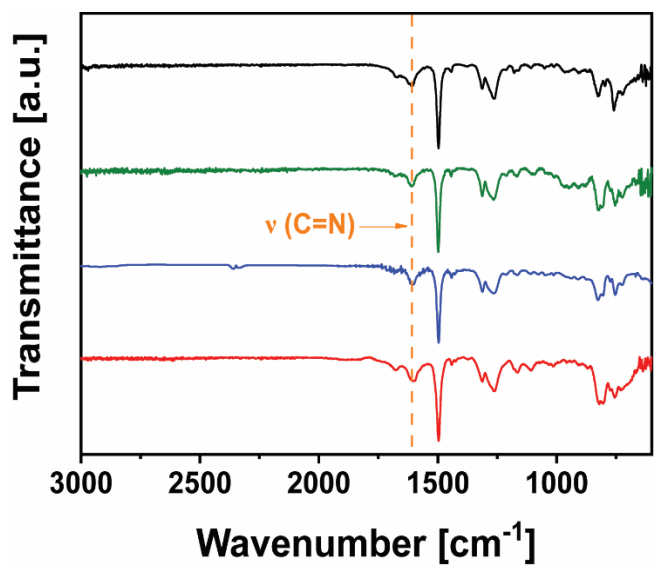

Figure S30. FT-IR spectra of the COFs W-A-H (black), W-A-Cl (green), W-A-Br (blue) and W-A-I (red).

## 11. Solid-state $^{13}$C NMR of COFs

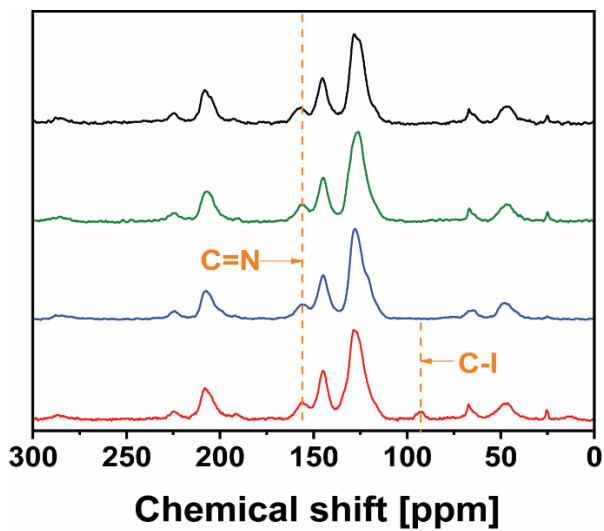

Figure S31. $^{13}$C NMR spectra of the COFs W-A-H (black), W-A-Cl (green), W-A-Br (blue) and W-A-I (red).

## 12. TGA analysis

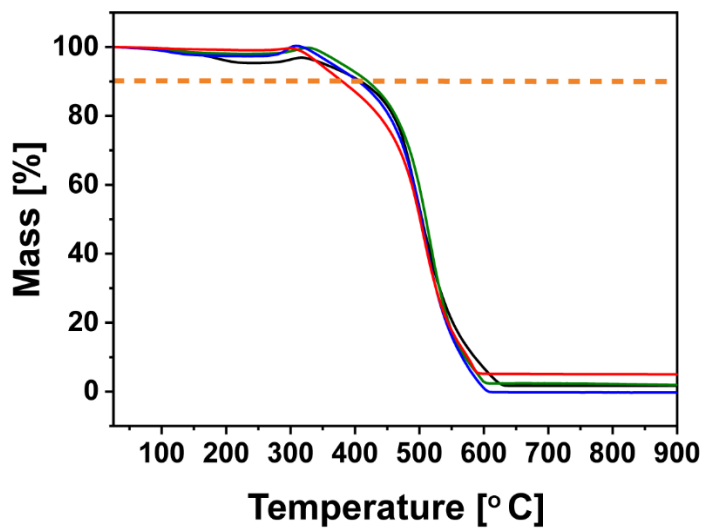

Figure S32. TGA analysis of the COFs W-A-H (black), W-A-Cl (green), W-A-Br (blue) and W-A-I (red), with decomposition temperatures determined at 10% mass loss.

## 13. Experimental optical properties

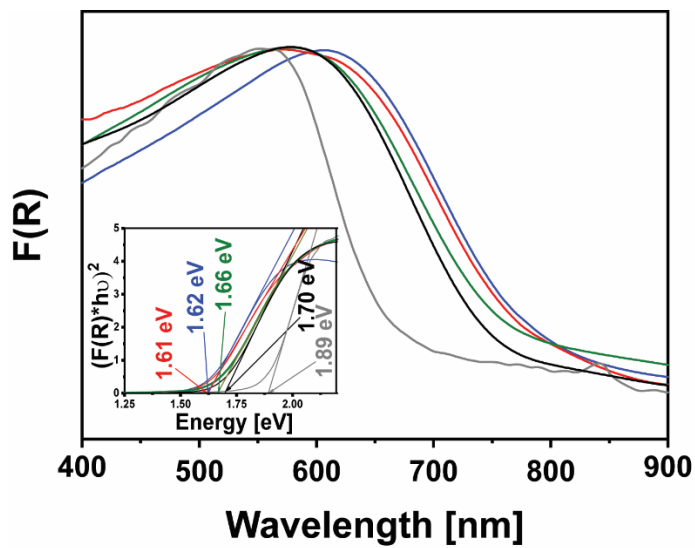

Figure S33. Optical absorption spectra (Kubelka Munk function F(R)) and Tauc plots of the COFs W-TA (grey), W-A-H (black), W-A-Cl (green), W-A-Br (blue), W-A-I (red).

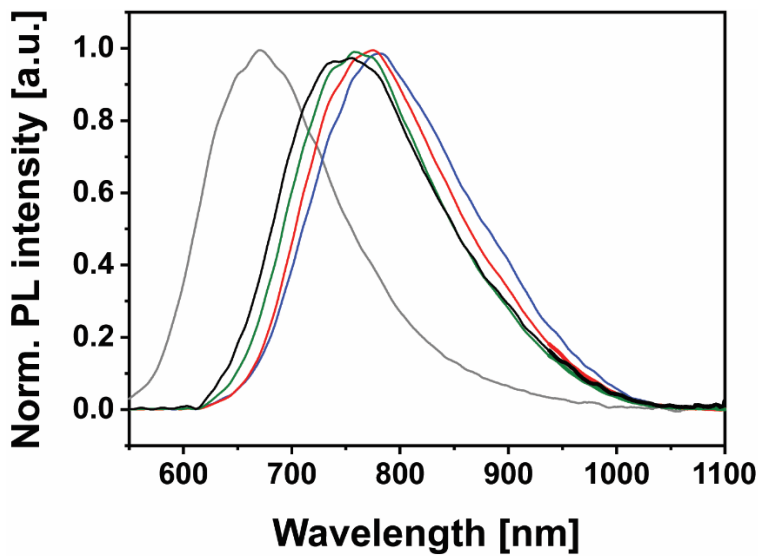

Figure S34. PL of the COFs W-TA (grey), W-A-H (black), W-A-Cl (green), W-A-Br (blue), W-A-I (red).

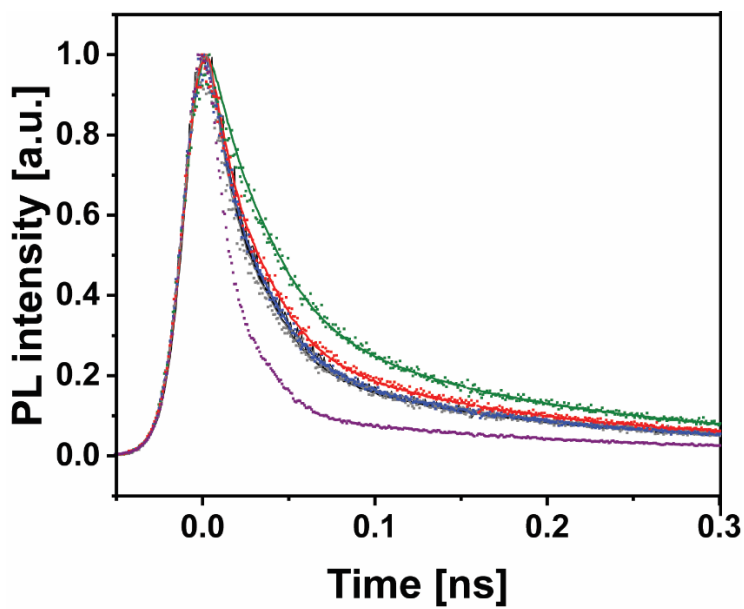

Figure S35. Time-resolved PL lifetime measurements of the COFs W-TA (grey), W-A-H (black), W-A-Cl (green), W-A-Br (blue), W-A-I (red). The purple curve represents the instrument response function (IRF).

## 14. Calculated optical properties

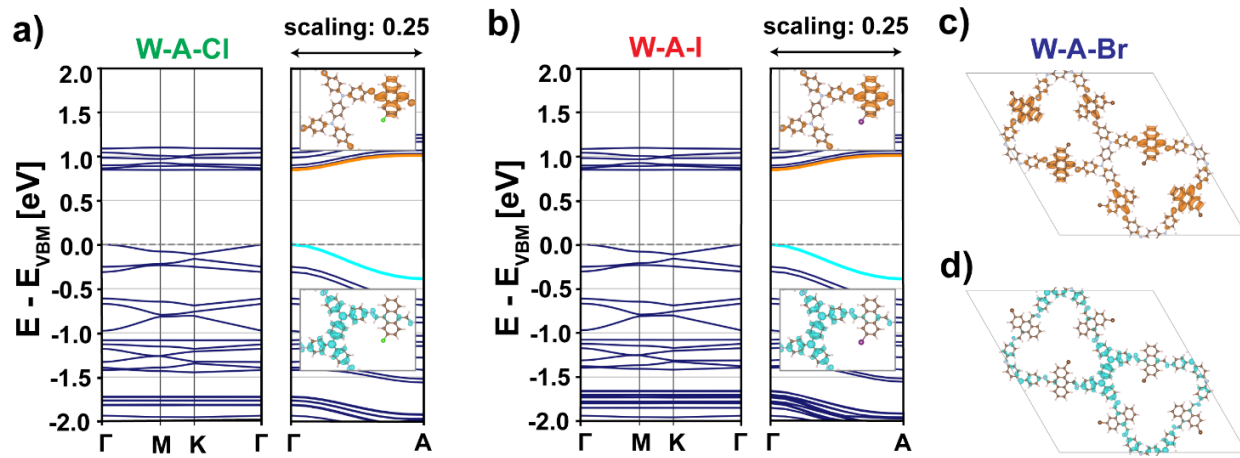

Figure S36. Electronic band structures of (a) W-A-Cl and (b) W-A-I COFs with insets of the partial charge densities of LUMO (orange) and HOMO (blue) bands. Partial charge densities of (c) LUMO band and (d) HOMO band of W-A-Br.

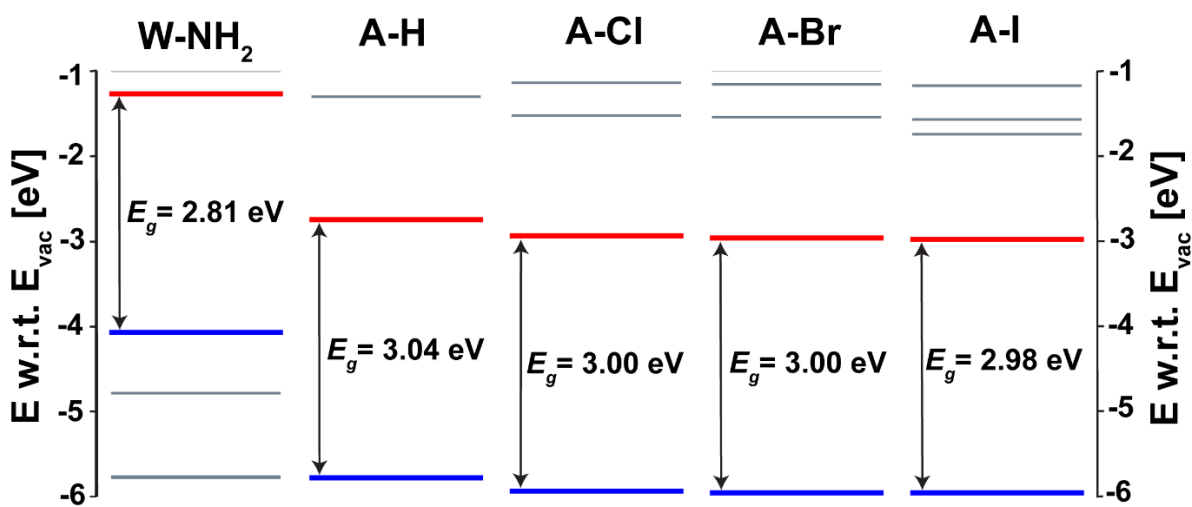

Figure S37. Kohn-Sham eigenvalues of W-NH$_2$, and (non-)halogenated anthracene (A-X, X = H, Cl, Br, I) building blocks with highlighted HOMO (blue) and LUMO (red).

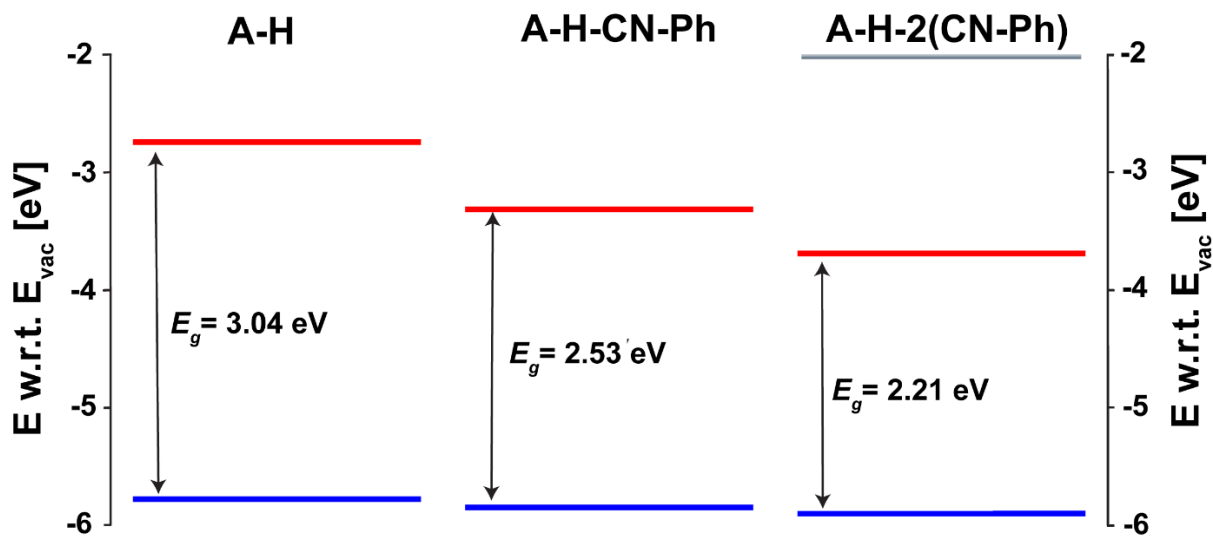

Figure S38. Kohn-Sham eigenvalues of the systematic extension of anthracene (A-H) toward a combined anthracene-Wurster fragment of W-A-H (A-H-2(CN-Ph)).

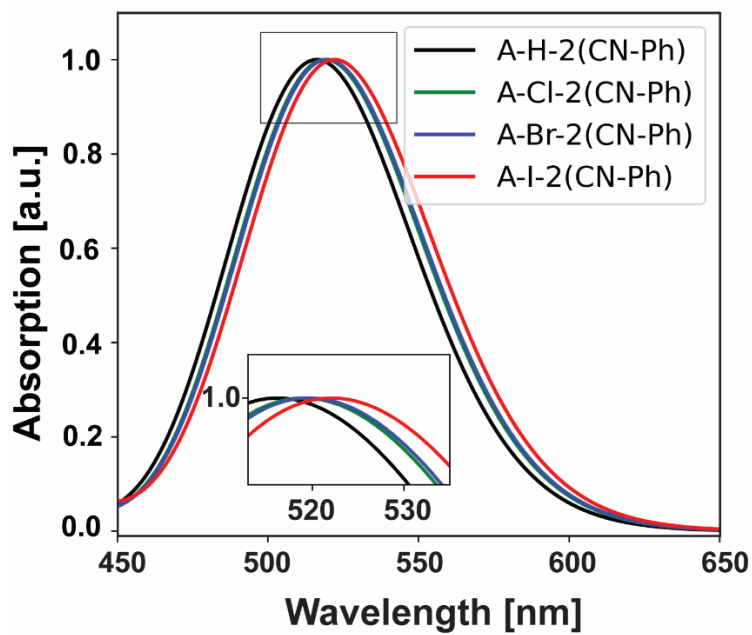

Figure S39. Theoretical absorption spectrum of (non-)halogenated extensions of anthracene towards a combined Wurster-anthracene fragment; inset shows close-up of the absorption maxima.

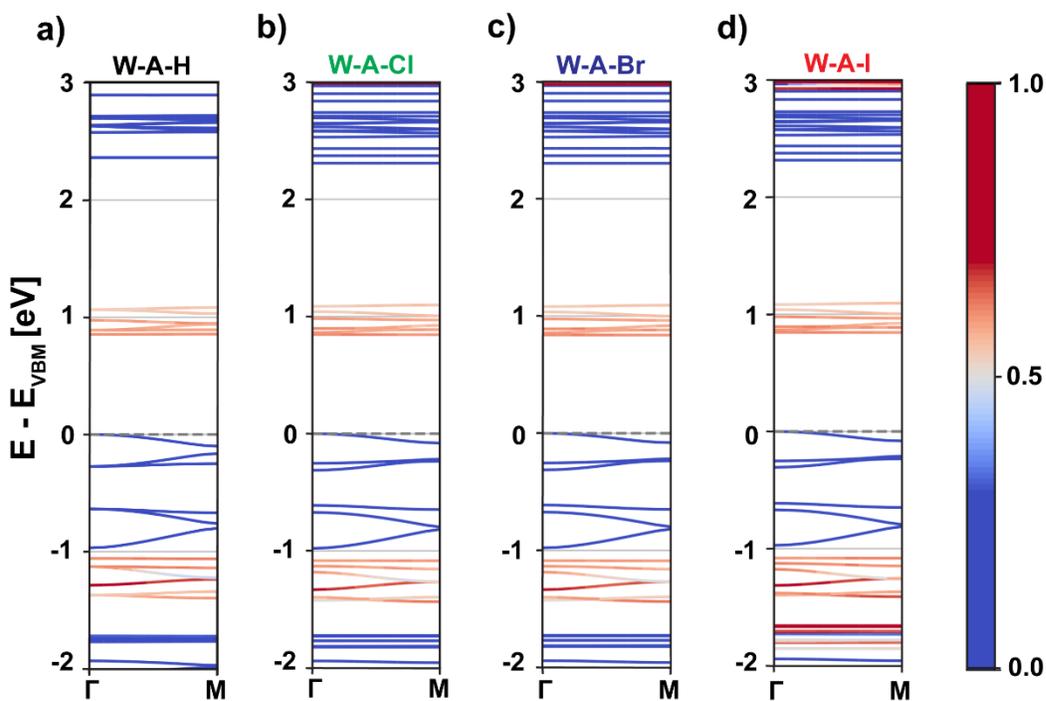

Figure S40. Projected band structure of (a) W-A-H, (b) W-A-Cl, (c) W-A-Br, and (d) W-A-I orbitals located on the (halogenated) anthracene atoms. Dark red color indicates a band with exclusive contribution from (halogenated) anthracene, whereas dark blue corresponds to a band formed by Wurster orbitals.